\documentclass[sigconf, screen, review=false]{acmart}
\usepackage{algorithm}
\usepackage{algpseudocode}
\usepackage{balance}
\usepackage{tikz}
\usetikzlibrary{positioning}
\usetikzlibrary{shapes.geometric}
\usetikzlibrary{shapes.misc}
\usetikzlibrary{calc}

\usepackage{makecell}
\usepackage{booktabs}
\usepackage{multirow}
\usepackage{array}
\usepackage{enumitem}
\usepackage{tcolorbox}

\newcommand{\THISWORK}{{\fontfamily{lmss}\selectfont
MoE-Lens}}
\AtBeginDocument{%
  }

\setcopyright{none} 
\copyrightyear{2025}
\acmYear{2025}
\acmDOI{XXXXXXX.XXXXXXX}

\author{Yichao Yuan}
\affiliation{%
  \institution{University of Michigan}
  \city{Ann Arbor}
  \state{Michigan}
  \country{USA}
}
\email{yichaoy@umich.edu}

\author{Lin Ma}
\affiliation{%
  \institution{University of Michigan}
  \city{Ann Arbor}
  \state{Michigan}
  \country{USA}
}
\email{linmacse@umich.edu}

\author{Nishil Talati}
\affiliation{%
  \institution{University of Michigan}
  \city{Ann Arbor}
  \state{Michigan}
  \country{USA}
}
\email{talatin@umich.edu}

\begin{document}
% \begin{textblock}{16}(3,1)
% {\normalsize \normalfont \textit{Authors’ version; .} }
% \end{textblock}

\title{\THISWORK: Towards the Hardware Limit of High-Throughput MoE LLM Serving Under Resource Constraints}
% \subtitle{\normalsize{MICRO 2025 Submission
    % \textbf{\#302} -- Confidential Draft -- Do NOT Distribute!!}}
%\author{\normalsize{ISCA 2025 Submission
 %   \textbf{\#NaN} -- Confidential Draft -- Do NOT Distribute!!}}



%%%%%% -- PAPER CONTENT STARTS-- %%%%%%%%

\begin{abstract}
Mixture of Experts (MoE) LLMs, characterized by their sparse activation patterns, offer a promising approach to scaling language models while avoiding proportionally increasing the inference cost.
However, their large parameter sizes present deployment challenges in resource-constrained environments with limited GPU memory capacity, as GPU memory is often insufficient to accommodate the full set of model weights.
Consequently, typical deployments rely on CPU-GPU hybrid execution: the GPU handles compute-intensive GEMM operations, while the CPU processes the relatively lightweight attention mechanism.
This setup introduces a key challenge: \textit{how to effectively optimize resource utilization across CPU and GPU?}
Prior work has designed system optimizations based on performance models with limited scope. 
Specifically, such models do not capture the complex interactions between hardware properties and system execution mechanisms.
Therefore, previous approaches neither identify nor achieve the hardware limit.

This paper presents \THISWORK, a high-throughput MoE LLM inference system designed through holistic performance modeling for resource-constrained environments. 
Our performance model thoroughly analyzes various fundamental system components, including CPU memory capacity, GPU compute power, and workload characteristics, to understand the theoretical performance
upper bound of MoE inference.
Furthermore, it captures the system execution mechanisms, including workload scheduling and the effects of paged KV cache, to identify the key hardware bottlenecks and accurately predict the achievable throughput. 
Informed by our performance model, \THISWORK\ introduces an inference system featuring a resource-aware scheduler for prefill and decode phases, an execution engine that overlaps their computation, a data transfer mechanism for model weights, and an optimized CPU-based attention implementation. 
Evaluated on diverse MoE models and datasets, \THISWORK\ outperforms the state-of-the-art MoE-Lightening by 4.6$\times$ on average (up to 25.5$\times$), with our theoretical model predicting performance with an average 94\% accuracy.
\end{abstract}

%\begin{CCSXML}
%<ccs2012>
%  <concept_id>00000000.0000000.0000000</concept_id>
%  <concept_desc>Do Not Use This Code, Generate the Correct Terms for Your Paper</concept_desc>
%  <concept_significance>500</concept_significance>
%  %<concept_id>00000000.00000000.00000000</concept_id>
%  <concept_desc>Do Not Use This Code, Generate the Correct Terms for Your Paper</concept_desc>
%  <concept_significance>300</concept_significance>
%  %<concept_id>00000000.00000000.00000000</concept_id>
%  <concept_desc>Do Not Use This Code, Generate the Correct Terms for Your Paper</concept_desc>
%  <concept_significance>100</concept_significance>
 % <concept_id>00000000.00000000.00000000</concept_id>
%  <concept_desc>Do Not Use This Code, Generate the Correct Terms for Your Paper</concept_desc>
%  <concept_significance>100</concept_significance>
%</ccs2012>
%\end{CCSXML}

%\ccsdesc[500]{Do Not Use This Code~Generate the Correct Terms for Your Paper}
%\ccsdesc[300]{Do Not Use This Code~Generate the Correct Terms for Your Paper}
%\ccsdesc{Do Not Use This Code~Generate the Correct Terms for Your Paper}
%\ccsdesc[100]{Do Not Use This Code~Generate the Correct Terms for Your Paper}


\maketitle

\section{Introduction}
The emergence of Mixture-of-Experts (MoE) models~\cite{dai2024deepseekmoeultimateexpertspecialization, shazeer2017outrageouslylargeneuralnetworks, deepseekai2025deepseekv3technicalreport} has marked a significant evolution in the design of Large Language Models (LLMs). 
In contrast to dense models~\cite{grattafiori2024llama3herdmodels, bai2023qwentechnicalreport} that activate the full set of model parameters for every input, MoE models introduce sparsity by routing each input through a small subset of expert networks. 
This design enables MoE models to scale up the total parameter count substantially without proportionally increasing the number of floating-point operations (FLOPs) per inference step. 
As a result, MoE-based LLMs have demonstrated strong empirical performance across a wide range of tasks~\cite{deepseekai2025deepseekr1incentivizingreasoningcapability, dai2024deepseekmoeultimateexpertspecialization}, while maintaining manageable compute requirements for inference.

However, the practical deployment of MoE models is challenging due to their high memory capacity demand to store the model weights. 
Although only a few experts are activated per token, all expert parameters must reside in memory to allow flexible routing decisions at runtime. 
This leads to substantial memory pressure that often exceeds the capacity of a GPU. 
For instance, recent models such as DeepSeek-V3/R1~\cite{deepseekai2025deepseekv3technicalreport, deepseekai2025deepseekr1incentivizingreasoningcapability} and Mixtral-8x22B~\cite{mistral_mixtral8x22b_2025} size hundreds of gigabytes, significantly outpacing the requirements of dense models with similar FLOPs. 
These memory capacity demands hinder the use of MoE models in \textit{resource-constrained environments}, such as low-cost servers, where the available GPU memory capacity is far less than the model size. 
% This poses a major barrier to their \textit{local deployment}, due to cost or privacy concerns, in environments such as personal computers and internal servers in small organizations.
% This poses a major barrier to their \textit{local deployment}, motivated by cost and privacy concerns, in environments such as personal computers and internal servers in small enterprises.

A key technique for enabling MoE inference in resource constrained environments is \textit{CPU offloading}~\cite{sheng2023flexgenhighthroughputgenerativeinference, cao2024moe}.
In this approach, the model weights and Key-Value (KV) cache are stored in CPU memory and transferred to the GPU on demand during inference. 
As a result, CPU-GPU IO becomes a critical bottleneck.
Prior work has sought to mitigate this bottleneck through improved scheduling strategies, including pipelining~\cite{sheng2023flexgenhighthroughputgenerativeinference}, attention offloading~\cite{cao2024moe, kamahori2025fiddlercpugpuorchestrationfast}, and model-aware prefetching~\cite{fang2025klotskiefficientmixtureofexpertinference}. 
While these techniques have led to notable, hardware utilization remains low even in state-of-the-art systems, leaving room for significant performance improvements.
For example, we find only 16.5\% of GPU utilization for MoE-Lightning~\cite{cao2024moe} during the generation stage.
This raises important questions: \textit{\textbf{what is the upper bound on achievable performance, and how can the system achieve such upper bound?}}
%How does this workload stress system components beyond CPU-GPU PCIe links?
%What key design principles are essential to approach this limit?} 

In this paper, we present \THISWORK, a high-throughput MoE inference framework designed for resource-constrained environments, achieving up to 25.5$\times$ and an average of 4.6$\times$ speedup over state-of-the-art MoE-Lightening~\cite{cao2024moe}. 
\THISWORK's design contains three stages. 
First, unlike prior work that relies on 
limited-scope performance models for system optimizations, \THISWORK\ employs a two-stage \textbf{\textit{holistic performance model}} that considers factors beyond CPU-GPU I/O bandwidth. It not only identifies theoretical performance upper bounds but also accurately predicts the execution time.
In the third stage, we propose a system design guided by this model and jointly optimize the execution pipeline and sequence-level scheduling to bring the system closer to hardware limits.
Similar to prior works~\cite{sheng2023flexgenhighthroughputgenerativeinference, cao2024moe, fang2025klotskiefficientmixtureofexpertinference, xu2025moegenhighthroughputmoeinference}, our focus is \textit{offline, batching processing} inference tasks, such as model evaluation~\cite{liang2023holisticevaluationlanguagemodels}, data wrangling~\cite{narayan2022foundationmodelswrangledata}, form processing~\cite{chen2021spreadsheetcoderformulapredictionsemistructured}, LLM for relational analytics~\cite{liu2025optimizingllmqueriesrelational}, and synthetic data generation~\cite{grattafiori2024llama3herdmodels}, where maximizing inference throughput directly reduces total job completion time.

% \THISWORK's performance model accounts for critical factors that are often overlooked in prior work, including CPU-side architectural resources and inference scheduling strategies. 
\THISWORK’s two-stage performance modeling accounts for critical factors that represent missed opportunities in prior works~\cite{cao2024moe, sheng2023flexgenhighthroughputgenerativeinference, fang2025klotskiefficientmixtureofexpertinference} to show how they influence throughput.
% It offers a comprehensive understanding of how these factors interact and influence end-to-end inference performance.
In the first stage, the model analyzes the theoretical performance
upper bound of MoE inference based on the fundamental system components.
It identifies \textit{CPU memory capacity}, an element overlooked by prior work, as a primary limiting factor and quantifies how prompt and generation lengths impact the memory utilization. 
The second stage captures how system execution mechanisms, including workload scheduling and paged KV cache, affect memory/compute utilization and overall system performance.
By integrating all these dimensions, our model accurately predicts end-to-end wall-clock inference time for systems operating near hardware limits.

In the third stage, \THISWORK\ introduces a high-throughput MoE LLM inference system design informed by our holistic performance model that significantly outperforms state-of-the-art solutions. 
The system maximizes hardware utilization by addressing key inefficiencies in CPU-side resource usage and balancing compute across the prefill and decode stages.
To this end, we introduce resource-aware scheduling that enables effective prefill/decode overlapping, reducing idle time and smoothing workload distribution.
We also propose a novel pipeline design, VSLpipe, which includes a contiguous data mover to maximize CPU-GPU bandwidth utilization during weight transfers.
Our hand-optimized CPU attention kernel using instrinsics fully leverages the vector units of modern CPUs, preventing the CPU compute throughput from becoming a bottleneck and improving GPU utilization.

% {
% \color{blue}
Evaluated on diverse models and datasets,
\THISWORK\ achieves on average 4.6$\times$, up to 25.5$\times$, speedup over the state-of-the-art solution MoE-Lightning~\cite{cao2024moe}.
% It also demonstrates a strong alignment between the performance model's predictions and actual execution times, with an average accuracy of 94\%. 
The results show the importance of holistic performance modeling and architecture-aware design decisions for high-throughput MoE inference in resource-constrained environments.
%These results underscore the importance of making architecture-aware design decisions and prefill/decode overlapping scheduling strategy, as highlighted by our performance model, for high-throughput MoE inference in resource-constrained environments.
% effectiveness of the insights concluded from \THISWORK's performance model, such as overlapping the prefill and decode stages for balanced resource utilization and a large effective execution batch size.
% }
In summary, \THISWORK\ makes the following contributions.
\begin{itemize}[nosep, leftmargin=*]
    \item A holistic performance model for MoE inference in resource-constrained environments capturing complex interactions between hardware properties and system execution mechanisms.
    \item An accurate throughput predictor for MoE LLM inference under hardware constraints.
    \item A system design informed by the model, featuring resource-aware phase scheduling, CPU-side attention execution, and efficient weight/KV cache transfer.
    \item \THISWORK: an architecture-aware, end-to-end CPU-GPU hybrid system with theoretical underpinnings that achieves an average 4.6$\times$ throughput improvement over the state-of-the-art.
\end{itemize}

\section{Background}
% \subsection{Mixture-of-Expert (MoE) LLMs}
\textbf{Mixture-of-Expert (MoE) LLMs.}
MoE LLMs achieve strong benchmark performance while reducing compute needs compared to dense LLMs with similar parameter counts, primarily composed of attention and MoE layers. For architectural details, see~\cite{cao2024moe, fang2025klotskiefficientmixtureofexpertinference}. 
A defining trait of modern MoE models is their large size: hundreds of GBs~\cite{databricks2024dbrxinstruct, deepseekai2025deepseekr1incentivizingreasoningcapability, deepseekai2025deepseekv3technicalreport}—which exceeds standard GPU memory~\cite{cao2024moe, sheng2023flexgenhighthroughputgenerativeinference, fang2025klotskiefficientmixtureofexpertinference}.

\noindent
\textbf{Concepts in LLM Model Inference.}
% One key module in LLM Models is the attention module, which calculates \textit{query}, \textit{key}, and \textit{value} vectors from the hidden representation and mixes the value vectors based on the query and key vectors.
% The key and value vectors are cached for efficiency, called \textit{KV Cache}.
One key module in LLM Models is the attention module, where \textit{key} and \textit{value} vectors are calculated and cached in \textit{KV Cache}.
\textit{Group Query Attention (GQA)} is a commonly used attention variant in MoE models, which allows a group of query vectors to share a single pair of key and value vectors, thereby reducing the size of the KV cache.
The LLM inference consists of two stages: the \textit{prefill stage}, typically \textit{compute-bound}, where the initial prompt is processed in parallel, and the \textit{decode stage}, typically \textit{memory-bound}~\cite{splitwise2024}, where tokens are generated sequentially in an auto-regressive manner.
\noindent
\textbf{Resource-Constrained LLM Inference.}
LLM inference in resource-constrained environments prioritizes high-throughput batch processing on systems where the GPU memory is significantly smaller than the model’s total parameter size, while the CPU has sufficient memory or disk capacity to store model weights.
This setting differs from traditional LLM serving systems~\cite{pagedattn, splitwise2024, qin2024mooncakekvcachecentricdisaggregatedarchitecture} optimized for latency-sensitive applications like chatbots and code completion, where low response time is critical.
Instead, resource-constrained inference systems, typically equipped with GPUs like T4 (16GB) or L4 (24GB)~\cite{cao2024moe, sheng2023flexgenhighthroughputgenerativeinference, fang2025klotskiefficientmixtureofexpertinference}, prioritize overall throughput and can afford to trade off latency.
Since these GPUs lack sufficient memory to store the full model, weights must be streamed from CPU memory over PCIe, introducing substantial overhead.
MoE-Lightning~\cite{cao2024moe} addresses this by offloading attention computation to the CPU, forming a CPU–GPU hybrid system.
This design avoids transferring the large KV cache to GPU, which is essential as growing GPU parallelism leads to KV sizes exceeding memory capacity.
Moreover, because attention has low arithmetic intensity~\cite{cao2024moe, dao2022flashattentionfastmemoryefficientexact}, the CPU can execute it efficiently while the GPU focuses on compute-intensive MoE layers.

\section{Motivation}
While MoE-Lightning~\cite{cao2024moe}, the state-of-the-art MoE inference system for resource-constrained environments, leverages a performance model to guide system design and achieves substantial speedups, an important question remains: \textbf{\textit{does the state-of-the-art fully harness the capabilities of the underlying hardware?}}

\subsection{Limited-Scope of Performance Modeling in Prior Work} \label{sec:limitation_moelightening}
MoE-Lightning introduced the Hierarchical Roofline Model (HRM) to address the CPU-GPU IO bottleneck, achieving notable gains by offloading decode-stage attention to the CPU and avoiding frequent KV cache transfers.
While effective, HRM’s focus is limited to arithmetic intensity and IO bandwidth, overlooking two crucial factors that influence performance ceilings: CPU memory capacity and the characteristics of input requests, such as prompt and generation lengths.
These factors directly affect how much weight transfer overhead can be amortized and how well pipelining can hide data movement latency.
As shown in Table~\ref{tab:motivation-plan}, typical execution plans generated by MoE-Lightning result in underutilized CPU memory, revealing inefficiencies in resource allocation.
Moreover, sustaining high concurrency requires not just fast IO, but also ample memory bandwidth and compute throughput on the CPU side: elements HRM does not model. 
This presents a \textit{significant opportunity} to rethink scheduling and architecture-aware execution strategies that better align with hardware constraints.
% A more comprehensive performance model and execution design can close the remaining gap to hardware limits by leveraging these underexplored dimensions.
% MoE-Lightning formulates its \textit{Hierarchical Roofline Model (HRM)} centering around the CPU-GPU IO bottleneck.
% Based on HRM's analysis, MoE-Lightning offloads the attention operation during the decode stage to the CPU, avoiding transferring the large KV cache through the narrow CPU-GPU IO channel and significantly improving the system efficiency.
% While achieving encouraging results, we argue that the HRM is limited in its scope to CPU-GPU IO and operations' arithmetic intensity, overlooking two critical factors to model the hardware limits: the CPU memory capacity and the input requests' characteristics, \textit{i.e.,} the prompt and generation length.
% Thus, there is a significant opportunity for further improving the system performance and approaching the hardware limits.
% For instance, for different prefill ($p$) and generation ($g$) lengths, Table~\ref{tab:motivation-plan} shows CPU memory usage from typical execution plans generated by MoE-Lightning.
% The low utilization underscores inefficiencies in resource allocation.
\begin{table}[t]
    \scriptsize
    \centering
    \begin{tabular}{cccc}
        \toprule
        Prefill Length & Generation Length & CPU Memory (GB) & CPU Memory Utilization \\
        \midrule
        98  & 32  & 265 & 52.0\% \\
        98  & 64  & 265 & 56.2\% \\
        926 & 128 & 265 & 35.0\% \\
        \bottomrule
    \end{tabular}
    \caption{CPU memory utilization for execution plans generated by MoE-Lightening~\cite{cao2024moe}, showcasing under-utilization.}
    \label{tab:motivation-plan}
    \vspace{-1cm}
\end{table}

\noindent
\textbf{\textit{Motivated Approach.}}
% The limitations motivate the \textit{Stage 1 Model} in \THISWORK's three-step approach, where we demonstrate the theoretical performance upper bound for processing requests with different prompt and generation length, and quantify the effect of CPU memory capacity on that limit, going beyond analysis limited to operator-level CPU-GPU IO bottlenecks.
These limitations motivate the development of the \textit{Stage 1 Model} in \THISWORK's three-step approach, which extends beyond operator-level IO analysis to capture a more complete picture of system performance.
This model establishes a theoretical upper bound on throughput for processing requests with varying prompt and generation lengths, incorporating the often-overlooked impact of CPU memory capacity.
% By explicitly quantifying how memory constraints affect achievable concurrency and weight reuse, the model reveals new optimization opportunities that prior work like MoE-Lightning could not capture.

% The limitations motivate \THISWORK's \textit{holistic performance model}, which provides a principled understanding of how \textit{CPU-side architectural resources}, specifically the CPU memory capacity, impacts system performance when processing requests with different prompts and generation lengths.
% This model demonstrates the theoretical performance upper bound for processing requests with different prompt and generation length, and quantifies the effect of CPU memory capacity on that limit.

% The limitations motivate our \textit{white-box, analytical-driven approach}, which provides a principled understanding of how \textit{CPU-side architectural resources} impact system performance.
% {
% \color{blue}
% By making the demands and performance dynamics of CPU resources explicit, our analysis yields practical insights for optimizing system and architectural design under resource-constrained MoE inference. 
% These insights enable designers to make \textit{robust and explainable} decisions, avoiding the pitfalls of opaque, black-box based methods.
% }
% By directly analyzing the key bottlenecks of throughput-oriented, resource-constrained MoE inference, we derive practical insights for optimizing system and architecture design.
% By following simple yet effective design rules, designers can make \textit{robust and explainable} decisions, avoiding the pitfalls of opaque models.

\subsection{Resource Utilization Imbalance between Prefill and Decode Stages.}
\label{sec:motivation-overlapping}
\begin{figure}[t]
    \centering
    \includegraphics[width=\linewidth]{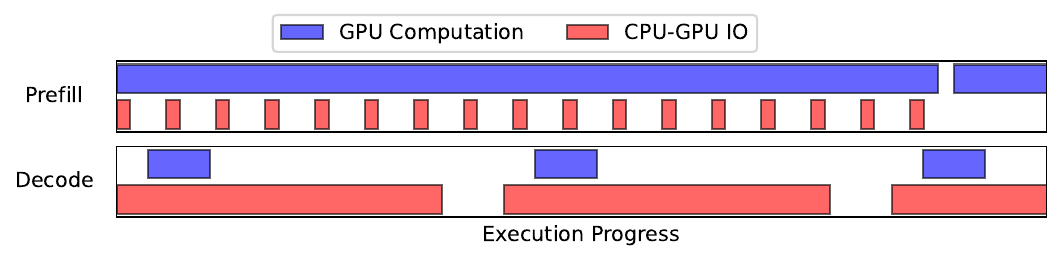}
    \vspace{-0.7cm}
    \caption{Sample of an execution timeline of GPU computation and CPU-GPU IO during the prefill and decode stages of MoE-Lightning.}
    \label{fig:profile-illustration}
    \vspace{-0.5cm}
\end{figure}

% To approach the throughput upper bound, it is crucial to take the \textit{workload heterogeneity} for LLM inference into account, which causes a resource utilization imbalance across the prefill and decode stages.
% As shown in Figure~\ref{fig:profile-illustration}, the prefill stage fully utilizes GPU computation but underutilizes other resources, such as CPU-GPU IO bandwidth. 
% In contrast, the decode stage suffers from low GPU utilization because model weights must be transferred from CPU to GPU due to the limited GPU memory capacity.
% Existing systems for MoE inference in resource-constrained environments treat the prefill and decode stages separately. 
% While this simplifies scheduling, it leads to resource utilization imbalance between the two stages.
% We profile MoE-lightning~\cite{cao2024moe} using a prompt length of 98 tokens and a generation length of 32, and the overall execution status is like Figure~\ref{fig:profile-illustration}, where clear imbalance between the prefill and decode stage can be observed.
% The results show that, in the prefill stage, only 23.9\% of the time CPU-GPU IO is active, while the decode stage achieves just 16.5\% GPU utilization.
To approach the throughput upper bound, it is essential to account for \textit{workload heterogeneity} in LLM inference, which causes imbalanced resource utilization between the prefill and decode stages.
As shown in Figure~\ref{fig:profile-illustration}, the prefill stage fully utilizes GPU compute but leaves CPU-GPU IO bandwidth underutilized, while the decode stage suffers from low GPU utilization due to on-demand weight transfers from CPU to GPU, constrained by limited GPU memory.
Existing MoE inference systems in resource-constrained settings handle these stages separately, simplifying scheduling but exacerbating the imbalance.
Profiling MoE-Lightning~\cite{cao2024moe} with a 98-token prompt and 32-token generation reveals this inefficiency: CPU-GPU IO is active only 23.9\% of the time during prefill, while GPU utilization drops to 16.5\% during decode.

\noindent
\textbf{\textit{Motivated Approach.}}
% By co-scheduling prefill and decode sequences, we improve resource utilization across both stages, ensuring more effective use of available GPU resources.
% The observation highlights the importance of \textit{overlapping prefill and decode execution} to approach the hardware limit.
% We formalize this intuition in the \textit{Stage 2 Model} in \THISWORK's three-step approach, where we focus on incorporating critical execution factors to make the model resource and workload aware, while still representing the hardware limit.
% By considering the effect of prefill/decode overlapped scheduling, and other execution factors, we make the \textit{Stage 2 Model} align with real systems execution such that it can be practical in guiding real system design and accurate in estimating the end-to-end execution time of a MoE-inference system that is approaching the hardware limit. 
This observation motivates the need to \textit{overlap prefill and decode execution} to approach the hardware limit. 
\THISWORK\ formalizes this in the \textit{Stage 2 Model}, which incorporates key execution factors to make the model both resource- and workload-aware while still reflecting hardware constraints.
By accounting for overlapped scheduling and system-level interactions, the \textit{Stage 2 Model} aligns closely with real system behavior, enabling practical guidance for system design and accurate estimation of end-to-end execution time in MoE inference.

\section{Overview of \THISWORK}
\label{sec:thiswork_overview}
\begin{figure}
    \centering
    \includegraphics[width=0.8\linewidth]{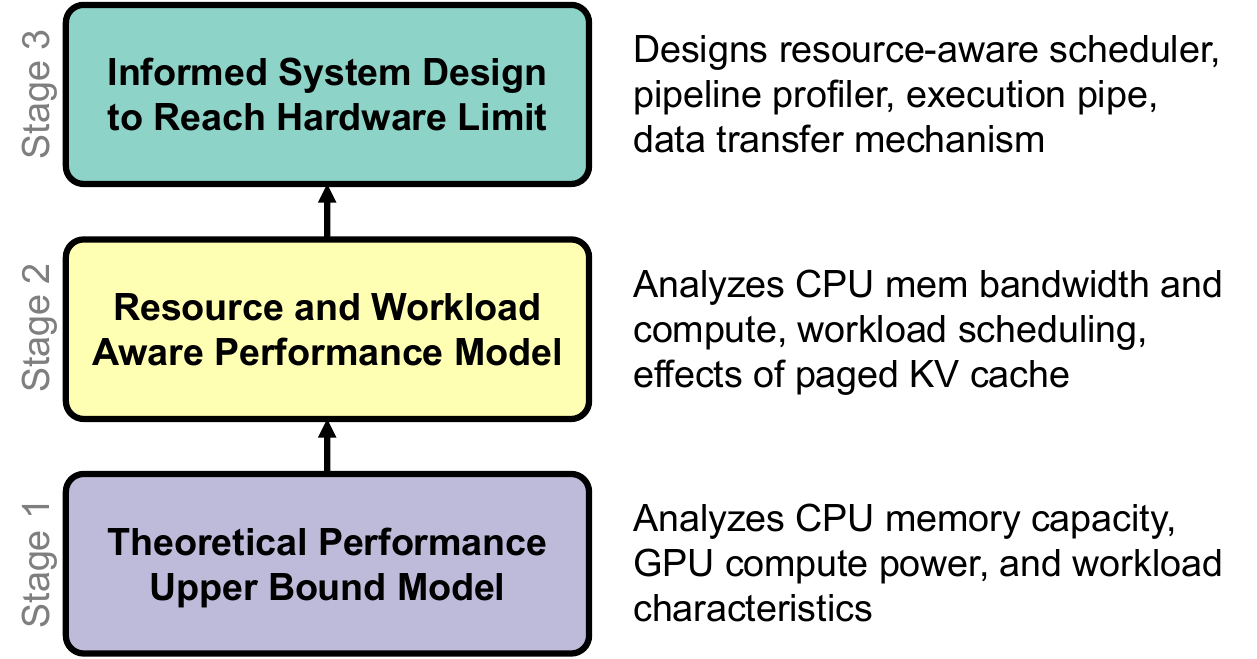}
    \caption{Overview of \THISWORK\ that combines a theoretical performance upper bound, resource and workload aware performance model, and an informed system design to reach hardware limits.}
    \label{fig:approach_overview}
    \vspace{-0.5cm}
\end{figure}

In this work, we aim to answer the following critical questions:
\begin{itemize}[nosep, leftmargin=*]
    \item \textit{What is the throughput upper bound for a machine?}
    \item \textit{How to systematically approach that performance upper bound?}
\end{itemize}
% \item How to get a concrete design that maximizes the hardware potential?
In \THISWORK, we take a three-step approach as shown in Figure~\ref{fig:approach_overview}, building up an increasingly detailed understanding of MoE inference in a resource-constrained environment, and eventually reach a concrete design that reaches the hardware limit.
In \textit{Stage 1}, we develop a model to understand the theoretical performance upper bound of MoE inference based on the fundamental system's architectural components (\S\ref{sec:perf-model-mem-capacity}, \S\ref{sec:perf-model-PME}).
We further analyze the requirements on CPU memory bandwidth and compute throughput, which support the performance upper bound (\S\ref{sec:perf-model-attn}) and the benefits of prefill/decode overlapped scheduling (\S\ref{sec:perf-model-sched}).
Holistically considering all factors above, and request batch size and paged KV cache, we derive our resource and workload aware performance model in \textit{Stage 2} from an implementation viewpoint (\S\ref{sec:perf-model-e2e}), providing realistic insights for concert design and predict system performance (with 94\% accuracy).
One important property of our \textit{Stage 2} model is that it \textit{converges} to the performance upper bound with an increasing batch size, thus still modeling the hardware limits but pricing in the physical execution factors.
Finally in \textit{Stage 3}, we provide a system design based on the insights from modeling and the guidance of the \textit{Stage 2} model, which adapts to the variance of the real execution environments while achieving the high hardware utilization outlined by the \textit{Stage 2} model (\S\ref{sec:system-design}).

\section{\THISWORK\ Performance Model}
\label{sec:perf-model}
{
% \color{blue}
% In this section, we provide an in-depth analysis of the strategy taken by MoE-Lightning~\cite{cao2024moe}, a state-of-the-art MoE inference model designed for a resource-constrained environment, which falls short in answering the following critical questions:
% In this section, we presents \THISWORK's holistic performance model, which aims at answering the following critical questions:
% In this section, we analyze the theoretical performance upper bound of resource-constrained MoE inference and further derive a realistic, resource and workload aware performance model taking other major execution factors into account.% \THISWORK's holistic performance model, which aims at answering the following critical questions:
This section describes the details of our \textit{Stage 1} and \textit{2} models.
% \begin{enumerate}[nosep, leftmargin=*]
%     \item What is the throughput upper bound one can achieve for a certain machine configuration?
%     \item How to systematically approach that performance upper bound?
    % \item How can we accurately predict the performance of a real MoE inference system in resource-constrained environments, whose performance is further subject to high-level scheduling strategy and system implementation details?
% \end{enumerate}
% \subsection{Limitations of Black-Box Optimizations}
}
% \THISWORK\ follows the established practice in terms of offloading attention from GPU to CPU in resource-constrained environments(\S\ref{sec:background-offloading}).
% While prior works identify CPU-GPU IO bandwidth as the primary bottleneck for MoE inference in resource-constrained environments, they neglect its relationship with the rest of the system resources.
% Specifically, one can still saturate GPU compute when weights loaded through off-device IO are reused sufficiently, i.e. the cost of moving weights from CPU to GPU is amortized by the computation for many tokens. \textcolor{red}{LM: what the negligence by prior work here, as stated in the previous sentence? This sentence did not talk about negligence.}

% \THISWORK's performance model highlights the impact of previously overlooked CPU-side architectural resources.
{
\color{blue}
% To answer these questions, \THISWORK's holistic performance model highlights the impact of previously overlooked CPU-side architectural resources.
% First, we identify \textit{CPU memory capacity} as a limiting factor for system performance (\S\ref{sec:perf-model-mem-capacity}), and derive the system theoratical performance upper bound for requests with different prompt and generation length (\S\ref{sec:perf-model-PME}).
% Then, we identify the CPU memory bandwidth and compute throughput requirements needed to support the theoretical performance upper bound (\S\ref{sec:perf-model-attn}), and illustrate the benefit of prefill/decode overlapping scheduling over system throughput (\S\ref{sec:perf-model-sched}).
% % Based on our discussion on the impact of CPU-resource utilization, we then dissect the benefit of prefill/decode overlapping scheduling over system throughput (\S\ref{sec:perf-model-sched}).
% Finally, considering all the above factors holistically, as well as the request batch size and paged KV cache as additional execution factors, in \S\ref{sec:perf-model-e2e}, we build a realistic performance model that not only converges to the performance upper bound, but also is practical enough to guide concrete implementation.
% % FIXME: 94\%
% The realistic performance model achieves 94\% accuracy in predicting the performance of \THISWORK's system design.
}
% We identify \textit{CPU memory capacity} as a limiting factor (\S\ref{sec:perf-model-mem-capacity}, \S\ref{sec:perf-model-PME}), and provide in-depth analysis of the influence of \textit{CPU memory bandwidth} and \textit{CPU compute throughput} for attention computation (\S\ref{sec:perf-model-attn}).
% Based on our discussion on the impact of CPU-resource utilization, we then dissect the benefit of prefill/decode overlapping scheduling over system throughput (\S\ref{sec:perf-model-sched}).

\subsection{CPU Memory Capacity as a Limiting Factor}
\label{sec:perf-model-mem-capacity}
As the CPU memory stores the KV cache, its capacity directly determines the number of sequences that can be processed in parallel. 
A large number of active sequences allows a large number of tokens to reuse the weights loaded from the CPU for computation, amortizing the cost of moving the weights from the CPU to the GPU.
\textit{An important question is \textbf{how much CPU memory is necessary to fully utilize the GPU?}}
% Compared to CPU-GPU IO bandwidth, it is much easier to scale in a local deployment like a PC.

Let $N_e$ denote the number of experts, $N_k$ the number of top-$k$ experts selected per token token, $h$ the model dimension, $h_i$ the intermediate dimension of the expert networks (typically, $h_i = mh$, where $m > 1$), $s$ the GQA group size and $n$ as the number of tokens processed in parallel.
The GEMM arithmetic-to-IO intensity, which is the amount of GEMM computation divided by the amount of weight data accessed, of a MoE model is
\begin{equation}
I = n\frac{6N_khh_i + 4h^2 + 4\frac{h^2}{s}}{6N_ehh_i + 4h^2 + 4\frac{h^2}{s}} = n\frac{6mN_k+ 2 + \frac{2}{s}}{6mN_e+2 +\frac{2}{s}} \approx n\frac{N_k}{N_e}
\end{equation}

Here, $\frac{N_k}{N_e}$ reflects the sparsity of the MoE layer.
Let $C_{GPU}$ be the GPU's GEMM throughput and $B_{IO}$ the GPU-CPU IO bandwidth.
To saturate the GPU computation power, we require
\begin{equation}
\label{eqn:sat_GPU}
I \geq \frac{C_{GPU}}{B} \Leftrightarrow
n \geq \frac{C_{GPU}}{B}\frac{N_e}{N_k}
\end{equation}

% For example, with model \texttt{Mixtral8x7B} running on NVidia A40 GPU ($C_{GPU} = 1.5e14$, $B=32$, $N_e=8$ and $N_k=2$), \textit{19200 tokens} must be processed in parallel to saturate the GPU compute.
% Table~\ref{tab:gpu_kvcache_size} shows the amount of tokens and corresponding KV cache size necessary to saturate the compute resources of different GPUs.
% Even if we assume a throughput-oriented system design, where we have considerable volume of pending sequences to process, achieving this parallelism is challenging as it requires \textbf{\textit{high amount of CPU memory capacity}} to store the KV cache.
% Each token needs some memory to store its KV cache for its sequences' past context, and their aggregated size can easily exceed the available CPU memory capacity, especially in the resource-constraint environments.
% For example, processing 500 token sequences demand \textit{1.2TB} of CPU memory to store KV cache, which is disproportional to the single GPU we are using.
% As GPUs grow more powerful, this demand escalates further, as shown in Table \ref{tab:gpu_kvcache_size}.

For example, saturating the compute of a single NVIDIA A40 GPU ($C_{\text{GPU}}$ = 150 TFLOPS, $B=32$, $N_e=8$, $N_k=2$) when running \texttt{Mixtral-8x7B} requires processing \textit{19,200 tokens} in parallel.
Table~\ref{tab:gpu_kvcache_size} quantifies the number of tokens and corresponding KV cache size, assuming a 512-token sequence length (sum of prompt and generation length), needed to saturate different GPUs.
Even under a throughput-optimized system design, where abundant pending sequences are assumed, achieving this level of parallelism is challenging due to the \textbf{large CPU memory capacity} required to hold the KV cache.
Each token contributes to the cumulative KV cache footprint, which can quickly exceed the available CPU memory in resource-constrained settings.
For instance, supporting 512-token sequences would require \textit{1.2TB} of CPU memory for KV cache: disproportionate for a single GPU.
This memory demand grows with increasing GPU compute capabilities, as illustrated in Table~\ref{tab:gpu_kvcache_size}.

\begin{table}[t]
    \centering
    \scriptsize
    \begin{tabular}{lcccccc}
        \hline
        & \multicolumn{3}{c}{Sequence Length 256} & \multicolumn{3}{c}{Sequence Length 512} \\
        \cline{2-7}
        NVIDIA GPU
         & A40 & L40 & A100  & A40 & L40 & A100 \\
        \hline
        \makecell[l]{BF16 Throughput (in T FLOPS)} & 150 & 181 & 312 & 150 & 181 & 312\\
        \makecell[l]{\# of Tokens to Saturate GPU compute} & 19.2k & 23.2k & 40.0k & 19.2k & 23.2k & 40.0k\\
        \makecell[l]{KVCache Size to Saturate (in GB)}  & 614 & 741 & 1277 & 1228 & 1482 & 2554 \\
        \hline
    \end{tabular}
    \caption{KV Cache Size Needed to Saturate GPU compute.}
    \label{tab:gpu_kvcache_size}
    \vspace{-1cm}
\end{table}

\begin{tcolorbox}[width=0.48\textwidth]
\textbf{\textit{Takeaway:}} \textit{CPU memory capacity for KV cache storage is a limiting factor to fully utilize GPU compute resources.}
\end{tcolorbox}
% \noindent
% \textbf{Design Decision Implications.}
% Fully utilizing CPU memory capacity is critical to achieve high-throughput for MoE inference in resource-constrained environments.

\subsection{Stage 1 Model: Parallelism-Memory Efficiency Analysis}
\label{sec:perf-model-PME}
% While the CPU memory capacity limits the number of parallel tokens, not all tokens require the same amount of memory.
% During the prefill stage, all tokens in the prompt can be processed in parallel, 
% amortizing the memory capacity cost per token computation.
% During the decode stage, tokens are generated autoregressively, and an entire sequence's KV cache only allows computing a single next token.
% As a result, tokens at the prefill stage are more efficient in utilizing memory capacity than the ones at the decode stage.
% \textit{A key question is what is the theoretical performance upper bound one can achieve for batch requests with different prompt and generation length, given a certain machine configuration?}
While CPU memory capacity constrains the number of tokens that can be processed in parallel, the memory footprint per token varies across stages of inference.  
During the prefill stage, all tokens in the prompt can be processed simultaneously, effectively amortizing the memory cost across multiple token computations.  
In contrast, the decode stage operates autoregressively: each sequence's KV cache enables computing only a single token at a time.  
Consequently, prefill tokens offer higher memory efficiency compared to decode tokens, in terms of parallel computation per unit of memory.
This leads to:  
\textbf{\textit{what is the theoretical upper bound on system throughput for a batch of requests with varying prompt and generation lengths, under a fixed hardware configuration?}}

We introduce \textit{Parallelism-Memory Efficiency} ($PME$), a metric quantifying how effectively a sequence translates memory capacity into the number of tokens that can be processed in parallel to saturate GPU resources.
% For a sequence $s_{(i, t)}$, at a specific time $t$ during processing, with prompt length $p_i$ and current processed length $c_{(i, t)}$ at $t$, it's \textit{instantaneous PME} is:
% $$
% \text{PME}^i_{t} = \frac{\text{Parallel Tokens}}{c_{(i, t)}} = \frac{\text{max}(c_{(i, t)} - p_i, 0) + 1}{c_{(i, t)}}
% $$
% The higher the $PME_i$, the more memory efficient a sequence to saturate the GPU for the computation at time $t$.
% Taking the entire generation process of a sequence into account, we define \textit{lifetime PME} for a sequence $s_i$ with generation length $g_i$ as 
The \textit{PME} for a sequence $s$ with prompt length $p$ and generation length $g$ is
\begin{equation}
\begin{split}
\text{PME} &= \frac{\sum_\text{gen. steps} \text{Parallel Tokens}}{\sum_\text{gen. steps} \text{Sequence KV Cache Size}} \\
&= \frac{p + g}{\sum^{g}_{j=0}(p +j)} = \frac{2(p + g)}{(2p + g)g}
\end{split}
\end{equation}
Here, the denominator is the sum of the memory capacity a sequence occupies across its entire generation lifetime.
We further define the time to transfer the weight of a model from CPU to GPU as $\delta = \frac{\text{Model Size}}{B_{IO}}$.
For a batch of sequences with average prefill length $p$ and generation length $g$, its theoretical maximum inference throughput (tokens/sec) can be estimated as:
\begin{equation}
T_{max} = min(\frac{\text{PME} \cdot M}{\delta}, T_{GPU}),
\end{equation}
where $M$ is the size of KV cache in number of tokens, and $T_{GPU}$ is the maximum throughput of a GPU in number of tokens per second.
\begin{figure}
    \centering
    \includegraphics[width=\linewidth]{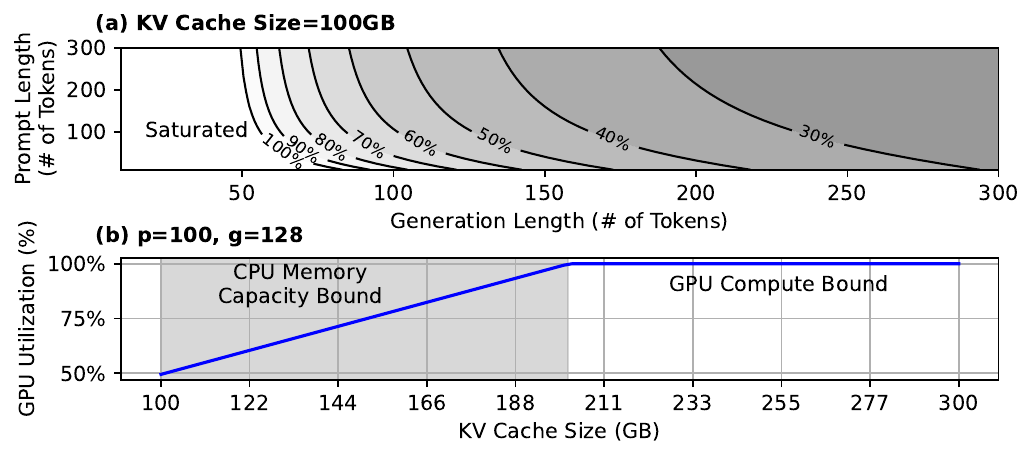}
    \vspace{-0.7cm}
    \caption{Visualization of the maximum GPU utilization $\frac{T_{max}}{T_{GPU}}$. (a) Maximum GPU utilization when running \texttt{Mixtral8x7B} on \texttt{A40} with 100GB KV cache. (b) For the same model and GPU, the maximum GPU utilization when $p = 100$ and $g=128$.}
    \label{fig:pme}
    \vspace{-0.5cm}
\end{figure}
% Figure~\ref{fig:pme}(a) depicts how the theoretical maximum GPU utilization changes for different $p$ and $g$ for a 100GB KV cache.
% The longer the sequence, the lower the theoretical maximum GPU utilization.
% For sequences with a certain length, the longer the prompt portion, the higher the theoretical maximum GPU utilization.
% Figure~\ref{fig:pme}(b) shows a \textit{roofline} model for theoretical maximum GPU utilization.
% As the KV cache increases, the system transitions from \textit{CPU Memory Capacity Bound} to \textit{GPU Bound}, where the GPU is fully utilized.
% \textcolor{blue}{In the CPU memory capacity-bound case, the system performance increases as the KV cache size increases, while the GPU is underutilized. A machine's limited CPU memory capacity limits the scaling of KV cache size and thus performance. In the GPU-bound case, the GPU is fully utilized, and further increasing the KV cache size brings no additional speedup. }
Figure~\ref{fig:pme}(a) illustrates how the theoretical maximum GPU utilization varies with prompt length ($p$) and generation length ($g$) under a 100GB KV cache budget. 
Longer sequences lead to lower theoretical GPU utilization, while a higher prompt-to-generation ratio improves utilization for a given sequence length.
Figure~\ref{fig:pme}(b) presents a roofline model of theoretical GPU utilization.
As KV cache capacity increases, the system transitions from a CPU memory capacity-bound regime to a GPU-bound regime.  
In the memory-bound regime, limited CPU memory constrains the number of parallel sequences, leading to underutilized GPU compute and throughput that scales with available KV cache.  
Once the system becomes GPU-bound, the GPU is fully saturated, and further increases in KV cache capacity yield diminishing returns in performance.

% The above formula can be used to estimate whether a batch of requests can potentially saturate a certain GPU/CPU combination or not, given a perfect schedule.
% For example, when running \texttt{Mixtral8x7B} on A40, a batch of sequences wth $p=100$ and $g=32$ can reach the GPU computation limitatoin when the KV cache size is 100GB (763k tokens), but if $p=100$ and $g=64$ it can saturate at most around 77\% of the GPU computation.

\begin{tcolorbox}[width=0.48\textwidth]
\textbf{\textit{Takeaway:}} \textit{Prompt and generation length jointly determine the theoretical upper bound on achievable GPU utilization.}
\end{tcolorbox}

\subsection{CPU Memory Bandwidth and Compute Throughput Requirements}
\label{sec:perf-model-attn}
% \textcolor{blue}{Do not start with prior}
% Prior work~\cite{cao2024moe} has demonstrated the necessity of offloading the KV Cache and decode attention operation from the GPU to the CPU.
In a hybrid CPU-GPU MoE inference system, the CPU hosts the KV cache and computes decode attention.
To avoid bottlenecking the overall execution, the CPU needs to process the attention at a certain throughput, which stresses its memory bandwidth and compute throughput.
\textit{It is crucial to understand: \textbf{what is the desired CPU memory bandwidth and compute throughput}?}

% Following a typical deployment scenario, the CPU attention computation, CPU-GPU weight transfer, and GPU computation is fully overlapped in the execution pipeline.
% For each LLM inference iteration, both the weights and the KV cache are read once from the CPU memory to get transferred either to the GPU or the CPU cores.
% Therefore, the CPU memory bandwidth requirement is the sum of the bandwidth requirement of accessing KV cache $B_{KV}$ and the bandwidth requirement of transferring weights from CPU to GPU $B_{IO}$.
In a typical deployment scenario, CPU attention computation, CPU-GPU weight transfer, and GPU computation are fully overlapped in the execution pipeline.
During each LLM inference iteration, both the model weights and the KV cache are read once from CPU memory, either for transfer to the GPU or for use by the CPU cores.
As a result, the total CPU memory bandwidth requirement is the sum of the bandwidth needed to access the KV cache ($B_{KV}$) and the bandwidth  to transfer weights from CPU to GPU ($B_{IO}$).
% \begin{align*}
% B_{Mem} &= B_{KV} + B_{IO} = \frac{M_{kvcache}}{M_{weight}}B_{IO} + B_{IO} \\
% &= (M_{weight} + M_{kvcache})\frac{B_{IO}}{M_{weight}} = M \frac{B_{IO}}{M_{weight}}
% \end{align*}
\begin{equation} \label{eq:mem_bandwidth}
B_{Mem} = B_{KV} + B_{IO} = \frac{M}{M_{weight}/B_{IO}} = \frac{M}{M_{weight}}B_{IO}
\end{equation}
Here, $M$ and $M_{weight}$ are the total memory capacity and the memory capacity reserved for the model weight, respectively.
The required CPU compute throughput for attention is related to the memory access bandwidth for the KV cache.
Assuming that the CPU up-converts the KV cached stored in BF16 to FP32 for computation, we have
\begin{equation}
T_{CPU} = 2 \cdot s \cdot I_{\text{cpu\_attn}} \cdot B_{KV}
\end{equation}
where $s$ is the GQA group size, and $I_{\text{cpu\_attn}}$ is the arithmetic intensity of attending to one query in the KV cache. 
$I_{\text{cpu\_attn}}$ is inherently small for the flash attention based implementation, as its primary operations are vector doc product and \texttt{saxpby}.

% \noindent
% \textbf{Design Decision Implications.}
In resource-constrained environments, the CPU memory capacity is usually limited.
% {
% \color{blue}
% While the CPU memory bandwidth increases as KV cache size increases, it is usually not a bottleneck for a CPU that supports multiple memory channels.
% For example, when the KV cache is twice as large as the model size, which is reaching 200GB for \texttt{Mixtral8x7B}, the $B_{Mem}$ is around three times the PCIe bandwidth (60GB/s), which is lower than the bandwidth provided by 4-channel DDR4@3200Mhz ($3 \times 25.6$.
% }
% While the CPU memory bandwidth requirement increases with an increase in the KV cache size, it is usually not a bottleneck for a CPU that supports multiple memory channels (more than 100GB/s bandwidth).
% For example, consider an extreme case when the KV cache is twice as large as as the model weight.
% For \texttt{Mixtral8x7B}, this would mean a KV cache size of 200GB.
% Using Equation~(\cite{eq:mem_bandwidth}), the memory bandwidth to ensure that that does not become a bottleneck is three times the PCIe bandwidth, i.e., $B_{Mem}=60GB/s$, which is reasonable for a CPU with multiple memory channels to support.
While the CPU memory bandwidth requirement increases with the size of the KV cache, it typically does not become a bottleneck on modern CPUs equipped with multiple memory channels: often supporting bandwidths exceeding 100GB/s. 
For instance, consider an extreme case where the KV cache is twice the size of the model weights.
In the case of \texttt{Mixtral8x7B}, this corresponds to a 200GB KV cache. 
According to Equation~(\ref{eq:mem_bandwidth}), the memory bandwidth required to avoid becoming a performance bottleneck is approximately three times the PCIe bandwidth, \textit{i.e.,} $B_{\text{Mem}} = 60\,\text{GB/s}$. 
This is well within the capabilities of modern CPUs.
% For example, if the KV cache is twice as large as the model size, which is reaching 200GB for \texttt{Mixtral8x7B}, the $B_{Mem}$ is around three times the PCIe bandwidth (60GB/s), lower than what the CPUs can provide.
% If we assume the memory allocated to KV cache is twice as large as the model size, $B_{Mem}$ is around three times the PCIe bandwidth (60GB/s), usually not a bottleneck for CPUs that support multiple memory channels (more than 100GB/s).
On the other hand, sustaining GPU utilization requires the CPU attention computation to deliver throughput on the order of hundreds of GFLOPs. Achieving this performance necessitates efficient use of vector units on modern CPUs.
We provide a detailed analysis of this challenge and our implementation in \S\ref{sec:sys-cpu-attn}.

\begin{tcolorbox}[width=0.48\textwidth]
\textbf{\textit{Takeaway:}} \textit{Fully utilizing CPU vector units and memory bandwidth through multiple channels is essential for high-throughput MoE inference.}
\end{tcolorbox}

% When using the flash attention algorithm, the major computation carried out on the KV cache is vector dot product on the K cache with the query vectors, and the \texttt{saxpby} operation over the V cache and the query vectors.
% The 

% Also, although decode attention is usually more memory-dense than compute-dense, the CPUs exhibit constrained computational capacity.
% Therefore, without careful optimizations on the CPU attention kernels, the CPU side attention computation can become the bottleneck of the overall execution, despite the adequet CPU memory bandwidth.

\subsection{Understanding the Effect of Prefill - Decode Overlap}
\label{sec:perf-model-sched}
In \S\ref{sec:motivation-overlapping}, we briefly explain the motivation behind overlapping the prefill and decode stages.
\textit{
% An important research question is: \textbf{how does prefill-decode overlap effectively improves the CPU memory usage?}}
Another important benefit of this strategy is: \textbf{prefill-decode overlap effectively improves the CPU memory usage.}}
Overlapping allows the decoding process of some sequences to start earlier than others.
This reduces the peak memory consumption of a batch of sequences in KV cache, as at each forward pass, some sequences finish execution and release their memory in the KV cache.
For an inference batch with on average $p$ prefill tokens and $g$ generation tokens, when using prefill/decode overlapping, we effectively enlarge the KV cache capacity, $C_{KV}$, to
\begin{equation}
C_{KV, eff.} = \frac{p+g}{p + \frac{g}{2}} C_{KV}
\end{equation}
compared to making the prefill/decode two isolated stages, without severely overflowing the KV cache.
The memory usage per sequence can be estimated using the average sequence length $p + g/2$ in the KV cache instead of the maximum sequence length $p+g$.
% \textcolor{blue}{We use the average sequence length $p + g/2$ in the KV cache, instead of the maximum sequence length $p+g$, to estimate the batch size (the denominator)}.
It is critical for resource-constrained MoE inference, where the memory capacity is a key limiter for large execution batch size.

\begin{tcolorbox}[width=0.48\textwidth]
\textbf{\textit{Takeaway:}} \textit{Besides balancing hardware resource usage, prefill/decode overlapping effectively enlarges the KV cache.}
\end{tcolorbox}
% For a sequence with a certain prefill length, its density decreases when more tokens are generated.
% Mixing sequences with different progress further smooths the average density of sequences as the sequences with different generation progress have different parallelism-memory densities.
% Besides, it reduces the peak memory consumption of a batch of sequences in KVcache, as at each forward pass some sequences finish execution and release their memory in the KV cache.
% \begin{figure}
%     \centering
%     \includegraphics[width=\linewidth]{micro58-latex-template/Figures/batch-size-effect.png}
%     \caption{Caption}
%     \label{fig:enter-label}
% \end{figure}

\subsection{Stage 2 Model: Modeling Performance in Realistic Setting}
\label{sec:perf-model-e2e}
% In a realistic deployment scenario, the number of sequences in a batch request is not infinite and is bounded by a large number $K$.
% Besides, the KV cache is usually organized in blocks~\cite{pagedattn}, with a block size of $b$ token slots and $N$ blocks in total.
% In this section, we build our resource and workload aware performance model (\textit{Stage 2} in \S\ref{sec:thiswork_overview}).
% It assumes an ideal pipelined execution scenario ignoring variance in inputs and system performance, and follow the \textit{Stage 1} analysis but further augmenting it with CPU resources consideration, prefill/decode overlapped scheduling strategy, the request batch size $K$, and paged KV cache.
% The model aims at accurately predicting the end-to-end execution time and being practical enough to guide real system designs (\textit{Stage 3}), while still converges to the \textit{Stage 1} upper-bound to be aligned with the hardware limits. 
In a realistic deployment scenario, the number of sequences in a batch request is not unbounded, but limited to a large finite value \( K \). 
Additionally, the KV cache is typically organized into blocks~\cite{pagedattn}, each containing \( b \) token slots, with a total of \( N \) blocks. 
In this section, we present our resource- and workload-aware performance model (\textit{Stage 2} in \S\ref{sec:thiswork_overview}). 
This model builds upon the idealized analysis in \textit{Stage 1}, while incorporating additional considerations such as CPU resource constraints, prefill/decode overlapped scheduling strategies, bounded request batch size \( K \), and paged KV cache organization.
The goal is to provide a performance model that accurately predicts end-to-end execution time and remains practical for guiding real system designs (\textit{Stage 3}), while still converging to the hardware-limited upper bound established in \textit{Stage 1}.
% To achieve this, it augments the \textit{Stage 1} analysis in \S\ref{sec:perf-model-PME} with real system execution factors, the request batch size $K$, and paged KV cache.

% \textcolor{blue}{should be able to link back to 5.2, this is just including the pipeline effect!}
% For modeling purposes, we assume the inference engine is processing a batch of $K$ sequences with prefill length $p$ and generation length $g$.
% For the Kv cache, we assume it is organized in blocks, with block size $b$ and total $N$ blocks.

We define $\delta = \frac{\text{Model Size}}{B_{IO}}$ same as \S\ref{sec:perf-model-PME}.
When the number of tokens processed in parallel is less than the GPU's limit $T_{GPU}$, each inference iteration takes time $\delta$ to finish as IO becomes a bottleneck.
In each iteration, we schedule $q$ sequences for prefill, where $q$ is calculated as
\begin{equation}
q = \frac{N}{\sum_{i=0}^g\lceil\frac{p + i}{b}\rceil}
\end{equation}
% Such a schedule ensures that we maximize the KV cache usage without subscribing it.
% For each inference iteration, $q (p + g)$ tokens are released and generated, due to the completion of the oldest sequences and sequences' generation.
% Note that in the naive scheduling where the prefill and decode stages are separated, the number of parallel decode tokens is $\frac{N}{p + g}$.
% More active sequences can be decoded in parallel as
Such a scheduling strategy ensures maximal utilization of the KV cache without over-committing it.
In each inference iteration, \( q(p + g) \) tokens are released and generated, due to the completion of older sequences and the progression of generation.
In contrast, under a naive scheduling approach that strictly separates the prefill and decode stages, the number of parallel decode tokens is limited to \( \frac{N}{p + g} \).
% By overlapping prefill and decode, the system can support a greater number of active sequences in parallel, thereby improving overall throughput and resource efficiency.
More active sequences can be decoded in parallel as
\begin{equation}
gq = g  \frac{N}{\sum_{i=0}^g\lceil\frac{p + i}{b}\rceil} > g\frac{N}{g(p+g)} = \frac{N}{p + g}
\end{equation}
% In the same iteration, the same number of tokens are processed by either prefilling or decoding, and the number of tokens being processed in parallel is $q (p + g)$.
\noindent
The generation throughput for a batch with $K$ sequences is
\begin{equation}
T_1 = \frac{K \cdot g}{(\frac{K}{q} + g)\delta} = \frac{K}{K + gq} \frac{gq}{\delta}
\end{equation}
\noindent
where $\frac{K}{K + gq}$ is the slowdown factor due to the epilogue of pipelining and $\frac{gq}{\delta}$ is the ideal decoding throughput with $gq$ as the number of sequences active for decoding.

When $T_{GPU} < q (p + g)$, the inference is bottlenecked by GPU computation instead of CPU memory capacity.
The total execution time is dominated by the speed we prefill sequences.
During execution, both the decode tokens and prefill tokens consume GPU computation.
The prefill throughput in the middle of the software pipeline is
\begin{equation}
T_{prefill} = T_{GPU}\frac{p}{p+g}
\end{equation}
The prologue of the software pipeline takes $g$ iterations, where the average prefill throughput is $\frac{T_{prefill} + T_{GPU}}{2}$.
Consequently, $K \cdot p - \frac{T_{prefill} + T_{GPU}}{2}g$ tokens are processed in the main pipeline stages.
The total number of iterations for the software pipeline is
\begin{equation}
It = 2 * g + (K \cdot p - \frac{T_{prefill} + T_{GPU}}{2} g) / T_{prefill}
\end{equation}
The end-to-end throughput is
\begin{equation}
T_2 = \frac{k \cdot g}{It \cdot \delta}
\end{equation}
An inference process is either bounded by computation or memory capacity, so the throughput will be
\begin{equation}
T = min(T_1, T_2)
\end{equation}

As shown in \S\ref{sec:eval-perf-model-acc}, this model achieves on average 94\% accuracy against real execution time.

\begin{figure}
    \centering
    \includegraphics[width=\linewidth]{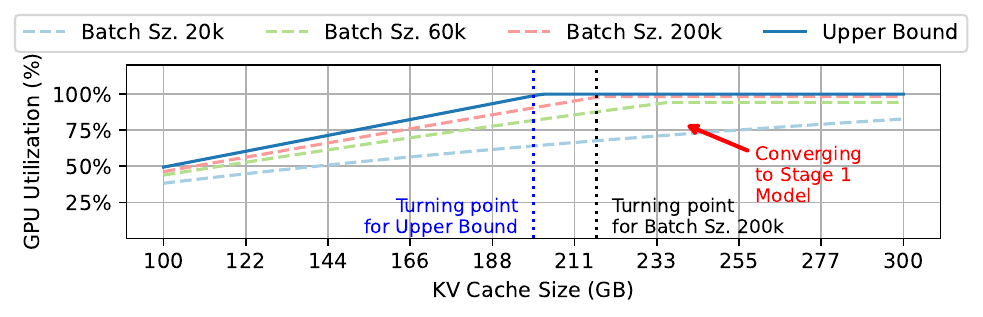}
    \vspace{-0.7cm}
    \caption{Predicted GPU utilization under different request batch sizes, with $p=100$ and $g=128$.}
    \label{fig:real-effect}
    \vspace{-0.3cm}
\end{figure}

{
% \color{blue}
\noindent
\textbf{Impact of real system execution factors.}
% We present the predicted GPU utilization under different request batch sizes in Figure~\ref{fig:real-effect}. 
% A larger request batch size both improves the throughput benefits brought by a larger KV cache and the maximum achievable throughput, by amortizing the pipeline prologue/epligue overhead.
% By comparing the line of batch size 200k with the performance upper bound, we can notice that paged KV cache shifts the turning point to the right, effectively increasing the memory capacity demand to achieve the same GPU utilization.
We present the predicted GPU utilization across different request batch sizes in Figure~\ref{fig:real-effect}.
Increasing the batch size improves both the throughput gains from a larger KV cache and the maximum achievable throughput as pipeline prologue and epilogue overhead is amortized.
Comparing the curve for a batch size of 200k against the theoretical upper bound, we observe that the use of paged KV cache shifts the turning point to the right, effectively increasing the memory capacity requirement to reach the same level of GPU utilization.
\textit{As the batch size theoretically approaches infinity and the KV cache block size approaches 1, the Stage 2 model converges to the Stage 1 theoretical upper bound, delivering the same result.}
}

\begin{tcolorbox}[width=0.48\textwidth]
\textbf{\textit{Takeaway:}} \textit{Limited number of sequences in a request batch and paged KV cache reduces the effectiveness of a large KV cache size compared to the theoretical analysis.}
\end{tcolorbox}

\section{\THISWORK\ System Design}
\label{sec:system-design}
\begin{figure}
    \centering
\includegraphics[width=\linewidth]{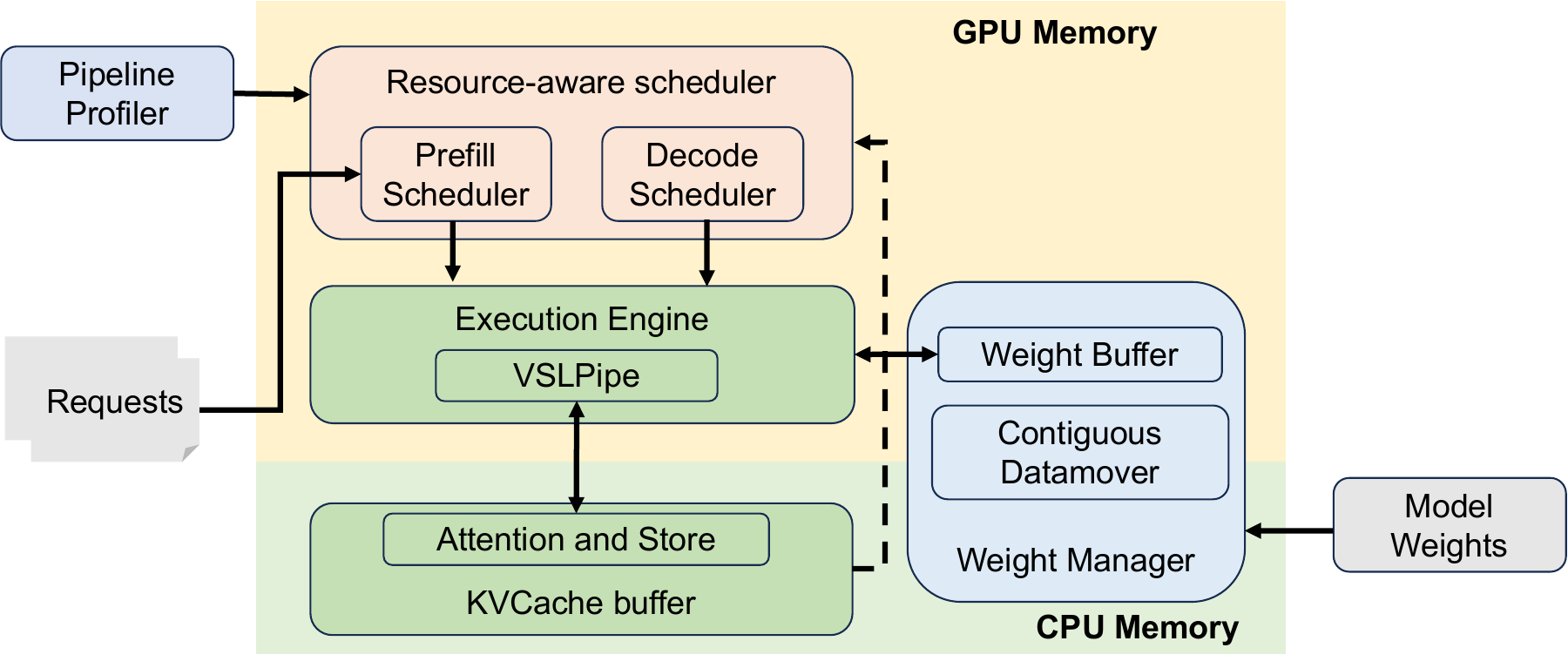}
    \vspace{-0.7cm}
    \caption{System overview of \THISWORK.}
    \vspace{-0.5cm}
    \label{fig:high-level}
\end{figure}

% Informed by insights from the proposed performance model (\S\ref{sec:perf-model}), \THISWORK\ presents a system design that closely approaches the performance limit projected by the \textit{Stage 2} model, while being adaptive to variations in input sequence lengths and underlying hardware performance.
Building on the insights from our architecture-aware performance model (\S\ref{sec:perf-model}), \THISWORK\ introduces a system design that closely tracks the theoretical performance limit projected by the \textit{Stage 2} model while remaining adaptive to real-world variations in input sequence lengths and hardware performance.

% The goal: realize the performance we predict, and also reversely prove our modeling is correct.
% \begin{enumerate}[nosep, leftmargin=*]
    % \item We need to fully utilize the KV cache capacity -> design decision: schedule as many sequences as we can greedily as long as there are space in the kvcache
    % \item Overlapping prefill and decode is important to achieve good performance -> use pipeline profiler to set a limit on the speed we process the prefill sequences: if we finish them all at once, it degenerates to two stages
    % \item to support prefill decode overlap, we need a pipeline that supports this, while not comprosing the good IO utilization which is achieved by prior work
%     \item to achieve the predicted performance, the CPU computation should be good enough, but the auto vec cannot do that. so we hand-optimize with intrinsic.
% \end{enumerate}

\subsection{Design Overview}

% Figure \ref{fig:high-level} illustrates the key components of \THISWORK\ system.
% When \THISWORK\ is deployed on a machine for a certain model, the \textbf{Pipeline Profiler} will profile the number of tokens to saturate the GPU computation becomes a bottleneck, which is later used by the scheduler to avoid overwhelming the pipeline.
% The inference requests are sent to the \textbf{Resource-Aware Scheduler}.
% This overlaps the prefill and decode stages of sequences, and makes prefill and decode scheduling decisions based on the information about available CPU/GPU architecture resources.
% At the core of the execution engine is \THISWORK's \textbf{VSLPipe}, abbreviation for \textit{Versatile Pipeline}, which efficiently processes prefill and decode requests together in resource-constrained environments.
% It uses weight tensors in \textit{Weight Buffer} managed by \textit{Weight Manager}, which leverages \textbf{Contiguous Data Mover} to eliminate the IO bubbles for weight transfer.
% The attention operation is fully offloaded to the CPU, and our optimized flash attention CPU kernel ensures that the CPU-side computation does not bottleneck the entire system throughput.
Figure~\ref{fig:high-level} illustrates the core components of the \THISWORK\ system.  
When deployed on a given machine for a specific model, the \textbf{Pipeline Profiler} identifies the number of tokens required to saturate the GPU compute, guiding the scheduler to prevent pipeline overcommitment.  
Incoming inference requests are handled by the \textbf{Resource-Aware Scheduler}, which overlaps prefill and decode stages and makes scheduling decisions based on the current availability of CPU and GPU resources.  
At the heart of the execution engine is \THISWORK's \textbf{VSLPipe}—short for \textit{Versatile Pipeline}—which efficiently co-processes prefill and decode requests in resource-constrained environments.  
It utilizes weight tensors stored in the \textit{Weight Buffer}, managed by the \textit{Weight Manager}, which coordinates with the \textbf{Contiguous Data Mover} to eliminate IO stalls during weight transfers.  
All attention computations are offloaded to the CPU, and our optimized CPU-based flash attention kernel ensures this stage does not become a throughput bottleneck.

\subsection{Resource-Aware Scheduler}
% The goals of the resource-aware scheduler is to properly schedule sequences' execution such that we can (a) fully utilize the CPU memory capacity and (b) achieves prefill/decode overlapping, based on our findings in performance model (\S\ref{sec:perf-model-PME} and \S\ref{sec:perf-model-e2e}).
% Compared to the analytical models, the system implementation needs to be adaptive to deal with varying sequence lengths and varying hardware performance.
% The resource-aware scheduler comprises two components, the \textbf{Prefill Scheduler} and \textbf{Decode Scheduler}.
% They manage the progression of sequences through prefill and decode stages while fully utilizing the limited memory capacity for KV caching in resource-constrained environments.
The goal of the Resource-Aware Scheduler is to schedule sequence execution in a way that (a) fully utilizes available CPU memory capacity, and (b) enables effective prefill/decode overlap, as the models in \S\ref{sec:perf-model-PME} and \S\ref{sec:perf-model-e2e} have shown the importance of CPU memory capacity and scheduling strategy in saturing the hardware.  
Unlike analytical models, the system implementation must adapt dynamically to variations in sequence lengths and hardware performance.  
The scheduler consists of two components: the \textbf{Prefill Scheduler} and the \textbf{Decode Scheduler}.  
Together, they orchestrate the flow of sequences through the prefill and decode stages, ensuring efficient use of memory for KV caching under resource constraints.
New sequences are initially queued in the Prefill Scheduler, awaiting prompt processing.
In the steady state, there are enough prefill and decode sequences to process.
Once a sequence completes the prefill stage, control is handed off to the Decode Scheduler for token generation.
During execution, both schedulers may issue sequences, whether in the prefill or decode stage, for parallel processing.  
To maximize efficiency, both schedulers run on the GPU and maintain their scheduling states directly in GPU memory.

% The new sequences are first pushed into the prefill scheduler, pending for processing the prompts.
% After a sequence is prefilled, its control is transferred to the decode scheduler for further generation.
% During scheduling, both schedulers may schedule sequences that are in either the prefill stage or the decode stage for parallel execution.
% Both schedulers execute on the GPU, maintaining states on the GPU for efficiency. 

% \textcolor{red}{NT: mention that in the steady state, there will be enough number of tokens to process in both queues.}

% \begin{itemize}[nosep, leftmargin=*]
%     \item \textit{Prefill Scheduler. } 
%     This scheduler maintains a queue of pending sequences with their associated token IDs and sequence lengths.
    % When a new sequence is scheduled for prefill, it is assigned a \textit{physical sequence ID}, which is used to index execution pipeline resources such as the KV cache page table. 
    % Meanwhile, its original \textit{logical sequence ID} is retained for indexing the sequence metadata, like sequence length and output buffer entries.
    % It uses a pointer to mark the head of the queue for the sequences that have not been prefilled yet.
    % When new sequences are scheduled for prefill, they will be allocated with a \textit{physical sequence ID}.
    
%     \item \textit{Decode Scheduler.} 
%      This scheduler manages sequences that have completed the prefill stage.
%      It tracks both logical and physical sequence IDs along with their sequence lengths, ensuring efficient scheduling and execution.
% \end{itemize}

\begin{figure}
    \centering
    \includegraphics[width=\linewidth]{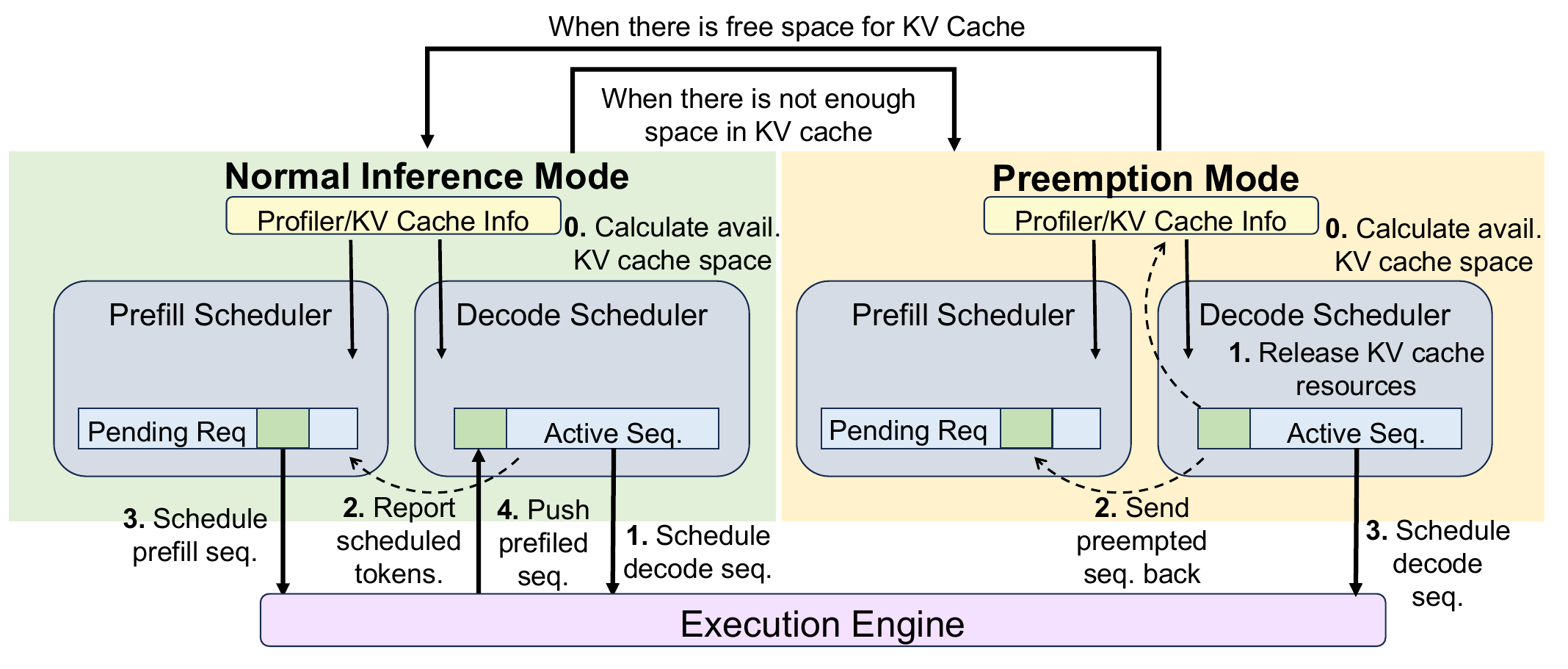}
    \vspace{-0.7cm}
    \caption{Operation of the Resource-Aware Scheduler.}
    \label{fig:io-scheduler}
    \vspace{-0.5cm}
\end{figure}
Figure~\ref{fig:io-scheduler} illustrates the two modes of interaction between the Prefill Scheduler and the Decode Scheduler.
Based on the availability of KV cache blocks, the system dynamically switches between: \textit{Normal Inference Mode}, where both schedulers operate concurrently without interference, and  
\textit{Preemption Mode}, where old decode sequences are prioritized by temporarily preempting new decode sequences to free up memory.

\begin{figure}
    \centering
    \includegraphics[width=\linewidth]{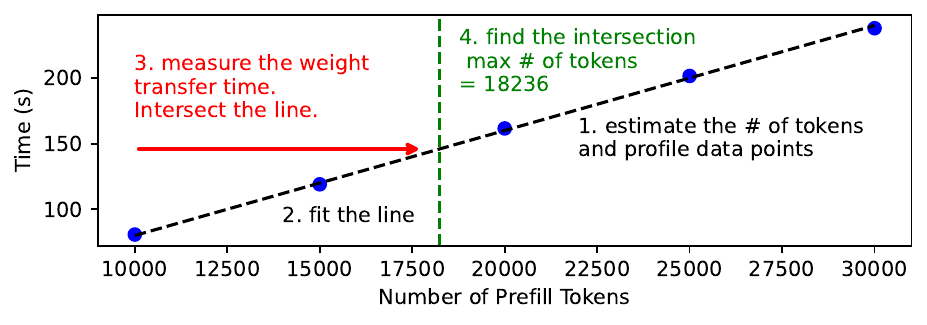}
    \vspace{-0.7cm}
    \caption{An Example for the threshold $n_{\text{real}}$ searching process of \THISWORK's pipeline profiler, which runs \texttt{Mixtral8x7B} on \texttt{A40}.}
    \label{fig:profiler}
    \vspace{-0.6cm}
\end{figure}

\noindent
\textbf{Normal Inference Mode.}
Before scheduling begins, the Decode Scheduler estimates the number of KV cache blocks required to decode the next token for its managed sequences.  
If a sufficient number of KV blocks are available, the system enters normal inference mode.
In this mode, both the Prefill Scheduler and Decode Scheduler can schedule sequences concurrently.  
The Decode Scheduler first schedules all sequences currently in the decode stage onto the execution engine.  
Next, the Prefill Scheduler reads the number of active decode sequences and calculates how many additional prefill tokens it can schedule without exceeding the pipeline capacity, as measured by the Pipeline Profiler (§\ref{sec:pipeline_profiler}).  
It then schedules prefill requests from the head of its queue, staying within this threshold.  
After the inference pass, newly prefilled sequences are handed off to the Decode Scheduler for subsequent token generation.  
As decode completes, the Decode Scheduler performs garbage collection to reclaim resources such as KV cache blocks.

\noindent
\textbf{Preemption Mode.}
If the Decode Scheduler detects insufficient KV cache blocks to schedule all decode-stage sequences, the system enters preemption mode.
In this mode, a subset of active decode sequences are preempted to free up KV cache resources for other sequences that are ready to generate the next token.  
The KV cache entries and other associated resources from the preempted sequences are reclaimed and reassigned to the remaining decode sequences.
Meanwhile, the Prefill Scheduler halts the scheduling of any new incoming sequences.  
Instead, it accommodates the preempted decode sequences, treating them as newly arrived sequences that must go through the prefill stage again.
This effectively re-inserts them into the execution pipeline from the beginning, but with the advantage that their earlier progress has already been partially completed.
% \textcolor{blue}{To make room for the sequences in the decode stage for further generation, some sequences at the decode stage are preempted, and their resources like KV cache entries are released.}

% In this mode, the prefill scheduler halts the scheduling of new sequences but instead accommodates preempted sequences from the decode scheduler. 
% The decode scheduler preempts the newly enqueued sequences to make room in KVCache. 
% After locating the target sequences, it collects the generated token IDs from the output buffer and sends both the sequence IDs and their token lists to the prefill scheduler.
% The prefill scheduler overwrites the existing entries in the prefill scheduler queue and moves the queue head pointer to accommodate the preempted sequences at the head of the prefill scheduler.
% Effectively, the preempted sequences are treated as new sequences pending prefilling.

% We list the complete sequence scheduling procedure of \THISWORK\ in Algo. \ref{alg:llm_scheduler}.

\noindent
\textbf{Discusstion.}
A key design in our scheduler is to alleviate preemption overhead with overlap.
When a sequence is preempted, its KV cache is evicted and must be re-prefilled.
Without overlapping prefill and decode, this recomputation can only begin after the current decode stage completes, incurring additional latency.
In contrast, overlapping prefill with decode allows the recomputation to proceed concurrently, effectively hiding its cost and minimizing impact on overall execution time.
% After a sequence is preempted, its KV cache is deleted and needs to be re-prefilled into the KV cache.
% Without prefill-decode overlapping, such recomputation can only happen after the decoding process of the current batch finishes, which increases the overall execution time.
% On the other hand, the cost of recomputation can be hidden when it is overlapped with the decoding process.
%as the recomputation, similar to prefilling, has high PME.

% \begin{algorithm}
% \caption{\THISWORK's Scheduling Algorithm}
% \label{alg:llm_scheduler}
% \begin{algorithmic}[1]
% \State Initialize the pipeline profiler, kvcache, and execution engine
% \State push income sequences to the prefill scheduler

% \While {$\neg$ sched\_prefill.done() $\land$ $\neg$ sched\_decode.done() }

% \If{sched\_decode.need\_preemption()}
% \State descheduled\_seqs $\gets$ sched\_decode.deschedule()
% \State prefill\_sched.deschedule(descheduled\_seq)
% \State decode\_seqs $\gets$ sched\_decode.schedule()
% \State prefill\_seqs $\gets$ empty
% \Else
% \State decode\_seqs $\gets$ sched\_decode.schedule()
% \State prefill\_meta $\gets$ len(decode\_seqs), kvcache.free\_blocks(), profiler.exec\_lim()
% \State prefill\_seqs $\gets$ prefill\_sched.schedule(prefill\_meta)
% \EndIf

% \State prefilled\_seqs $\gets$ exec\_engine.forward(decode\_seqs, prefill\_seqs)

% \State decode\_sched.push(prefilled\_seqs)
% \State done\_seqs $\gets$ decode\_sched.garbage\_collect()
% \State prefill\_sched.release(done\_seqs)

% \EndWhile

% \end{algorithmic}
% \end{algorithm}

\subsection{Pipeline Profiler} \label{sec:pipeline_profiler}
% To avoid draining the prefill sequences fast and deviating from the prefill/decode overlapping strategy we presented in \S\ref{sec:perf-model-e2e}, \THISWORK\ includes a pipeline profiler to estimate the maximum number of tokens, $n_{real}$, until the CPU computation is saturated, \textit{i.e.,} the measured value of $n$ in Equation~\ref{eqn:sat_GPU}.
% If the resource-aware scheduler schedules more tokens than $n_{real}$, the CPU-GPU weight transfer is no longer a bottleneck, but the GPU computation.
% The resource-aware scheduler schedules as many sequences as possible, but their total number of tokens is always below $n_{real}$.
% This policy ensures that the available GPU compute throughput is fully utilized, while preventing the scheduler from excessively scheduling sequences in the prefill stage beyond the GPU's capability. 
% \textcolor{red}{
To maintain effective prefill/decode overlap as modeled in \S\ref{sec:perf-model-e2e}, \THISWORK\ uses a Pipeline Profiler to estimate the token threshold \( n_{\text{real}} \) at which GPU-side GEMM becomes the bottleneck, \textit{i.e.,} the measured \( n \) in Equation~\ref{eqn:sat_GPU}.
The resource-aware scheduler ensures that the total number of scheduled tokens stays below \( n_{\text{real}} \). 
This avoids prematurely exhausting prefill sequences and dimensing the effects of prefill/decoce overlapping, thereby sustaining high GPU utilization.
% }
% The pipeline profiler first estimates $n$ based on Equation~\ref{eqn:sat_GPU}.
% Then, it sweeps around that number of tokens to prefill pipelines with different numbers of tokens, and measures the GPU computation time. 
% Based on the collected data, it fits a line to describe the relationship between the number of tokens and the GPU computation time.
% Then, it measures how long it takes to transfer a layer of weight to the GPU.
% The maximum number of parallel tokens can be calculated using the slope and the weight transfer time.
% An example of finding the maximum number of tokens for \texttt{Mixtral8x7B} on A40 is presented in Figure~ \ref{fig:profiler}.
The pipeline profiler estimates \( n \) using Equation~\ref{eqn:sat_GPU}, then varies the number of prefilled tokens and measures the corresponding GPU computation time.
As shown by an example for \texttt{Mixtral8x7B} on the A40 in Figure~\ref{fig:profiler}, the profiler fits a line to capture the relationship between token count and GPU time.
It also measures the time required to transfer a layer of weights to the GPU.
It calculates the maximum number of parallel tokens using the line's slope and weight transfer time.

\subsection{Execution Engine}
To achieve prefill/decode overlapping without compromising CPU-GPU IO efficiency, as outlined in the performance model (\S\ref{sec:perf-model-e2e}), \THISWORK's execution engine features a novel CPU-GPU hybrid pipelined schedule, \textit{VSLPipe}.
It is designed to maximize throughput when executing a batch of requests with both prefill and decode sequences.

% To achieving the prefill/decode overlapping, while not compromising the CPU-GPU IO efficiency as outlined by the performance model in \S\ref{sec:perf-model-e2e}, \THISWORK's execution engine includes a novel CPU-GPU pipelined schedule, \textit{VLSPipe}.
% It is designed to maximize the throughput of executing a batch of requests with both prefill and decode sequences.

\noindent
\textbf{VSLPipe Compute Graph Division.}
\begin{figure}
    \centering
    \includegraphics[width=\linewidth]{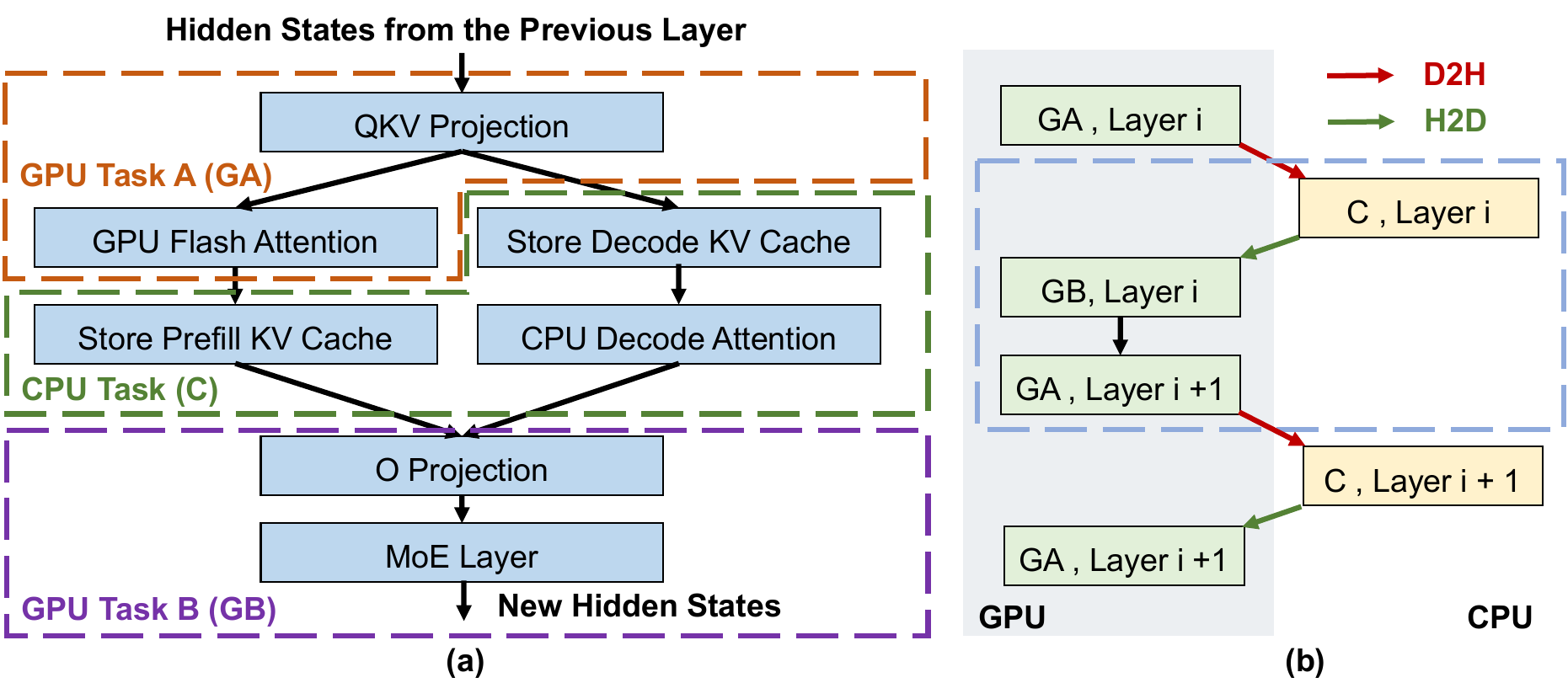}
    \vspace{-0.7cm}
    \caption{Compute graph division of \THISWORK's VSLPipe.}
    \label{fig:pipe-comp-graph-div}
    \vspace{-0.5cm}
\end{figure}
Figure~\ref{fig:pipe-comp-graph-div}(a) shows how VSLPipe restructures the compute graph of a MoE transformer layer.
The per-token \textit{QKV projection} and \textit{GPU Flash Attention} are grouped into \textit{GPU Task A (GA)}, producing KV values for both prefill (post-attention) and decode (pre-attention) sequences.
These KV values are offloaded to the CPU, where \textit{CPU Task (C)} stores them in the KV cache and performs \textit{Decode Attention} for decode tokens.
The attention results for decode tokens are transferred back to the GPU and combined with prefill outputs.
Finally, \textit{GPU Task B (GB)}, which includes the \textit{O projection} and MoE layer, is applied to all tokens.
\begin{figure*}[h]
    \centering
    \hspace*{-7mm}
\newcommand{\myfont}{\fontsize{5pt}{6.4pt}\selectfont} 
\newcommand{\myfontb}{\fontsize{7pt}{8.4pt}\selectfont} 
    \begin{tikzpicture}[
         task/.style={
         chamfered rectangle, chamfered rectangle xsep=1cm, chamfered rectangle angle=30, draw,
         font=\myfont,
         minimum height=10pt,  % 基于当前字体高度的相对单位
         inner ysep=0pt
        },
        gpu/.style={task, fill=green!30, minimum width=1.2cm},
        cpu/.style={task, fill=blue!30, minimum width=1.3cm},
        io/.style={task, fill=red!30},
        timeline/.style={->, thick},
        label/.style={text width=2cm, align=right},
    ]
    
    % % 定义时间轴
    % \draw[timeline] (0,0) -- (12,0);
    % \foreach \x in {0,1,...,11} {
    %     \draw (\x,0) -- (\x,-0.1) node[below] {\x};
    % }
    
    % 添加标签
    \node[label] (Lg) at (-1.5,2) {GPU};
    \node[label, anchor=south, yshift=-1cm] (Lc) at (Lg) {CPU};
    \node[label, anchor=south, yshift=-1cm] (Lio) at (Lc) {Data Mover};
    
    % prologue
    \node[gpu, anchor=west, xshift=.3cm] (g1) at (Lg.east) {$GA_{0}(\alpha)$};
    \node[cpu, anchor=west] (c1) at (g1.east |- Lc.east) {$C_0$($\alpha$)};
    \node[io, anchor=west, xshift=.3cm, minimum width=2.5cm] (io1) at (Lio.east) {Prologue Weight Loading};

    %% sync Phase
    \draw[dashed, blue, thick, shorten <=-2mm, shorten >=-2mm] (g1.east |- g1.north) -- (g1.east |- c1.south);
    \node[above, yshift=2mm, font=\myfontb] at (g1.east |- g1.north) {Sync Phase};

    %% prologue phase 2
    \node[gpu, anchor=west] (g2) at (g1.east) {$GA_{0}(\beta)$};

    %% sync stage
    \draw[dashed, red, thick, shorten <=-2mm, shorten >=-1.8mm] (io1.east |- g1.north) -- (io1.east |- io1.south);
    \node[above, yshift=2mm, font=\myfontb] at (io1.east |- g1.north) {Sync Stage};

    \draw[black, thick] ($(io1.west |- io1.west) - (-0.05cm, 0.5cm)$) -- node[pos=0.5, fill=white, inner sep=1pt, font=\myfontb] {Prologue}  ($(io1.east |- io1.east) - (0.05cm, 0.5cm)$);

    % Main Pipeline Stage 1
    \node[gpu, anchor=west, minimum width=1.2cm] (g3) at (io1.east |- g2.east) {$GB_{0}(\alpha)$};
    \node[gpu, anchor=west] (g4) at (g3.east) {$GA_{1}(\alpha)$};
    \node[cpu, anchor=west] (c2) at (io1.east |- Lc.east) {$C_0$($\beta$)};
    \node[io, anchor=west, minimum width=5.1cm] (io2) at (io1.east) {Pipeline Weight Loading Stage 0};
    
    %% sync Phase
    \draw[dashed, blue, thick, shorten <=-2mm, shorten >=-2mm] (g4.east |- g4.north) -- (g4.east |- c2.south);
    \node[above, yshift=2mm, font=\myfontb] at (g4.east |- g4.north) {Sync Phase};

    %% phase 2
    \node[gpu, anchor=west, minimum width=1.2cm] (g5) at (g4.east) {$GB_{0}(\alpha)$};
    \node[gpu, anchor=west] (g6) at (g5.east) {$GA_{1}(\alpha)$};
    \node[cpu, anchor=west] (c3) at (g4.east |- Lc.east) {$C_1$($\alpha$)};

    %%% a bubble
    % \draw[rounded corners, anchor=west] (g6.east |- g6.south) rectangle (io2.east |- g6.north);
    
    %% sync stage
    \draw[dashed, red, thick, shorten <=-2mm, shorten >=-1.8mm] (io2.east |- g6.north) -- (io2.east |- io2.south);
    \node[above left, yshift=2mm, font=\myfontb] at (io2.east |- g6.north) {Sync Stage};

    \node[font=\myfontb, align=center] at ($(io2.east |- c3.east) + (0.25cm, 0)$) {$\cdots$};
    
    %% sync stage
    \draw[dashed, red, thick, shorten <=-2mm, shorten >=-1.8mm] 
    ($(io2.east |- g6.north) + (.5cm, 0)$) -- ($(io2.east |- io2.south) + (0.5cm, 0)$);
    \node[above right, yshift=2mm, font=\myfontb] at ($(io2.east |- g6.north) + (.5cm, 0)$) {Sync Stage};

    % Main Pipeline Stage 2
    \node[io, anchor=west, minimum width=5.1cm] (io3) at ($(io2.east |- Lio.east) + (.5cm, 0)$) {Pipeline Weight Loading Stage $N-2$};
    
    \node[gpu, anchor=west, minimum width=1.2cm] (g6) at (io3.west |- g2.east) {$GB_{-2}(\alpha)$};
    \node[gpu, anchor=west] (g7) at (g6.east) {$GA_{-1}(\alpha)$};
    \node[cpu, anchor=west] (c3) at (io3.west |- Lc.east) {$C_{-2}$($\beta$)};
    
    %% sync Phase
    \draw[dashed, blue, thick, shorten <=-2mm, shorten >=-2mm] (g7.east |- g4.north) -- (g7.east |- c2.south);
    \node[above, yshift=2mm, font=\myfontb] at (g7.east |- g4.north) {Sync Phase};

    %% phase 2
    \node[gpu, anchor=west, minimum width=1.2cm] (g8) at (g7.east) {$GB_{-2}(\beta)$};
    \node[gpu, anchor=west] (g9) at (g8.east) {$GA_{-1}(\beta)$};
    \node[cpu, anchor=west] (c4) at (g7.east |- Lc.east) {$C_{-1}$($\alpha$)};
    
    %% sync stage
    \draw[dashed, red, thick, shorten <=-2mm, shorten >=-1.8mm] (io3.east |- g1.north) -- (io3.east |- io3.south);
    \node[above, yshift=2mm, font=\myfontb] at (io3.east |- g1.north) {Sync Stage};

    \draw[black, thick] ($(io2.west |- io2.west) - (-0.05cm, 0.5cm)$) -- node[pos=0.5, fill=white, inner sep=1pt, font=\myfontb] {Main Pipeline Stages}  ($(io3.east |- io3.east) - (0.05cm, 0.5cm)$);

    % epilogue
    \node[io, anchor=west, minimum width=2.5cm] (io4) at (io3.east) {Epilogue Weight Loading};
    \node[gpu, anchor=west] (g10) at (io4.west |- Lg.east) {$H(\alpha)$};
    \node[cpu, anchor=west] (c5) at (io4.west |- Lc.east) {$C_{-1}$($\beta$)};

    %% sync Phase
    \draw[dashed, blue, thick, shorten <=-2mm, shorten >=-2mm] (c5.east |- g4.north) -- (c5.east |- c2.south);
    \node[above, yshift=2mm, font=\myfontb] at (c5.east |- g4.north) {Sync Phase};
    
    \node[gpu, anchor=west] (g11) at (c5.east |- g10.east) {$H(\beta)$};
    
    \draw[black, thick] ($(io4.west |- io4.west) - (-0.05cm, 0.5cm)$) -- node[pos=0.5, fill=white, inner sep=1pt, font=\myfontb] {Epilogue}  ($(io4.east |- io4.east) - (0.05cm, 0.5cm)$);
    
    \node[below, xshift=2mm, font=\myfontb, anchor=west] at ($(Lio.east |- Lio.south) + (0, -7mm)$) {$GA_x$, $GB_x$, and $C_x$ are the GPU Task A, GPU Task B, and CPU Task, defined in Figure~\ref{fig:pipe-comp-graph-div}, at layer $x$. $H$ means $GB$ of the last layer and the model head.};

    % % 添加时间线
    % \draw[dashed] (g1.east) -- (0.5, -3);
    % \draw[dashed] (g2.south) -- (2.5, -3);
    % \draw[dashed] (g3.south) -- (4.5, -3);
    % \draw[dashed] (g4.south) -- (6.5, -3);
    % \draw[dashed] (g5.south) -- (8.5, -3);
    
    % \draw[dashed] (c1.south) -- (1.5, -2.2);
    % \draw[dashed] (c2.south) -- (3.5, -2.2);
    % \draw[dashed] (c3.south) -- (5.5, -2.2);
    % \draw[dashed] (c4.south) -- (7.5, -2.2);
    % \draw[dashed] (c5.south) -- (9.5, -2.2);
    
    % % 添加箭头表示依赖关系
    % \draw[->, thick, dotted] (g1.east) -- (c1.west);
    % \draw[->, thick, dotted] (c1.east) -- (i1.west);
    % \draw[->, thick, dotted] (g2.east) -- (c2.west);
    % \draw[->, thick, dotted] (c2.east) -- (i2.west);
    % \draw[->, thick, dotted] (g3.east) -- (c3.west);
    % \draw[->, thick, dotted] (c3.east) -- (i3.west);
    % \draw[->, thick, dotted] (g4.east) -- (c4.west);
    % \draw[->, thick, dotted] (c4.east) -- (i4.west);
    % \draw[->, thick, dotted] (g5.east) -- (c5.west);
    % \draw[->, thick, dotnner ted] (c5.east) -- (i5.west);
    
    \end{tikzpicture}
    \vspace{-0.7cm}
    \caption{\THISWORK's VSLPipe execution timeline.}
    \vspace{-0.3cm}
    \label{fig:execution-engine}
\end{figure*}
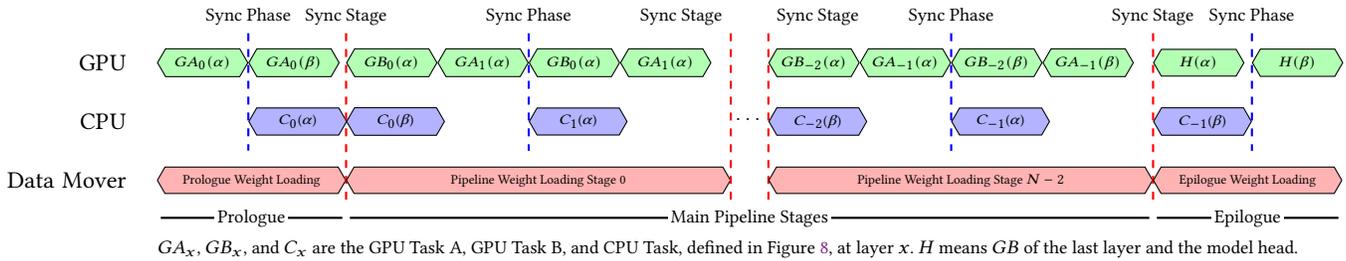

The CPU and GPU tasks from different layers can be regrouped to better delineate the boundaries between CPU and GPU computation, as illustrated in Figure~\ref{fig:pipe-comp-graph-div}(b).
Specifically, the CPU task from layer $i$, GPU Task B from layer $i$, and GPU Task A from layer $i+1$ are combined into a single \textit{execution stage}.
Each stage consists of two \textit{execution phases}: a CPU-only phase followed by a GPU-only phase.
Between these phases, either a \textit{Device-to-Host (D2H)} or \textit{Host-to-Device (H2D)} transfer occurs: KV values are offloaded to the CPU after GPU computation, and decode attention results are later loaded back to the GPU. 
Importantly, the amount of computation-related data transferred is relatively small compared to the cost of weight transfers.
The total data transferred per stage is bounded by $2n(d + \frac{2d}{s})$, which represents only a small fraction of the overhead when the number of tokens is large enough to saturate the GPU.

\noindent
\textbf{VSLPipe Scheduling.}
% After the execution engine receives the prefill and decode jobs, it makes two partitions, $\alpha$ and $\beta$, such that the number of decode tokens and prefill tokens is balanced between the two partitions.
% To maximize CPU-GPU utilization, \textit{VSLPipe} employs a software-pipelined execution strategy, as illustrated in Figure~\ref{fig:execution-engine}. 
% The execution consists of a \textit{pipeline prologue}, $N-1$ \textit{main pipeline stages}, and a \textit{pipeline epilogue} for a model with $N$ layers. 
% Each stage, including the prologue and epilogue, is divided into two phases.
% Within each phase, CPU-side attention computations for one partition and GPU-side GEMM operations for the other are executed concurrently.
% At the end of each phase, the CPU and GPU synchronize, exchanging computed query vectors and key-value (KV) cache data. 
% The attended query vectors are sent from the CPU to the GPU, while the query to be attended and KV cache are offloaded from the GPU to the CPU.
Upon receiving prefill and decode jobs, the execution engine partitions them into two groups, $\alpha$ and $\beta$, balancing the number of decode and prefill tokens in each.
To maximize CPU-GPU utilization, \textit{VSLPipe} applies a software-pipelined execution strategy, as shown in Figure~\ref{fig:execution-engine}.
The pipeline comprises a \textit{prologue}, $N - 1$ \textit{main stages}, and an \textit{epilogue} for a model with $N$ layers.
Each stage is divided into two phases.
In each phase, CPU-side attention computations for one partition run concurrently with GPU-side GEMM operations for the other.
At the end of each phase, the CPU and GPU synchronize to exchange intermediate results: attended query vectors are transferred from the CPU to the GPU, while fresh query vectors and KV cache entries are offloaded from the GPU to CPU.

% The amount of computation-related data transfer is relatively small compared to the weight transfer overhead. The total data transferred per computation is bounded by: $ 2n(d + \frac{2d}{s}) $.
% For $n=18500$, this amounts to approximately 200MB -- only a small fraction of the per-layer weight size.
% At the start of each stage, weights required for the next stage are prefetched by the \textit{Contiguous Data Mover}.
% The data mover operates asynchronously with the computation threads and is synchronized only at the end of each stage, not at the end of each phase.
% This ensures the data mover to schedule transfers independently and fully utilize the CPU-GPU bandwidth, thereby alleviating the bandwidth bottleneck.
% We detail the design of the data mover in the next section.
The computation-related data transfer is relatively small compared to the overhead of weight transfers.
The total data transferred per computation is bounded by $2n(d + \frac{2d}{s})$.
For instance, with $n = 18500$, this amounts to roughly 200MB: only a small fraction of the per-layer weight size.
To mitigate the CPU-GPU IO bottleneck, weights for the next execution stage are prefetched at the beginning of each stage by the \textit{Contiguous Data Mover}.
This data mover runs asynchronously with the computation threads and synchronizes only at the stage boundaries: not at the end of each individual phase.
This design allows it to schedule transfers independently and fully utilize the available CPU-GPU bandwidth.
We describe the data mover’s design in detail in the next section.

\subsection{Weight Layout, Weight Buffer, and Contiguous Data Mover}
% \subsubsection{How is the weight and weight buffer organized}
% The weight buffer can hold the weight for two layers.
% \THISWORK\ treat the weights of each layer as two parts, the layer-wise weights, which include attention's projection matrix and normalization weights, and expert weights.
% The weights of the MoE models are stored in CPU memory as pinned memory for fast CPU-GPU IO transfer.
% During inference, the weight values are loaded to the weight buffer on-the-fly.
\THISWORK\ treats the weights of each layer as two components: layer-wise weights, which include the attention projection matrices and normalization parameters, and expert weights specific to the MoE layers. 
All weights are stored in pinned CPU memory to enable efficient CPU-GPU IO transfers. 
During inference, weights are dynamically loaded into the GPU's weight buffer on demand.
The size of the weight buffer is two times the model weight size divided by the number of layers.
Given that a MoE model usually has tens of layers, the weight buffer is only a few percent of the original model size.

\noindent
\textbf{Contiguous Data Mover.}
% Instead of embedding the data movement API calls inside the execution pipeline, \THISWORK\ implements a separate \textit{Contiguous Data Mover} that is dedicated to IO operation running on a separate thread.
% The pipeline pushes the weight transfer requests to the data mover at layer-wise granularity, and the data mover performs the data transfer internally at finer granularity for overall efficiency.
% The data mover first re-groups the data it needs to transfer into small packets, and each time only issues one packet transfer request to the runtime. 
% This is because the computation pipeline also need to use the CPU-GPU IO transfer for some PyTorch operations as well as synchronizing attention inputs and outputs.
% If all weight transfer requests are dumped to the runtime at once, it causes head-of-line blocking for the computation IO requests and blocks their execution.
% Empirically, we set the packet size to 100MB to balance the data transfer speed and the degree of interference between computation IO and weight transfer IO.
% The contiguous data mover is implemented in C++ as a Pytorch extension.
Rather than embedding data movement API calls within the execution pipeline, \THISWORK\ implements a dedicated Contiguous Data Mover, running on a separate thread, to handle CPU-GPU IO.
The execution pipeline pushes weight transfer requests to the data mover at \textit{layer-wise granularity}, while the data mover internally performs fine-grained transfers for efficiency.
It first partitions the requested weights into small packets and issues one packet transfer at a time to the runtime.
This strategy prevents contention with other CPU-GPU transfers used by the compute pipeline, such as those triggered by PyTorch operations or attention-related data synchronization.
Issuing all weight transfers at once would lead to head-of-line blocking, delaying latency-sensitive compute transfers. 
Empirically, a 100MB packet size strikes a good balance between transfer throughput and minimizing interference.
The data mover is implemented in C++ as a PyTorch extension.

\subsection{CPU Decode Attention}
\label{sec:sys-cpu-attn}
% \THISWORK\ fully offload the KVcache and decode computation to CPU.
% To avoid the becoming a bottleneck in the overall execution, the CPU memory needs to process the attention at a certain throughput.
% For a complete forward pass, the weights are fully read from the CPU memory to be transferred to the GPU.
% The KV Cache also needs to be read once for attention computation.
% The CPU-side memory bandwidth requirement is therefore
% $$
% \frac{M_{kvcache}}{M_{weight}}B + B = (1 + \frac{M_{kvcache}}{M_{weight}})B
% $$

% The larger the KVcache size, the higher the memory bandwidth we need.
% Also, although decode attention is usually more memory-dense than compute-dense, the CPUs exhibit constrained computational capacity.
% Therefore, without careful optimizations on the CPU attention kernels, the CPU side attention computation can become the bottleneck of the overall execution, despite the adequet CPU memory bandwidth.

\begin{figure}
    \centering
    \includegraphics[width=\linewidth]{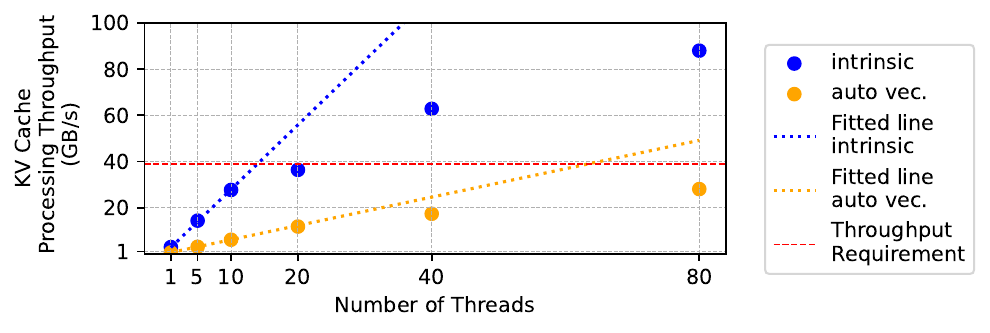}
    \vspace{-0.8cm}
    \caption{Performance comparison of decode flash attention implemented with intrinsic and auto vectorization. Trend lines are fitted using the first three data points to understand linear scaling. The throughput requirement is estimated assuming KV cache size is twice larger than the model size.}
    \vspace{-0.3cm}
    \label{fig:decode_attn}
\end{figure}

In \S\ref{sec:perf-model-attn}, we highlight the importance of efficient vector unit utilization for CPU-side decode attention to fully exploit the CPU-GPU hybrid system.
To this end, \THISWORK\ implements a high-performance decode attention kernel using hand-written SIMD intrinsics. 
Figure~\ref{fig:decode_attn} compares this implementation with an auto-vectorized baseline in terms of KV cache tokens attended per second.
Although both leverage the \texttt{AVX512} ISA on our test machine, they show stark performance differences.
The auto-vectorized version under-utilizes vector units, falling short of system throughput requirements.
In contrast, \THISWORK's hand-optimized kernel—featuring manual vectorization, loop unrolling, and data prefetching—achieves $4.7\times$ higher throughput in single-thread mode and $3.1\times$ with full thread utilization, exceeding the system target.
However, we observe that throughput gain saturates beyond 20 threads, likely due to memory controller contention.
% In \S\ref{sec:perf-model-attn}, we emphasize efficiently utilizing vector units for the CPU decode attention is required to achieve the full potential of the CPU-GPU hybrid system's hardware.
% For this purpose, \THISWORK\ implements a high-performance decode attention implementation with SIMD intrinsic.
% In Figure~\ref{fig:decode_attn}, we compare an auto-vectorized decode attention with \THISWORK's decode attention using hand-written intrinsic in terms of how much KV cache they can attend to per second.
% While both use the \texttt{AVX512} ISA available on our test machine, they demonstrate diverse characteristics. 
% The auto-vectorized version fails to fully leverage the vector units, delivering sub-optimal performance below the system requirement.
% By incorporating hand-tuning, like intrinsic, fetching, unrolling, etc, \THISWORK's implementation delivers $4.7\times$ higher throughput in single thread condition and $3.1\times$ when all hardware threads are fully utilized, surpassing the typical throughput requirement.
% On the other hand, we notice that \THISWORK scales slower when more than 20 threads are used, indicating potential contention in the memory controller.

 % In \THISWORK, we implement an optimized flash attention kernel with vector intrinsic, which merges the kv cache saving functionality for prefilling.

\section{Evaluation Methodology}
\textbf{Hardware Configuration and Test Environment.}
% All experiments are conducted on a dual-socket server with two Intel Platinum 8380 CPUs.
% Each socket is equipped with eight channel 750GB DDR4@3200MT/s memory and two NVidia A40 GPUs.
% The aggregated CPU memory bandwidth per socket measured is around 150GB/s.
% We use \texttt{numactl} to only use one socket and one GPU for our experiments.
% The A40 GPU has 48GB memory, thus we occupy 24-32GB memory with random tensors to mimic a GPU with 16-24GB memory like NVidia T4 and L4.
% We set different KV cache size of \THISWORK\ to mimic machine with different CPU memory capacity.
% Our experiments use the KV cache sizes from 70GB and 210GB, which represent machines with small and large CPU-side memory.
All experiments are conducted on a dual-socket server equipped with two Intel Platinum 8380 CPUs.
Each socket has eight DDR4-3200 memory channels, totaling 750GB capacity, and is connected to two NVIDIA A40 GPUs.
The measured aggregate CPU memory bandwidth per socket is approximately 150GB/s.
To ensure consistency and isolate the experimental environment, we use \texttt{numactl} to restrict execution to a single CPU socket and a single GPU.
The NVIDIA A40 GPU provides 48GB of memory; to simulate GPUs with more constrained memory capacities, such as the NVIDIA T4 or L4 (16–24GB), we allocate 24–32GB of GPU memory with random tensors, thereby reducing the effective available memory for LLM serving.
We configure the KV cache size in \THISWORK\ to emulate systems with varying CPU memory capacities.
Specifically, we experiment with KV cache sizes ranging from 70GB to 210GB, corresponding to machines with limited CPU-side memory.

\noindent
\textbf{Language Models.}
% Similar to prior works~\cite{}, we evaluate \THISWORK\ with three diverse MoE models: \texttt{Mixtral8x7B}, \texttt{Mixtral8x22B} and \texttt{DBRX}.
% The have 47B, 141B, and 132B parameters in BF16 and are 94GB, 282GB, 264GB in size.
% We limit the free GPU memory on A40 to 16GB for \texttt{Mixtral8x7B} and 24GB for \texttt{Mixtral8x22B} and \texttt{DBRX}.
% By running \texttt{Mixtral8x7B} with 70GB KV cache, we are mimicking a machine with around 164GB memory, and a 210GB KV cache means a machine with around 300GB memory.
% Similaly, running \texttt{Mixtral8x22B} and \texttt{DBRX} with 70GB kvcache means a machine with around 350GB memory, and a 210GB kvcache means a machine with around 500GB memory.
Following prior works~\cite{cao2024moe, fang2025klotskiefficientmixtureofexpertinference}, we evaluate \THISWORK\ using three diverse MoE models: \texttt{Mixtral8x7B}~\cite{mistralai2023mixtral}, \texttt{Mix-}\\\texttt{tral8x22B}~\cite{mistral_mixtral8x22b_2025}, and \texttt{DBRX}~\cite{databricks2024dbrxinstruct}.
These models have 47B, 141B, and 132B parameters in BF16, with model sizes of 94GB, 282GB, and 264GB, respectively.
We constrain the available GPU memory on the A40 to 16GB for \texttt{Mixtral8x7B}, and 24GB for both \texttt{Mixtral8x22B} and \texttt{DBRX}. 
Running \texttt{Mixtral8x7B} with a 70GB KV cache simulates a system with approximately 164GB of memory, while a 210GB KV cache corresponds to around 300GB.
Similarly, using a 70GB KV cache for \texttt{Mixtral8x22B} and \texttt{DBRX} simulates systems with roughly 350GB of memory, and a 210GB KV cache corresponds to about 500GB.

\noindent
\textbf{LLM Benchmark/Datasets.}
% We evaluate \THISWORK\ with diverse LLM benchmarks that are commonly used.
% Similar to prior work, we use replicates \texttt{MTBench}, a benchmark with 80 high-quality multi-turn questions in different domain, to form large batches for in-depth evaluation.
% We also set the maximum generation length to 32, 64, 128, and 256, similar to prior work.
% We further evaluate \THISWORK's performance on a RAG dataset and AIME dataset to understand its behavior under long prompt length and long generation cases.
% We summarize the prefill length, $p$, and generation length, $g$ information in Table \ref{tab:dataset_metrics}.
% To complete the evaluation in a reasonable amount of time, when KVCache size is 70GB, we set the request batch size to 25k for $g=32$ and $20k$ otherwise when running \texttt{MTBench}.
% In any other cases, we set the request batch size to $5gq$ according to the formula in \S\ref{sec:perf-model-sched}.
We evaluate \THISWORK\ using a diverse set of commonly used LLM model evaluation benchmarks.
Following prior work, we use a replicated version of \texttt{MTBench}~\cite{mtbenchBai2024}, which includes 80 high-quality multi-turn questions across various domains, to construct large batches for in-depth analysis.
We also vary the maximum generation length across 32, 64, 128, and 256 tokens, consistent with previous studies.
In addition, we assess \THISWORK’s performance on a RAG dataset~\cite{neuralbridge2024rag} and the AIME dataset~\cite{jia2024aime} to study its behavior under long-prompt and long-generation scenarios.
The prefill length ($p$) and generation length ($g$) settings are summarized in Table~\ref{tab:dataset_metrics}.
To ensure evaluations complete within reasonable time, we set the request batch size to 25k for $g=32$, and 20k otherwise, when running \texttt{MTBench} with a 70GB KV cache.
In all other settings, the request batch size is set to $5gq$, as defined in the performance model in \S\ref{sec:perf-model-sched}.

% \begin{table}[t]
%     \scriptsize
%     \centering
%     \begin{tabular}{lccc}
%         \hline
%         Dataset & $p_{avg}$ & $p_{max}$ & $g_{max}$ \\
%         \hline
%         MTbech & 98 & 450 & 32, 64, 128, 256 \\
%         RAG & 926 & 1843 & 128 \\
%         AIME 2024 & 128 & 410 & 512 \\
%         \hline
%     \end{tabular}
%     \caption{Workload Information}
%     \label{tab:dataset_metrics}
% \end{table}

\begin{table}[t]
    \scriptsize
    \centering
    \begin{tabular}{lcccc}
        \hline
        Dataset & \multicolumn{2}{c}{Prefill Length} & Max Generation Length & Category \\
        \cline{2-3}
        & Avg. & Max & & \\
        \hline
        MTBench~\cite{mtbenchBai2024} & 98 & 450 & 32, 64, 128, 256 & Multi-turn conversation \\
        RAG~\cite{neuralbridge2024rag} & 926 & 1843 & 128 & Retrieval-Augmented Q\&A \\
        AIME 2024~\cite{jia2024aime} & 128 & 410 & 512 & Math Problem Solving \\
        \hline
    \end{tabular}
    \caption{Model evaluation benchmarks used in evaluation.}
    \label{tab:dataset_metrics}
    \vspace{-0.95cm}
\end{table}

\noindent
\textbf{Baselines.} 
We compare \THISWORK\ with two state-of-the-art baselines: \texttt{MoE-Lightning}~\cite{cao2024moe} and \texttt{vllm}~\cite{pagedattn}.
We use open-source implementations~\cite{caoshiyi_artifacts_asplos25, vllm_github} of both baselines for comparison.
\begin{itemize}[nosep, leftmargin=*]
    \item \texttt{MoE-Lightning} is the state-of-the-art throughput-oriented MoE inference system for resource-constrained environments.
    We evaluate \texttt{MoE-Lightning} with two CPU memory settings.
    The normal setting set the CPU memory size profile of \texttt{MoE-Lightning} as the sum of model size and KV cache size, plus an additional 30GB for execution overhead.
    We further evaluate a large CPU memory setting where we set the CPU memory capacity to 1.25$\times$ of the sum of the model and KV cache size.
    \item \texttt{vllm} is a widely used LLM serving system based on the idea of paged attention.
    We use its CPU offload option to run models larger than GPU memory size.
    
\end{itemize}

\section{Evaluation Results}
\subsection{Overall Performance}
\begin{figure*}
    \centering
    \includegraphics[width=\linewidth]{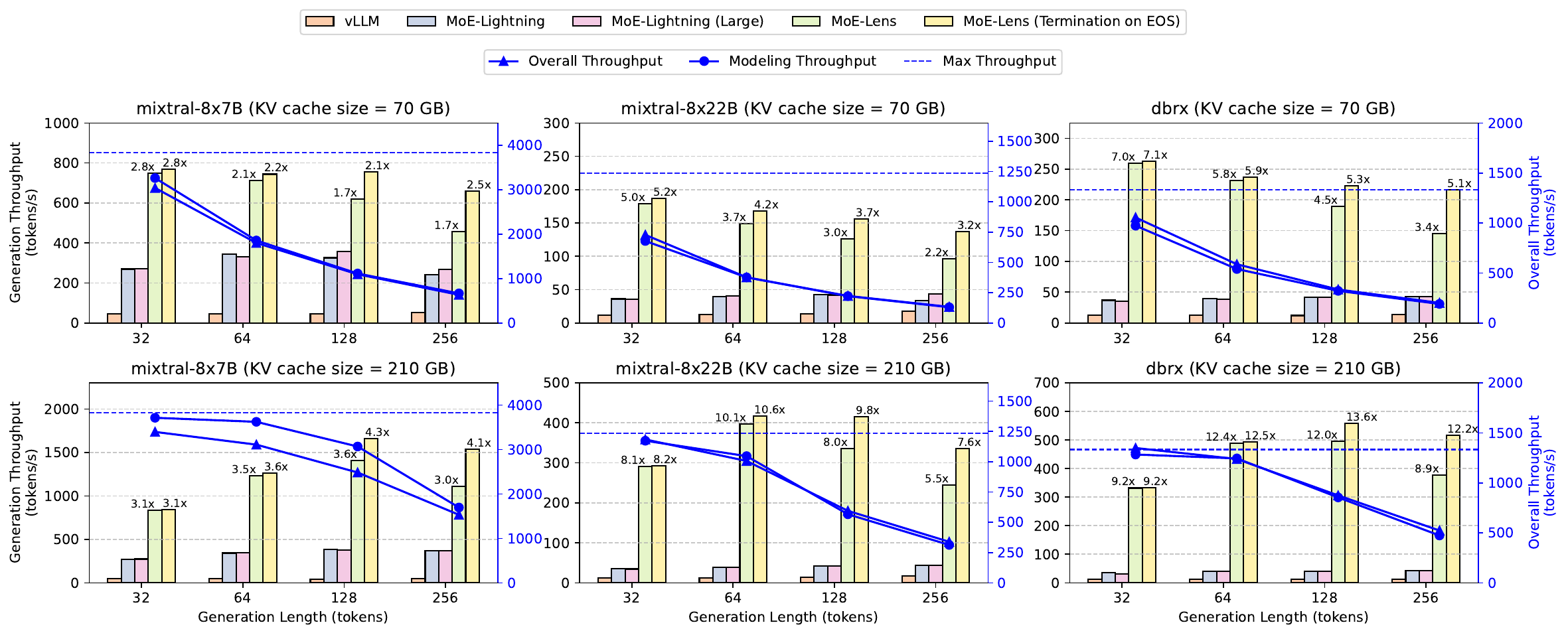}
    \vspace{-0.7cm}
    \caption{Overall performance of \THISWORK\ compared to baselines. The text annotations represents the speedup of \THISWORK\ against the best performance of MoE-Lightning. The secondary Y-axis shows the predicted and measured throughput of \THISWORK, validating our theoretical model.}
    \vspace{-0.3cm}
    \label{fig:throughput}
\end{figure*}
\textbf{System Generation Throughput.}
Figure~\ref{fig:throughput} compares the generation throughput (\textit{i.e.}, tokens generated per unit time) of vLLM, MoE-Lightening, and \THISWORK.
The vLLM baseline performs all computations, including GEMMs and attention, on the GPU while paging KV cache to and from CPU memory. 
Since model weights and KV cache exceed GPU memory capacity, vLLM is bottlenecked by the limited CPU–GPU PCIe bandwidth.
MoE-Lightening mitigates this by offloading attention to the CPU, reducing I/O traffic and improving throughput.
However, its modeling strategy overlooks key optimization opportunities, leaving substantial headroom before fully utilizing the hardware.

% Figure~\ref{fig:throughput} shows that \THISWORK\ improves an average throughput of XX$\times$ (up to XX$\times$) over MoE-Lightening.
% The improvement in \THISWORK\ is attributed to the fact that it identifies a missed opportunity in the modeling approach in prior work in terms of modeling the effect of CPU memory capacity.
% Using this critical parameter, the proposed approach is able to identify the optimal schedule for offloading the workload of LLM inference to CPU-GPU hybrid system.
% Specifically, our approach correctly identifies two critical design decisions: (1) the number of tokens that can be run concurrently on a GPU that maximizes CPU memory capacity for the allocation of KV cache, and (2) the prefill-decode overlap scheduling strategy on the GPU that even further increases the effective capacity of KV cache.
% MoE-Lightening, on the other hand, under-utilizes the memory capacity, as depicted in Table~\ref{tab:motivation-plan}, that results in reduced throughput as it does not effectively utilize the GPU hardware resources by allocating less number of tokens.
% This difference is further pronounced when the KV cache size increases to 210GB in Figure~\ref{fig:throughput}, where \THISWORK\ results in even higher speedup of XX$\times$ compared to XX$\times$ with a KV cache capacity of 210GB.
Figure~\ref{fig:throughput} shows that \THISWORK\ achieves an average throughput improvement of 4.6$\times$ (up to 12.4$\times$) over MoE-Lightening.
This gain stems from a key modeling insight overlooked in prior work: the role of CPU memory capacity in determining optimal workload scheduling.
By incorporating this constraint, \THISWORK\ identifies two critical decisions: (1) maximize the number of concurrent tokens on the GPU by fully utilizing CPU memory for KV cache usage, and (2) a prefill-decode overlap strategy that further expands effective KV cache capacity.
In contrast, MoE-Lightening underutilizes CPU memory, as shown in Table~\ref{tab:motivation-plan}, leading to fewer concurrent tokens and lower throughput.
This disparity becomes more pronounced with a 210GB KV cache, where \THISWORK\ delivers a higher average speedup of 6.4$\times$, compared to 3.2$\times$ with a 70GB cache.
The speedup is further improved to on average 5.3$\times$ when allowing \THISWORK\ to terminate generation when the EOS token is reached. 

With a large KV cache size (210GB), we observe that generation throughput increases with generation length up to a point.
For example, running \texttt{MTBench} on \texttt{Mixtral-8x7B}, throughput improves as generation length increases from 32 to 128 tokens.
However, at 256 generation tokens, KV cache becomes a bottleneck, causing throughput to drop.
This rise-then-drop pattern is consistent across all models at 210GB KV cache.
In contrast, with a 70GB KV cache, the rising trend disappears entirely, as even short generations saturate the limited cache, limiting throughput from the start.

% The overall system throughput decreases when the generation length is longer for a certain prompt length.
% As we analyzed in \S\ref{sec:perf-model-PME}, longer generation length decreases the PME of sequences, which means a higher demand on CPU memory capacity and lower throughput for a limited KV cache size.
% Note that when the KV cache is adequate, \THISWORK\ approaches the computational limit of a GPU, as shown be the blue dotted line in the Figure \ref{fig:throughput} when KV cache size is 210 and the $g_{max}$ is 32/64.
System throughput decreases with longer generation lengths for a fixed prompt length.
As analyzed in \S\ref{sec:perf-model-PME}, longer generations reduce the PME, increasing CPU memory demand and reducing throughput under limited KV cache.
When KV cache is sufficient, however, \THISWORK\ approaches around 90\% of GPU’s computational limit: illustrated by the blue line with triangle markers in Figure~\ref{fig:throughput} for $g_{\text{max}} = 32/64$ with 210GB KV cache.
% When KV cache is sufficient, however, \THISWORK\ approaches the GPU’s computational limit: illustrated by the blue dotted line in Figure~\ref{fig:throughput} for $g_{\text{max}} = 32/64$ with 210GB KV cache.

\begin{figure}
    \centering
    \includegraphics[width=\linewidth]{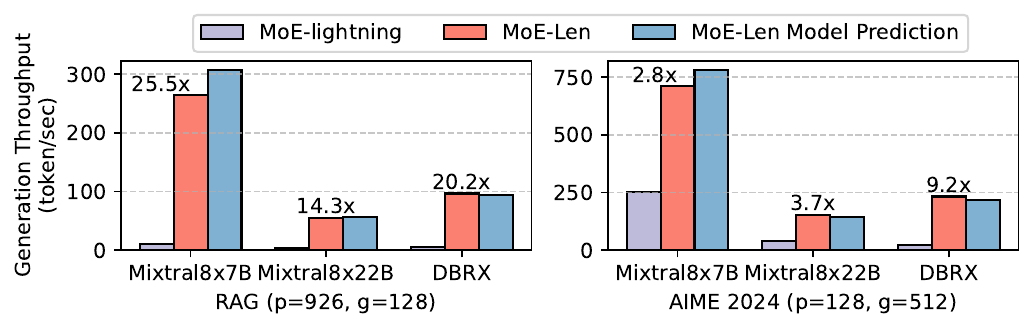}
    \vspace{-0.65cm}
    \caption{Performance comparison of \THISWORK\ and MoE-Lightning for \texttt{RAG} and \texttt{AIME2024} datasets.}
    \vspace{-0.6cm}
    \label{fig:aime_rag_lot}
\end{figure}

Figure~\ref{fig:aime_rag_lot} compares \THISWORK's performance on the prefill-heavy \texttt{RAG} and generation-heavy \texttt{AIME2024} datasets.
\THISWORK\ achieves up to 25.5$\times$ (19.4$\times$ avg) speedup over MoE-Lightning on \texttt{RAG}, and up to 9.9$\times$ (4.7$\times$ avg) on \texttt{AIME2024}, demonstrating consistently high throughput across workloads of diverse characteristics.

\noindent
\textbf{\THISWORK's Performance Model's Accuracy.}
\label{sec:eval-perf-model-acc}
% On average, \THISWORK's performance model (\S\ref{sec:perf-model-sched}) achieves 94\% accuracy in predicting throughput compared to the execution results, as demonstrated in Figures~\ref{fig:throughput} and \ref{fig:aime_rag_lot}.
% For the performance model, we estimate the CPU-GPU IO bandwidth $B_{IO}$ based on the time to transfer 1GB tensors from CPU to GPU, which is around 19.5GB/s on our test machine.
% \THISWORK\ achieves noticable less throughput than the performance model's prediction when running benchmarks on \texttt{Mixtral8x7B} with 210 GB KV cache.
% This is because the CPU attention computation completes memory bandwidth with CPU-GPU IO activities, increasing the weight transfer time.
% We elaborate the details in \S\ref{sec:attn-conflicts}.
On average, \THISWORK's performance model (\S\ref{sec:perf-model-sched}) predicts throughput with 94\% accuracy, as shown in Figures~\ref{fig:throughput} and \ref{fig:aime_rag_lot}.
We estimate CPU-GPU IO bandwidth $B_{IO}$ at 19.5GB/s, based on 1GB tensor transfers. 
Notably, \THISWORK\ falls short of the model’s prediction when running \texttt{Mixtral8x7B} with a 210GB KV cache, due to contention between CPU attention computation and IO, which delays weight transfers.
We analyze this conflict further in \S\ref{sec:attn-conflicts}.

\subsection{Detailed Execution Status Analysis}
\begin{figure*}
    \centering
    \includegraphics[width=\linewidth]{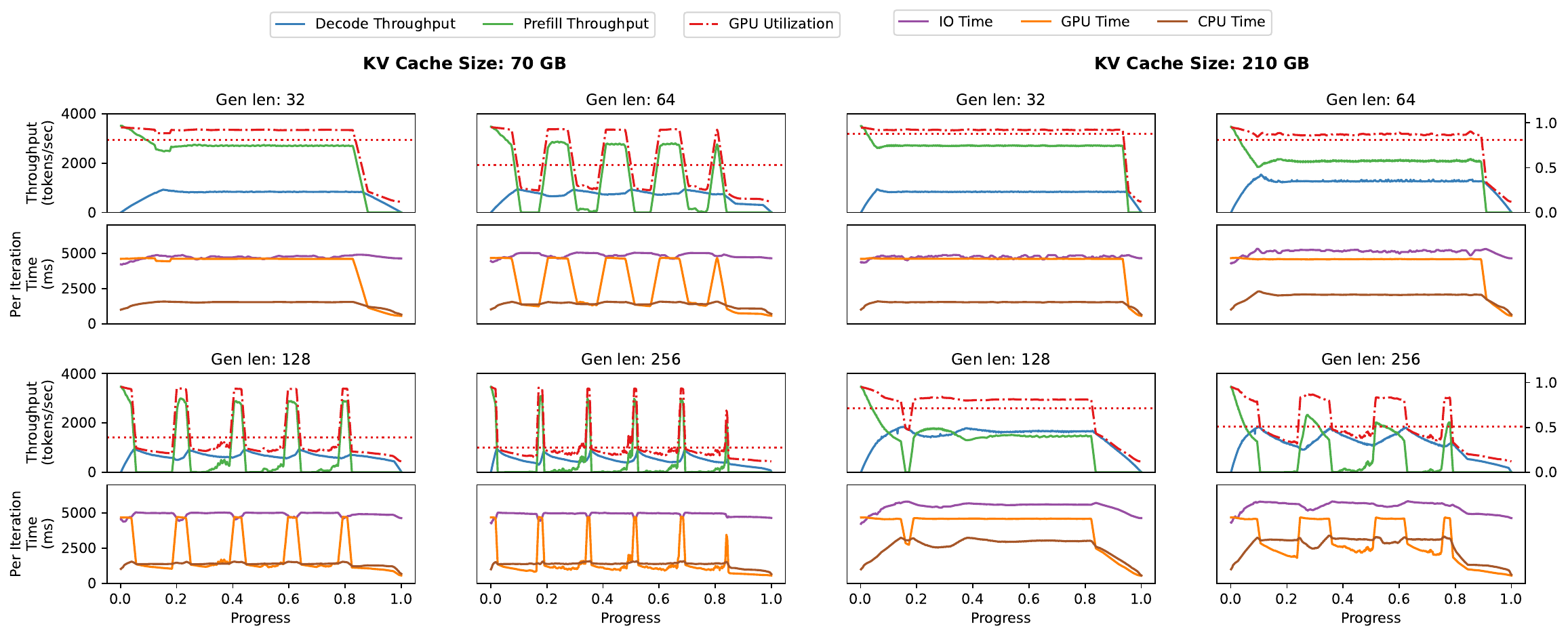}
    \vspace{-0.75cm}
    \caption{Execution status of \THISWORK\ for running \texttt{MTBench} on \texttt{Mixtral8x7B}, with different maximum generation length and KV cache size.
    The first and third row illustrate how the throughput and GPU utilization changes with workload progress.
    The second and the fourth rows show the corresponding IO, GPU computation, and CPU attention time for each inference pass.
    }
    \vspace{-0.2cm}
    \label{fig:exec-status}
\end{figure*}

\textbf{The Execution Dynamic of \THISWORK\ and the KV Cache Capacity Bottleneck.}
% Figure \ref{fig:exec-status} presents the detailed execution status of \THISWORK\ when running MTBench on \texttt{Mixtral8x7B}.
% In the more resource constrained 70GB kvcache case, kv cache capacity has significant impact on the overall system throughput and execution dynamic.
% When the maximum generation length is 32, a 70GB of kvcache is large enough to avoid thrashing sequences.
% As no sequences are preempted, the \THISWORK\ delivers steady and high decode and prefill throughput.
% In such scenario, the GPU utilization is closs to one, and the machine throughput are distributed to prefill and decode corroding to the ratio between the numbers of prefill tokens and decode tokens.
% When the maximum generation length increase, the amount of memory each sequence take also increases.
% The lack of kvcache blocks resulting in preemption of sequences, as a result, fluctuating decode and prefill throughput curves can be observed.
Figure~\ref{fig:exec-status} illustrates the detailed execution status of \THISWORK\ when running \texttt{MTBench} on \texttt{Mixtral8x7B}.
In the more resource-constrained 70GB KV cache setting, cache capacity significantly affects both overall system throughput and execution dynamics. 
When the maximum generation length is 32, the 70GB KV cache is sufficient to avoid sequence thrashing.
Since no sequences are preempted, \THISWORK\ achieves steady and high throughput in both prefill and decode phases.
In this case, GPU utilization approaches around 90\%, and system throughput is effectively distributed between prefill and decode phases, proportional to the ratio of prefill to decode tokens.
As the maximum generation length increases, each sequence consumes more memory.
Insufficient KV cache capacity causes sequence preemption, leading to observable fluctuations in the decode and prefill throughput curves.

% When the maximum generation length is 64, around half of the time the GPU cannot prefill new sequences due to lack of space.
% When the preemption happens, we can observe that the prefill throughput drops to zero.
% Then the decode throughput also begin to drop, as some sequence finish generation and their KV cache blocks are released.
% The deocde throughput keeps dropping until the speed of releasing memory is larger than the speed of generating new tokens.
% After that, empty slots appears and the \THISWORK's prefill schedulers schedules new sequence into the kv cache, increasing the number of sequences being decoded in parallel.
% The prefill and decode throughput starts to increase again.
When the maximum generation length is 64, the GPU is unable to prefill new sequences roughly half of the time due to insufficient KV cache capacity.
During these preemption events, the prefill throughput drops to zero.
As ongoing sequences complete their generation and release their KV cache blocks, the decode throughput also begins to decline.
This decline continues until the rate of KV cache release exceeds the rate of token generation.
Once enough KV cache blocks are freed, empty slots become available, allowing \THISWORK's prefill scheduler to admit new sequences into the cache.
This, in turn, increases the number of sequences being decoded in parallel, causing both prefill and decode throughput to rise again.

% This phenomenon becomes more extreme when the maximum generation length is 256.
% \THISWORK\ only perform perfilling for a tiny fraction of time, and the overall GPU utilization is much lower as only a limited number of sequences at decode stage is active, exhibiting low parallelism-memory capacity density.
% A similar trend can be observed from the per-forward-pass time for GPU computation, CPU computation and weight transfer IO.
% \THISWORK\ manages to match the GPU computation time and IO time when the amount of KV cache is sufficient, when maximum generation length is 32, but leaving the GPU idle when the generation length is 256 where the KV cache capacity becomes a bottleneck.
This phenomenon becomes more pronounced when the maximum generation length increases to 256.
In this case, \THISWORK\ performs prefilling for only a small fraction of the time, and the overall GPU utilization drops significantly.
This is due to a limited number of active sequences in the decode stage, resulting in low PME.
A similar trend is reflected in the per-forward-pass breakdown of GPU computation, CPU computation, and weight transfer I/O time.
When the KV cache capacity is sufficient (\textit{e.g.,} with a maximum generation length of 32), \THISWORK\ effectively balances GPU computation and I/O time.
However, when the generation length is 256, the KV cache becomes a bottleneck, leading to substantial GPU idle time.

\noindent
\textbf{The Effect of a Large KV Cache.}
% By comparing the cases with 70GB KV cache and 210GB KV cache, we can see the effect of a larger KV cahe on the overall throughput and execution dynamics.
% When the generation length is 32, a 70GB KV cache is enough and further increasing the KV Cache size does not have any impact on its performance.
% The effect of a larger KV cache becomes obvious when the generation length is greater or equal to 64.
% When the generation length is 64 for KV cache capacity of 210GB, we can observe a smoother decode and prefill throughput, indicating that less preemption is happening with abundant KV cache capacity.
% This results in an overall higher throughput and GPU utilization.
% Similar trend is happening on the cases of maximum generation length 128 and 256.
% We can only observe a slight degree of preemption at the early stage of execution, and the throughput curve quickly stabilized after that.
% As for the case whose maximum generation length is 256, we can also notice the period of time that no prefilling is happening is decreasing.
% In both cases, the throughput is significantly improved.
By comparing the cases with 70GB and 210GB KV cache, we observe the impact of larger KV cache capacity on overall throughput and execution dynamics. 
When the generation length is 32, the 70GB cache is sufficient, and increasing the cache size yields no further performance benefit.
However, the advantage of a larger KV cache becomes evident at generation lengths of 64 or more.
With 210GB of KV cache and a generation length of 64, decode and prefill throughput curves are noticeably smoother, reflecting fewer sequence preemptions due to ample cache capacity.
This leads to higher overall throughput and better GPU utilization.
A similar trend is observed for generation lengths of 128 and 256. 
Preemptions, if any, occur only in the early stages of execution, and the throughput stabilizes quickly.
In the 256-token case, the duration of prefill stalls significantly decreases, further enhancing throughput.
In both cases, the larger KV cache leads to substantial performance improvements.

\noindent
\textbf{The Bandwidth Competition between CPU Attention and CPU-GPU Weight Transfer.}
\label{sec:attn-conflicts}
% When the KV cache is large, the CPU attention need to scan through a large number of KV cache blocks concurrent to the CPU-GPU weight transfer for GPU computation.
% These two operations compete for CPU side resources like CPU memory bandwidth.
% When the maximum generation length is 256 and the KV cache size is 210, the CPU attention computation time is significant compared to the IO transfer time, which causes contention on the CPU side memory controller and slow down the weight transfer process.
% The time to transfer the weights one time from CPU to GPU increases from around 5s to 6s.
% The IO transfer time drops when the preemption occurs begins as fewer sequences are in the decode stage, which are using KVcache for CPU attention.
When the KV cache is large, the CPU attention computation must scan through a substantial number of KV cache blocks, concurrent with CPU-GPU weight transfers for GPU computation.
These two operations contend for shared CPU-side resources, particularly memory bandwidth. 
When the maximum generation length is 256 and the KV cache size is 210GB, CPU attention time becomes significant relative to IO transfer time.
This contention at the CPU memory controller slows down weight transfers, increasing the time to transfer weights from CPU to GPU from approximately 5 seconds to 6 seconds.
Interestingly, IO transfer time decreases when preemption begins—fewer sequences remain in the decode stage, reducing the demand on CPU attention and thus easing the pressure on the memory subsystem.

\section{Related Work}
\textbf{Resource-constrained LLM Inference.}
To enable LLM inference on resource-constrained hardware like PCs and low-end servers where the GPU memory capacity is limited, prior works focus on offloading GPU states to CPU~\cite{eliseev2023fastinferencemixtureofexpertslanguage, sheng2023flexgenhighthroughputgenerativeinference, song2024powerinferfastlargelanguage, xue2025moeinfinityefficientmoeinference, fang2025klotskiefficientmixtureofexpertinference, xu2025moegenhighthroughputmoeinference} and disks~\cite{alizadeh2024llmflashefficientlarge, sheng2023flexgenhighthroughputgenerativeinference, peng2024harnessingdramssdsustainable}.
% While some prior works mainly focuses on improving the scheduling and execution pipeline efficiency~\cite{sheng2023flexgenhighthroughputgenerativeinference, cao2024moe, eliseev2023fastinferencemixtureofexpertslanguage, xu2025moegenhighthroughputmoeinference},  some others leverage the activation sparsity~\cite{song2024powerinferfastlargelanguage} or expert activation patterns~\cite{fang2025klotskiefficientmixtureofexpertinference, xue2025moeinfinityefficientmoeinference}.
MoE-lightning~\cite{cao2024moe} is the state-of-the-art MoE inference system for resource-constrained environments. 
It provides a CPU-GPU IO-aware performance model, HRM, on offloading KV cache and attention computation to CPUs.
On comparison, \THISWORK\ provides a more comprehensive and fine-grained performance model that goes beyond CPU–GPU I/O considerations by accounting for CPU architecture resources and sequence-level scheduling dynamics, and a system significantly outperforms MoE-lightning.

\noindent
\textbf{Online LLM Inference Optimization.}
Online LLM systems with strict latency constraints are usually distributed~\cite{orca2022, llumnix2024, splitwise2024, qin2024mooncakekvcachecentricdisaggregatedarchitecture, stojkovic2024dynamollmdesigningllminference}.
Prior works focus on diverse optimization objectives, including but not limited to latency/throughput~\cite{qiao2024conserveharvestinggpuslowlatency, distserve,kim2025adordesignexplorationframework, juravsky2024hydragenhighthroughputllminference, zheng2024sglangefficientexecutionstructured, pagedattn}, cost~\cite{theaibrixteam2025aibrixscalablecosteffectivelarge, huang2024enovaautoscalingcosteffectivestable}, energy efficiency~\cite{stojkovic2025tapasthermalpowerawarescheduling, stojkovic2024dynamollmdesigningllminference, patel2024asplos}.
The online LLM serving system usually employs prefill decode disaggregation~\cite{splitwise2024, qin2024mooncakekvcachecentricdisaggregatedarchitecture} to achieve low latency for token generation. 
Techniques like continuous batching~\cite{daniel2023continuous} also let the new sequences prefill while others are decoding, but they do not discuss their dynamics for inference under resource-constrained environments, and focus on online serving.
In contrast, \THISWORK\ provides a detailed performance model to quantify the impact of prefill-decode overlap in resource-constrained environments, and demonstrates how to achieve this with weight transfer and attention computation on the CPU in a balanced manner. 
This differs fundamentally from continuous batching, whose techniques cannot be directly applied in our setting due to the differing coordination and resource constraints.

\noindent
\textbf{Algorithmic Techniques for LLM Acceleration.}
Another line of research in accelerating LLM inference focuses on trading off the generation accuracy with the LLM inference speed, such as quantization~\cite{MLSYS2024_42a452cb, MLSYS2024_5edb57c0, dettmers2023spqrsparsequantizedrepresentationnearlossless}, kv cache compression~\cite{cai2024pyramidkvdynamickvcache, zhang2023h2oheavyhitteroracleefficient}, developing attention variants~\cite{liu2024retrievalattentionacceleratinglongcontextllm, yuan2025nativesparseattentionhardwarealigned}, etc.
\THISWORK\ focuses on improving the hardware resource utilization for MoE inference in resource-constrained environments, without algorithmic changes.

% FlexGen~\cite{sheng2023flexgenhighthroughputgenerativeinference} first applies linear programming in finding an optimal execution policy for dense LLM models for resource-constrained environments.
% On the other hand, MoE inference systems Fiddle~\cite{kamahori2025fiddlercpugpuorchestrationfast}, MoE-Infinity~\cite{xue2025moeinfinityefficientmoeinference}, and Klotski~\cite{fang2025klotskiefficientmixtureofexpertinference} focus on leveraging the expert activation patterns to intelligently orchestrate the CPU-GPU IO for better performance.

\section{Conclusion}
% This paper uncovered a significantly opportunity to improve performance for resource-constraint batch-processing of MoE LLM inference.
% In a CPU-GPU hybrid execution, we showed how utilizing the CPU memory capacity is critical to improving end-to-end throughput as it translates to higher number of tokens processed on the GPU.
% Based on this insight, we built a theoretical performance model that accounts for architectural heterogeneity in CPU, GPU, and CPU-GPU interconnects, and workload heterogeneity in terms of prefill and decode stages.
% The model not only uncovered critical areas to improve but also predicted the overall performance with an average validation accuracy of 94\%.
% We further built an inference system using the insights from the theoretical analysis that achieved an average improvement of 4.6$
% \times$ (up to 25.5$\times$) over the state-of-the-art.
This paper uncovered a significant opportunity to improve performance in resource-constrained, batch-processing of MoE LLM inference.
In CPU–GPU hybrid execution, we showed that \textit{effectively utilizing CPU memory capacity} is key to boosting end-to-end throughput, as it enables more tokens to be processed on the GPU.
Building on this insight, we developed a theoretical performance model that captures architectural heterogeneity across CPU, GPU, and their interconnects, as well as workload heterogeneity between prefill and decode stages.
The model not only identifies critical performance bottlenecks but also predicts overall performance with an average validation accuracy of 94\%.
Using these insights, we designed an inference system that achieves a 4.6$\times$ average speedup (up to 25.5$\times$) over the state-of-the-art.

\begin{acks}
   This work was supported in part by Advanced Micro Devices, Inc. under the “Funding Academic Research” and the AMD AI \& HPC Cluster Program.
\end{acks}

%%%%%%% -- PAPER CONTENT ENDS -- %%%%%%%%

\balance
\bibliographystyle{ACM-Reference-Format}
\bibliography{sample-base}


\begin{thebibliography}{52}

%%% the Web address.

\ifx \showCODEN    \undefined \def \showCODEN     #1{\unskip}     \fi
\ifx \showDOI      \undefined \def \showDOI       #1{#1}\fi
\ifx \showISBNx    \undefined \def \showISBNx     #1{\unskip}     \fi
\ifx \showISBNxiii \undefined \def \showISBNxiii  #1{\unskip}     \fi
\ifx \showISSN     \undefined \def \showISSN      #1{\unskip}     \fi
\ifx \showLCCN     \undefined \def \showLCCN      #1{\unskip}     \fi
\ifx \shownote     \undefined \def \shownote      #1{#1}          \fi
\ifx \showarticletitle \undefined \def \showarticletitle #1{#1}   \fi
\ifx \showURL      \undefined \def \showURL       {\relax}        \fi
% invisible to TeX
\providecommand\bibfield[2]{#2}
\providecommand\bibinfo[2]{#2}
\providecommand\natexlab[1]{#1}
\providecommand\showeprint[2][]{arXiv:#2}

\bibitem[AI(2023)]%
        {mistralai2023mixtral}
\bibfield{author}{\bibinfo{person}{Mistral AI}.} \bibinfo{year}{2023}\natexlab{}.
\newblock \bibinfo{title}{{Mixtral-8x7B-Instruct-v0.1}}.
\newblock \bibinfo{howpublished}{\url{https://huggingface.co/mistralai/Mixtral-8x7B-Instruct-v0.1}}.
\newblock
\newblock
\shownote{Accessed: 2025-04-10}.


\bibitem[AI(2025)]%
        {mistral_mixtral8x22b_2025}
\bibfield{author}{\bibinfo{person}{Mistral AI}.} \bibinfo{year}{2025}\natexlab{}.
\newblock \bibinfo{title}{Mixtral-8x22B-Instruct-v0.1}.
\newblock \bibinfo{howpublished}{\url{https://huggingface.co/mistralai/Mixtral-8x22B-Instruct-v0.1}}.
\newblock
\newblock
\shownote{Accessed: 2025-04-10}.


\bibitem[Alizadeh et~al\mbox{.}(2024)]%
        {alizadeh2024llmflashefficientlarge}
\bibfield{author}{\bibinfo{person}{Keivan Alizadeh}, \bibinfo{person}{Iman Mirzadeh}, \bibinfo{person}{Dmitry Belenko}, \bibinfo{person}{Karen Khatamifard}, \bibinfo{person}{Minsik Cho}, \bibinfo{person}{Carlo C~Del Mundo}, \bibinfo{person}{Mohammad Rastegari}, {and} \bibinfo{person}{Mehrdad Farajtabar}.} \bibinfo{year}{2024}\natexlab{}.
\newblock \bibinfo{title}{LLM in a flash: Efficient Large Language Model Inference with Limited Memory}.
\newblock
\newblock
\showeprint[arxiv]{2312.11514}~[cs.CL]
\urldef\tempurl%
\url{https://arxiv.org/abs/2312.11514}
\showURL{%
\tempurl}


\bibitem[Bai et~al\mbox{.}(2024)]%
        {mtbenchBai2024}
\bibfield{author}{\bibinfo{person}{Ge Bai}, \bibinfo{person}{Jie Liu}, \bibinfo{person}{Xingyuan Bu}, \bibinfo{person}{Yancheng He}, \bibinfo{person}{Jiaheng Liu}, \bibinfo{person}{Zhanhui Zhou}, \bibinfo{person}{Zhuoran Lin}, \bibinfo{person}{Wenbo Su}, \bibinfo{person}{Tiezheng Ge}, \bibinfo{person}{Bo Zheng}, {and} \bibinfo{person}{Wanli Ouyang}.} \bibinfo{year}{2024}\natexlab{}.
\newblock \showarticletitle{MT-Bench-101: A Fine-Grained Benchmark for Evaluating Large Language Models in Multi-Turn Dialogues}. In \bibinfo{booktitle}{\emph{Proceedings of the 62nd Annual Meeting of the Association for Computational Linguistics (Volume 1: Long Papers)}}. \bibinfo{publisher}{Association for Computational Linguistics}, \bibinfo{pages}{7421–7454}.
\newblock
\urldef\tempurl%
\url{https://doi.org/10.18653/v1/2024.acl-long.401}
\showDOI{\tempurl}


\bibitem[Bai et~al\mbox{.}(2023)]%
        {bai2023qwentechnicalreport}
\bibfield{author}{\bibinfo{person}{Jinze Bai}, \bibinfo{person}{Shuai Bai}, \bibinfo{person}{Yunfei Chu}, \bibinfo{person}{Zeyu Cui}, \bibinfo{person}{Kai Dang}, \bibinfo{person}{Xiaodong Deng}, \bibinfo{person}{Yang Fan}, \bibinfo{person}{Wenbin Ge}, \bibinfo{person}{Yu Han}, \bibinfo{person}{Fei Huang}, \bibinfo{person}{Binyuan Hui}, \bibinfo{person}{Luo Ji}, \bibinfo{person}{Mei Li}, \bibinfo{person}{Junyang Lin}, \bibinfo{person}{Runji Lin}, \bibinfo{person}{Dayiheng Liu}, \bibinfo{person}{Gao Liu}, \bibinfo{person}{Chengqiang Lu}, \bibinfo{person}{Keming Lu}, \bibinfo{person}{Jianxin Ma}, \bibinfo{person}{Rui Men}, \bibinfo{person}{Xingzhang Ren}, \bibinfo{person}{Xuancheng Ren}, \bibinfo{person}{Chuanqi Tan}, \bibinfo{person}{Sinan Tan}, \bibinfo{person}{Jianhong Tu}, \bibinfo{person}{Peng Wang}, \bibinfo{person}{Shijie Wang}, \bibinfo{person}{Wei Wang}, \bibinfo{person}{Shengguang Wu}, \bibinfo{person}{Benfeng Xu}, \bibinfo{person}{Jin Xu}, \bibinfo{person}{An Yang}, \bibinfo{person}{Hao Yang},
  \bibinfo{person}{Jian Yang}, \bibinfo{person}{Shusheng Yang}, \bibinfo{person}{Yang Yao}, \bibinfo{person}{Bowen Yu}, \bibinfo{person}{Hongyi Yuan}, \bibinfo{person}{Zheng Yuan}, \bibinfo{person}{Jianwei Zhang}, \bibinfo{person}{Xingxuan Zhang}, \bibinfo{person}{Yichang Zhang}, \bibinfo{person}{Zhenru Zhang}, \bibinfo{person}{Chang Zhou}, \bibinfo{person}{Jingren Zhou}, \bibinfo{person}{Xiaohuan Zhou}, {and} \bibinfo{person}{Tianhang Zhu}.} \bibinfo{year}{2023}\natexlab{}.
\newblock \bibinfo{title}{Qwen Technical Report}.
\newblock
\newblock
\showeprint[arxiv]{2309.16609}~[cs.CL]
\urldef\tempurl%
\url{https://arxiv.org/abs/2309.16609}
\showURL{%
\tempurl}


\bibitem[Bridge(2024)]%
        {neuralbridge2024rag}
\bibfield{author}{\bibinfo{person}{Neural Bridge}.} \bibinfo{year}{2024}\natexlab{}.
\newblock \bibinfo{title}{{RAG Dataset 12000}}.
\newblock \bibinfo{howpublished}{\url{https://huggingface.co/datasets/neural-bridge/rag-dataset-12000}}.
\newblock
\newblock
\shownote{Accessed: 2025-04-10}.


\bibitem[Cai et~al\mbox{.}(2024)]%
        {cai2024pyramidkvdynamickvcache}
\bibfield{author}{\bibinfo{person}{Zefan Cai}, \bibinfo{person}{Yichi Zhang}, \bibinfo{person}{Bofei Gao}, \bibinfo{person}{Yuliang Liu}, \bibinfo{person}{Tianyu Liu}, \bibinfo{person}{Keming Lu}, \bibinfo{person}{Wayne Xiong}, \bibinfo{person}{Yue Dong}, \bibinfo{person}{Baobao Chang}, \bibinfo{person}{Junjie Hu}, {and} \bibinfo{person}{Wen Xiao}.} \bibinfo{year}{2024}\natexlab{}.
\newblock \bibinfo{title}{PyramidKV: Dynamic KV Cache Compression based on Pyramidal Information Funneling}.
\newblock
\newblock
\showeprint[arxiv]{2406.02069}~[cs.CL]
\urldef\tempurl%
\url{https://arxiv.org/abs/2406.02069}
\showURL{%
\tempurl}


\bibitem[Cao(2025)]%
        {caoshiyi_artifacts_asplos25}
\bibfield{author}{\bibinfo{person}{Shiyi Cao}.} \bibinfo{year}{2025}\natexlab{}.
\newblock \bibinfo{title}{{Artifact Evaluation Repository for ASPLOS 2025}}.
\newblock \bibinfo{howpublished}{\url{https://github.com/caoshiyi/artifacts/tree/asplos25}}.
\newblock
\newblock
\shownote{Accessed: 2025-04-11}.


\bibitem[Cao et~al\mbox{.}(2024)]%
        {cao2024moe}
\bibfield{author}{\bibinfo{person}{Shiyi Cao}, \bibinfo{person}{Shu Liu}, \bibinfo{person}{Tyler Griggs}, \bibinfo{person}{Peter Schafhalter}, \bibinfo{person}{Xiaoxuan Liu}, \bibinfo{person}{Ying Sheng}, \bibinfo{person}{Joseph~E Gonzalez}, \bibinfo{person}{Matei Zaharia}, {and} \bibinfo{person}{Ion Stoica}.} \bibinfo{year}{2024}\natexlab{}.
\newblock \showarticletitle{Moe-lightning: High-throughput moe inference on memory-constrained gpus}.
\newblock \bibinfo{journal}{\emph{arXiv preprint arXiv:2411.11217}} (\bibinfo{year}{2024}).
\newblock


\bibitem[Chen et~al\mbox{.}(2021)]%
        {chen2021spreadsheetcoderformulapredictionsemistructured}
\bibfield{author}{\bibinfo{person}{Xinyun Chen}, \bibinfo{person}{Petros Maniatis}, \bibinfo{person}{Rishabh Singh}, \bibinfo{person}{Charles Sutton}, \bibinfo{person}{Hanjun Dai}, \bibinfo{person}{Max Lin}, {and} \bibinfo{person}{Denny Zhou}.} \bibinfo{year}{2021}\natexlab{}.
\newblock \bibinfo{title}{SpreadsheetCoder: Formula Prediction from Semi-structured Context}.
\newblock
\newblock
\showeprint[arxiv]{2106.15339}~[cs.SE]
\urldef\tempurl%
\url{https://arxiv.org/abs/2106.15339}
\showURL{%
\tempurl}


\bibitem[Dai et~al\mbox{.}(2024)]%
        {dai2024deepseekmoeultimateexpertspecialization}
\bibfield{author}{\bibinfo{person}{Damai Dai}, \bibinfo{person}{Chengqi Deng}, \bibinfo{person}{Chenggang Zhao}, \bibinfo{person}{R.~X. Xu}, \bibinfo{person}{Huazuo Gao}, \bibinfo{person}{Deli Chen}, \bibinfo{person}{Jiashi Li}, \bibinfo{person}{Wangding Zeng}, \bibinfo{person}{Xingkai Yu}, \bibinfo{person}{Y. Wu}, \bibinfo{person}{Zhenda Xie}, \bibinfo{person}{Y.~K. Li}, \bibinfo{person}{Panpan Huang}, \bibinfo{person}{Fuli Luo}, \bibinfo{person}{Chong Ruan}, \bibinfo{person}{Zhifang Sui}, {and} \bibinfo{person}{Wenfeng Liang}.} \bibinfo{year}{2024}\natexlab{}.
\newblock \bibinfo{title}{DeepSeekMoE: Towards Ultimate Expert Specialization in Mixture-of-Experts Language Models}.
\newblock
\newblock
\showeprint[arxiv]{2401.06066}~[cs.CL]
\urldef\tempurl%
\url{https://arxiv.org/abs/2401.06066}
\showURL{%
\tempurl}


\bibitem[Daniel et~al\mbox{.}(2023)]%
        {daniel2023continuous}
\bibfield{author}{\bibinfo{person}{Cade Daniel}, \bibinfo{person}{Chen Shen}, \bibinfo{person}{Eric Liang}, {and} \bibinfo{person}{Richard Liaw}.} \bibinfo{year}{2023}\natexlab{}.
\newblock \bibinfo{title}{How Continuous Batching Enables 23x Throughput in LLM Inference While Reducing p50 Latency}.
\newblock \bibinfo{howpublished}{\url{https://www.anyscale.com/blog/continuous-batching-llm-inference}}.
\newblock
\newblock
\shownote{Accessed: 2025-04-11}.


\bibitem[Dao et~al\mbox{.}(2022)]%
        {dao2022flashattentionfastmemoryefficientexact}
\bibfield{author}{\bibinfo{person}{Tri Dao}, \bibinfo{person}{Daniel~Y. Fu}, \bibinfo{person}{Stefano Ermon}, \bibinfo{person}{Atri Rudra}, {and} \bibinfo{person}{Christopher Ré}.} \bibinfo{year}{2022}\natexlab{}.
\newblock \bibinfo{title}{FlashAttention: Fast and Memory-Efficient Exact Attention with IO-Awareness}.
\newblock
\newblock
\showeprint[arxiv]{2205.14135}~[cs.LG]
\urldef\tempurl%
\url{https://arxiv.org/abs/2205.14135}
\showURL{%
\tempurl}


\bibitem[Databricks(2024)]%
        {databricks2024dbrxinstruct}
\bibfield{author}{\bibinfo{person}{Databricks}.} \bibinfo{year}{2024}\natexlab{}.
\newblock \bibinfo{title}{{DBRX Instruct}}.
\newblock \bibinfo{howpublished}{\url{https://huggingface.co/databricks/dbrx-instruct}}.
\newblock
\newblock
\shownote{Accessed: 2025-04-10}.


\bibitem[DeepSeek-AI et~al\mbox{.}(2025a)]%
        {deepseekai2025deepseekr1incentivizingreasoningcapability}
\bibfield{author}{\bibinfo{person}{DeepSeek-AI}, \bibinfo{person}{Daya Guo}, \bibinfo{person}{Dejian Yang}, \bibinfo{person}{Haowei Zhang}, \bibinfo{person}{Junxiao Song}, \bibinfo{person}{Ruoyu Zhang}, \bibinfo{person}{Runxin Xu}, \bibinfo{person}{Qihao Zhu}, \bibinfo{person}{Shirong Ma}, \bibinfo{person}{Peiyi Wang}, \bibinfo{person}{Xiao Bi}, \bibinfo{person}{Xiaokang Zhang}, \bibinfo{person}{Xingkai Yu}, \bibinfo{person}{Yu Wu}, \bibinfo{person}{Z.~F. Wu}, \bibinfo{person}{Zhibin Gou}, \bibinfo{person}{Zhihong Shao}, \bibinfo{person}{Zhuoshu Li}, \bibinfo{person}{Ziyi Gao}, \bibinfo{person}{Aixin Liu}, \bibinfo{person}{Bing Xue}, \bibinfo{person}{Bingxuan Wang}, \bibinfo{person}{Bochao Wu}, \bibinfo{person}{Bei Feng}, \bibinfo{person}{Chengda Lu}, \bibinfo{person}{Chenggang Zhao}, \bibinfo{person}{Chengqi Deng}, \bibinfo{person}{Chenyu Zhang}, \bibinfo{person}{Chong Ruan}, \bibinfo{person}{Damai Dai}, \bibinfo{person}{Deli Chen}, \bibinfo{person}{Dongjie Ji}, \bibinfo{person}{Erhang Li},
  \bibinfo{person}{Fangyun Lin}, \bibinfo{person}{Fucong Dai}, \bibinfo{person}{Fuli Luo}, \bibinfo{person}{Guangbo Hao}, \bibinfo{person}{Guanting Chen}, \bibinfo{person}{Guowei Li}, \bibinfo{person}{H. Zhang}, \bibinfo{person}{Han Bao}, \bibinfo{person}{Hanwei Xu}, \bibinfo{person}{Haocheng Wang}, \bibinfo{person}{Honghui Ding}, \bibinfo{person}{Huajian Xin}, \bibinfo{person}{Huazuo Gao}, \bibinfo{person}{Hui Qu}, \bibinfo{person}{Hui Li}, \bibinfo{person}{Jianzhong Guo}, \bibinfo{person}{Jiashi Li}, \bibinfo{person}{Jiawei Wang}, \bibinfo{person}{Jingchang Chen}, \bibinfo{person}{Jingyang Yuan}, \bibinfo{person}{Junjie Qiu}, \bibinfo{person}{Junlong Li}, \bibinfo{person}{J.~L. Cai}, \bibinfo{person}{Jiaqi Ni}, \bibinfo{person}{Jian Liang}, \bibinfo{person}{Jin Chen}, \bibinfo{person}{Kai Dong}, \bibinfo{person}{Kai Hu}, \bibinfo{person}{Kaige Gao}, \bibinfo{person}{Kang Guan}, \bibinfo{person}{Kexin Huang}, \bibinfo{person}{Kuai Yu}, \bibinfo{person}{Lean Wang}, \bibinfo{person}{Lecong Zhang},
  \bibinfo{person}{Liang Zhao}, \bibinfo{person}{Litong Wang}, \bibinfo{person}{Liyue Zhang}, \bibinfo{person}{Lei Xu}, \bibinfo{person}{Leyi Xia}, \bibinfo{person}{Mingchuan Zhang}, \bibinfo{person}{Minghua Zhang}, \bibinfo{person}{Minghui Tang}, \bibinfo{person}{Meng Li}, \bibinfo{person}{Miaojun Wang}, \bibinfo{person}{Mingming Li}, \bibinfo{person}{Ning Tian}, \bibinfo{person}{Panpan Huang}, \bibinfo{person}{Peng Zhang}, \bibinfo{person}{Qiancheng Wang}, \bibinfo{person}{Qinyu Chen}, \bibinfo{person}{Qiushi Du}, \bibinfo{person}{Ruiqi Ge}, \bibinfo{person}{Ruisong Zhang}, \bibinfo{person}{Ruizhe Pan}, \bibinfo{person}{Runji Wang}, \bibinfo{person}{R.~J. Chen}, \bibinfo{person}{R.~L. Jin}, \bibinfo{person}{Ruyi Chen}, \bibinfo{person}{Shanghao Lu}, \bibinfo{person}{Shangyan Zhou}, \bibinfo{person}{Shanhuang Chen}, \bibinfo{person}{Shengfeng Ye}, \bibinfo{person}{Shiyu Wang}, \bibinfo{person}{Shuiping Yu}, \bibinfo{person}{Shunfeng Zhou}, \bibinfo{person}{Shuting Pan}, \bibinfo{person}{S.~S. Li},
  \bibinfo{person}{Shuang Zhou}, \bibinfo{person}{Shaoqing Wu}, \bibinfo{person}{Shengfeng Ye}, \bibinfo{person}{Tao Yun}, \bibinfo{person}{Tian Pei}, \bibinfo{person}{Tianyu Sun}, \bibinfo{person}{T. Wang}, \bibinfo{person}{Wangding Zeng}, \bibinfo{person}{Wanjia Zhao}, \bibinfo{person}{Wen Liu}, \bibinfo{person}{Wenfeng Liang}, \bibinfo{person}{Wenjun Gao}, \bibinfo{person}{Wenqin Yu}, \bibinfo{person}{Wentao Zhang}, \bibinfo{person}{W.~L. Xiao}, \bibinfo{person}{Wei An}, \bibinfo{person}{Xiaodong Liu}, \bibinfo{person}{Xiaohan Wang}, \bibinfo{person}{Xiaokang Chen}, \bibinfo{person}{Xiaotao Nie}, \bibinfo{person}{Xin Cheng}, \bibinfo{person}{Xin Liu}, \bibinfo{person}{Xin Xie}, \bibinfo{person}{Xingchao Liu}, \bibinfo{person}{Xinyu Yang}, \bibinfo{person}{Xinyuan Li}, \bibinfo{person}{Xuecheng Su}, \bibinfo{person}{Xuheng Lin}, \bibinfo{person}{X.~Q. Li}, \bibinfo{person}{Xiangyue Jin}, \bibinfo{person}{Xiaojin Shen}, \bibinfo{person}{Xiaosha Chen}, \bibinfo{person}{Xiaowen Sun}, \bibinfo{person}{Xiaoxiang
  Wang}, \bibinfo{person}{Xinnan Song}, \bibinfo{person}{Xinyi Zhou}, \bibinfo{person}{Xianzu Wang}, \bibinfo{person}{Xinxia Shan}, \bibinfo{person}{Y.~K. Li}, \bibinfo{person}{Y.~Q. Wang}, \bibinfo{person}{Y.~X. Wei}, \bibinfo{person}{Yang Zhang}, \bibinfo{person}{Yanhong Xu}, \bibinfo{person}{Yao Li}, \bibinfo{person}{Yao Zhao}, \bibinfo{person}{Yaofeng Sun}, \bibinfo{person}{Yaohui Wang}, \bibinfo{person}{Yi Yu}, \bibinfo{person}{Yichao Zhang}, \bibinfo{person}{Yifan Shi}, \bibinfo{person}{Yiliang Xiong}, \bibinfo{person}{Ying He}, \bibinfo{person}{Yishi Piao}, \bibinfo{person}{Yisong Wang}, \bibinfo{person}{Yixuan Tan}, \bibinfo{person}{Yiyang Ma}, \bibinfo{person}{Yiyuan Liu}, \bibinfo{person}{Yongqiang Guo}, \bibinfo{person}{Yuan Ou}, \bibinfo{person}{Yuduan Wang}, \bibinfo{person}{Yue Gong}, \bibinfo{person}{Yuheng Zou}, \bibinfo{person}{Yujia He}, \bibinfo{person}{Yunfan Xiong}, \bibinfo{person}{Yuxiang Luo}, \bibinfo{person}{Yuxiang You}, \bibinfo{person}{Yuxuan Liu}, \bibinfo{person}{Yuyang Zhou},
  \bibinfo{person}{Y.~X. Zhu}, \bibinfo{person}{Yanhong Xu}, \bibinfo{person}{Yanping Huang}, \bibinfo{person}{Yaohui Li}, \bibinfo{person}{Yi Zheng}, \bibinfo{person}{Yuchen Zhu}, \bibinfo{person}{Yunxian Ma}, \bibinfo{person}{Ying Tang}, \bibinfo{person}{Yukun Zha}, \bibinfo{person}{Yuting Yan}, \bibinfo{person}{Z.~Z. Ren}, \bibinfo{person}{Zehui Ren}, \bibinfo{person}{Zhangli Sha}, \bibinfo{person}{Zhe Fu}, \bibinfo{person}{Zhean Xu}, \bibinfo{person}{Zhenda Xie}, \bibinfo{person}{Zhengyan Zhang}, \bibinfo{person}{Zhewen Hao}, \bibinfo{person}{Zhicheng Ma}, \bibinfo{person}{Zhigang Yan}, \bibinfo{person}{Zhiyu Wu}, \bibinfo{person}{Zihui Gu}, \bibinfo{person}{Zijia Zhu}, \bibinfo{person}{Zijun Liu}, \bibinfo{person}{Zilin Li}, \bibinfo{person}{Ziwei Xie}, \bibinfo{person}{Ziyang Song}, \bibinfo{person}{Zizheng Pan}, \bibinfo{person}{Zhen Huang}, \bibinfo{person}{Zhipeng Xu}, \bibinfo{person}{Zhongyu Zhang}, {and} \bibinfo{person}{Zhen Zhang}.} \bibinfo{year}{2025}\natexlab{a}.
\newblock \bibinfo{title}{DeepSeek-R1: Incentivizing Reasoning Capability in LLMs via Reinforcement Learning}.
\newblock
\newblock
\showeprint[arxiv]{2501.12948}~[cs.CL]
\urldef\tempurl%
\url{https://arxiv.org/abs/2501.12948}
\showURL{%
\tempurl}


\bibitem[DeepSeek-AI et~al\mbox{.}(2025b)]%
        {deepseekai2025deepseekv3technicalreport}
\bibfield{author}{\bibinfo{person}{DeepSeek-AI}, \bibinfo{person}{Aixin Liu}, \bibinfo{person}{Bei Feng}, \bibinfo{person}{Bing Xue}, \bibinfo{person}{Bingxuan Wang}, \bibinfo{person}{Bochao Wu}, \bibinfo{person}{Chengda Lu}, \bibinfo{person}{Chenggang Zhao}, \bibinfo{person}{Chengqi Deng}, \bibinfo{person}{Chenyu Zhang}, \bibinfo{person}{Chong Ruan}, \bibinfo{person}{Damai Dai}, \bibinfo{person}{Daya Guo}, \bibinfo{person}{Dejian Yang}, \bibinfo{person}{Deli Chen}, \bibinfo{person}{Dongjie Ji}, \bibinfo{person}{Erhang Li}, \bibinfo{person}{Fangyun Lin}, \bibinfo{person}{Fucong Dai}, \bibinfo{person}{Fuli Luo}, \bibinfo{person}{Guangbo Hao}, \bibinfo{person}{Guanting Chen}, \bibinfo{person}{Guowei Li}, \bibinfo{person}{H. Zhang}, \bibinfo{person}{Han Bao}, \bibinfo{person}{Hanwei Xu}, \bibinfo{person}{Haocheng Wang}, \bibinfo{person}{Haowei Zhang}, \bibinfo{person}{Honghui Ding}, \bibinfo{person}{Huajian Xin}, \bibinfo{person}{Huazuo Gao}, \bibinfo{person}{Hui Li}, \bibinfo{person}{Hui Qu},
  \bibinfo{person}{J.~L. Cai}, \bibinfo{person}{Jian Liang}, \bibinfo{person}{Jianzhong Guo}, \bibinfo{person}{Jiaqi Ni}, \bibinfo{person}{Jiashi Li}, \bibinfo{person}{Jiawei Wang}, \bibinfo{person}{Jin Chen}, \bibinfo{person}{Jingchang Chen}, \bibinfo{person}{Jingyang Yuan}, \bibinfo{person}{Junjie Qiu}, \bibinfo{person}{Junlong Li}, \bibinfo{person}{Junxiao Song}, \bibinfo{person}{Kai Dong}, \bibinfo{person}{Kai Hu}, \bibinfo{person}{Kaige Gao}, \bibinfo{person}{Kang Guan}, \bibinfo{person}{Kexin Huang}, \bibinfo{person}{Kuai Yu}, \bibinfo{person}{Lean Wang}, \bibinfo{person}{Lecong Zhang}, \bibinfo{person}{Lei Xu}, \bibinfo{person}{Leyi Xia}, \bibinfo{person}{Liang Zhao}, \bibinfo{person}{Litong Wang}, \bibinfo{person}{Liyue Zhang}, \bibinfo{person}{Meng Li}, \bibinfo{person}{Miaojun Wang}, \bibinfo{person}{Mingchuan Zhang}, \bibinfo{person}{Minghua Zhang}, \bibinfo{person}{Minghui Tang}, \bibinfo{person}{Mingming Li}, \bibinfo{person}{Ning Tian}, \bibinfo{person}{Panpan Huang}, \bibinfo{person}{Peiyi
  Wang}, \bibinfo{person}{Peng Zhang}, \bibinfo{person}{Qiancheng Wang}, \bibinfo{person}{Qihao Zhu}, \bibinfo{person}{Qinyu Chen}, \bibinfo{person}{Qiushi Du}, \bibinfo{person}{R.~J. Chen}, \bibinfo{person}{R.~L. Jin}, \bibinfo{person}{Ruiqi Ge}, \bibinfo{person}{Ruisong Zhang}, \bibinfo{person}{Ruizhe Pan}, \bibinfo{person}{Runji Wang}, \bibinfo{person}{Runxin Xu}, \bibinfo{person}{Ruoyu Zhang}, \bibinfo{person}{Ruyi Chen}, \bibinfo{person}{S.~S. Li}, \bibinfo{person}{Shanghao Lu}, \bibinfo{person}{Shangyan Zhou}, \bibinfo{person}{Shanhuang Chen}, \bibinfo{person}{Shaoqing Wu}, \bibinfo{person}{Shengfeng Ye}, \bibinfo{person}{Shengfeng Ye}, \bibinfo{person}{Shirong Ma}, \bibinfo{person}{Shiyu Wang}, \bibinfo{person}{Shuang Zhou}, \bibinfo{person}{Shuiping Yu}, \bibinfo{person}{Shunfeng Zhou}, \bibinfo{person}{Shuting Pan}, \bibinfo{person}{T. Wang}, \bibinfo{person}{Tao Yun}, \bibinfo{person}{Tian Pei}, \bibinfo{person}{Tianyu Sun}, \bibinfo{person}{W.~L. Xiao}, \bibinfo{person}{Wangding Zeng},
  \bibinfo{person}{Wanjia Zhao}, \bibinfo{person}{Wei An}, \bibinfo{person}{Wen Liu}, \bibinfo{person}{Wenfeng Liang}, \bibinfo{person}{Wenjun Gao}, \bibinfo{person}{Wenqin Yu}, \bibinfo{person}{Wentao Zhang}, \bibinfo{person}{X.~Q. Li}, \bibinfo{person}{Xiangyue Jin}, \bibinfo{person}{Xianzu Wang}, \bibinfo{person}{Xiao Bi}, \bibinfo{person}{Xiaodong Liu}, \bibinfo{person}{Xiaohan Wang}, \bibinfo{person}{Xiaojin Shen}, \bibinfo{person}{Xiaokang Chen}, \bibinfo{person}{Xiaokang Zhang}, \bibinfo{person}{Xiaosha Chen}, \bibinfo{person}{Xiaotao Nie}, \bibinfo{person}{Xiaowen Sun}, \bibinfo{person}{Xiaoxiang Wang}, \bibinfo{person}{Xin Cheng}, \bibinfo{person}{Xin Liu}, \bibinfo{person}{Xin Xie}, \bibinfo{person}{Xingchao Liu}, \bibinfo{person}{Xingkai Yu}, \bibinfo{person}{Xinnan Song}, \bibinfo{person}{Xinxia Shan}, \bibinfo{person}{Xinyi Zhou}, \bibinfo{person}{Xinyu Yang}, \bibinfo{person}{Xinyuan Li}, \bibinfo{person}{Xuecheng Su}, \bibinfo{person}{Xuheng Lin}, \bibinfo{person}{Y.~K. Li},
  \bibinfo{person}{Y.~Q. Wang}, \bibinfo{person}{Y.~X. Wei}, \bibinfo{person}{Y.~X. Zhu}, \bibinfo{person}{Yang Zhang}, \bibinfo{person}{Yanhong Xu}, \bibinfo{person}{Yanhong Xu}, \bibinfo{person}{Yanping Huang}, \bibinfo{person}{Yao Li}, \bibinfo{person}{Yao Zhao}, \bibinfo{person}{Yaofeng Sun}, \bibinfo{person}{Yaohui Li}, \bibinfo{person}{Yaohui Wang}, \bibinfo{person}{Yi Yu}, \bibinfo{person}{Yi Zheng}, \bibinfo{person}{Yichao Zhang}, \bibinfo{person}{Yifan Shi}, \bibinfo{person}{Yiliang Xiong}, \bibinfo{person}{Ying He}, \bibinfo{person}{Ying Tang}, \bibinfo{person}{Yishi Piao}, \bibinfo{person}{Yisong Wang}, \bibinfo{person}{Yixuan Tan}, \bibinfo{person}{Yiyang Ma}, \bibinfo{person}{Yiyuan Liu}, \bibinfo{person}{Yongqiang Guo}, \bibinfo{person}{Yu Wu}, \bibinfo{person}{Yuan Ou}, \bibinfo{person}{Yuchen Zhu}, \bibinfo{person}{Yuduan Wang}, \bibinfo{person}{Yue Gong}, \bibinfo{person}{Yuheng Zou}, \bibinfo{person}{Yujia He}, \bibinfo{person}{Yukun Zha}, \bibinfo{person}{Yunfan Xiong},
  \bibinfo{person}{Yunxian Ma}, \bibinfo{person}{Yuting Yan}, \bibinfo{person}{Yuxiang Luo}, \bibinfo{person}{Yuxiang You}, \bibinfo{person}{Yuxuan Liu}, \bibinfo{person}{Yuyang Zhou}, \bibinfo{person}{Z.~F. Wu}, \bibinfo{person}{Z.~Z. Ren}, \bibinfo{person}{Zehui Ren}, \bibinfo{person}{Zhangli Sha}, \bibinfo{person}{Zhe Fu}, \bibinfo{person}{Zhean Xu}, \bibinfo{person}{Zhen Huang}, \bibinfo{person}{Zhen Zhang}, \bibinfo{person}{Zhenda Xie}, \bibinfo{person}{Zhengyan Zhang}, \bibinfo{person}{Zhewen Hao}, \bibinfo{person}{Zhibin Gou}, \bibinfo{person}{Zhicheng Ma}, \bibinfo{person}{Zhigang Yan}, \bibinfo{person}{Zhihong Shao}, \bibinfo{person}{Zhipeng Xu}, \bibinfo{person}{Zhiyu Wu}, \bibinfo{person}{Zhongyu Zhang}, \bibinfo{person}{Zhuoshu Li}, \bibinfo{person}{Zihui Gu}, \bibinfo{person}{Zijia Zhu}, \bibinfo{person}{Zijun Liu}, \bibinfo{person}{Zilin Li}, \bibinfo{person}{Ziwei Xie}, \bibinfo{person}{Ziyang Song}, \bibinfo{person}{Ziyi Gao}, {and} \bibinfo{person}{Zizheng Pan}.}
  \bibinfo{year}{2025}\natexlab{b}.
\newblock \bibinfo{title}{DeepSeek-V3 Technical Report}.
\newblock
\newblock
\showeprint[arxiv]{2412.19437}~[cs.CL]
\urldef\tempurl%
\url{https://arxiv.org/abs/2412.19437}
\showURL{%
\tempurl}


\bibitem[Dettmers et~al\mbox{.}(2023)]%
        {dettmers2023spqrsparsequantizedrepresentationnearlossless}
\bibfield{author}{\bibinfo{person}{Tim Dettmers}, \bibinfo{person}{Ruslan Svirschevski}, \bibinfo{person}{Vage Egiazarian}, \bibinfo{person}{Denis Kuznedelev}, \bibinfo{person}{Elias Frantar}, \bibinfo{person}{Saleh Ashkboos}, \bibinfo{person}{Alexander Borzunov}, \bibinfo{person}{Torsten Hoefler}, {and} \bibinfo{person}{Dan Alistarh}.} \bibinfo{year}{2023}\natexlab{}.
\newblock \bibinfo{title}{SpQR: A Sparse-Quantized Representation for Near-Lossless LLM Weight Compression}.
\newblock
\newblock
\showeprint[arxiv]{2306.03078}~[cs.CL]
\urldef\tempurl%
\url{https://arxiv.org/abs/2306.03078}
\showURL{%
\tempurl}


\bibitem[Eliseev and Mazur(2023)]%
        {eliseev2023fastinferencemixtureofexpertslanguage}
\bibfield{author}{\bibinfo{person}{Artyom Eliseev} {and} \bibinfo{person}{Denis Mazur}.} \bibinfo{year}{2023}\natexlab{}.
\newblock \bibinfo{title}{Fast Inference of Mixture-of-Experts Language Models with Offloading}.
\newblock
\newblock
\showeprint[arxiv]{2312.17238}~[cs.LG]
\urldef\tempurl%
\url{https://arxiv.org/abs/2312.17238}
\showURL{%
\tempurl}


\bibitem[Fang et~al\mbox{.}(2025)]%
        {fang2025klotskiefficientmixtureofexpertinference}
\bibfield{author}{\bibinfo{person}{Zhiyuan Fang}, \bibinfo{person}{Yuegui Huang}, \bibinfo{person}{Zicong Hong}, \bibinfo{person}{Yufeng Lyu}, \bibinfo{person}{Wuhui Chen}, \bibinfo{person}{Yue Yu}, \bibinfo{person}{Fan Yu}, {and} \bibinfo{person}{Zibin Zheng}.} \bibinfo{year}{2025}\natexlab{}.
\newblock \bibinfo{title}{Klotski: Efficient Mixture-of-Expert Inference via Expert-Aware Multi-Batch Pipeline}.
\newblock
\newblock
\showeprint[arxiv]{2502.06888}~[cs.LG]
\urldef\tempurl%
\url{https://arxiv.org/abs/2502.06888}
\showURL{%
\tempurl}


\bibitem[Grattafiori et~al\mbox{.}(2024)]%
        {grattafiori2024llama3herdmodels}
\bibfield{author}{\bibinfo{person}{Aaron Grattafiori}, \bibinfo{person}{Abhimanyu Dubey}, \bibinfo{person}{Abhinav Jauhri}, \bibinfo{person}{Abhinav Pandey}, \bibinfo{person}{Abhishek Kadian}, \bibinfo{person}{Ahmad Al-Dahle}, \bibinfo{person}{Aiesha Letman}, \bibinfo{person}{Akhil Mathur}, \bibinfo{person}{Alan Schelten}, \bibinfo{person}{Alex Vaughan}, \bibinfo{person}{Amy Yang}, \bibinfo{person}{Angela Fan}, \bibinfo{person}{Anirudh Goyal}, \bibinfo{person}{Anthony Hartshorn}, \bibinfo{person}{Aobo Yang}, \bibinfo{person}{Archi Mitra}, \bibinfo{person}{Archie Sravankumar}, \bibinfo{person}{Artem Korenev}, \bibinfo{person}{Arthur Hinsvark}, \bibinfo{person}{Arun Rao}, \bibinfo{person}{Aston Zhang}, \bibinfo{person}{Aurelien Rodriguez}, \bibinfo{person}{Austen Gregerson}, \bibinfo{person}{Ava Spataru}, \bibinfo{person}{Baptiste Roziere}, \bibinfo{person}{Bethany Biron}, \bibinfo{person}{Binh Tang}, \bibinfo{person}{Bobbie Chern}, \bibinfo{person}{Charlotte Caucheteux}, \bibinfo{person}{Chaya Nayak},
  \bibinfo{person}{Chloe Bi}, \bibinfo{person}{Chris Marra}, \bibinfo{person}{Chris McConnell}, \bibinfo{person}{Christian Keller}, \bibinfo{person}{Christophe Touret}, \bibinfo{person}{Chunyang Wu}, \bibinfo{person}{Corinne Wong}, \bibinfo{person}{Cristian~Canton Ferrer}, \bibinfo{person}{Cyrus Nikolaidis}, \bibinfo{person}{Damien Allonsius}, \bibinfo{person}{Daniel Song}, \bibinfo{person}{Danielle Pintz}, \bibinfo{person}{Danny Livshits}, \bibinfo{person}{Danny Wyatt}, \bibinfo{person}{David Esiobu}, \bibinfo{person}{Dhruv Choudhary}, \bibinfo{person}{Dhruv Mahajan}, \bibinfo{person}{Diego Garcia-Olano}, \bibinfo{person}{Diego Perino}, \bibinfo{person}{Dieuwke Hupkes}, \bibinfo{person}{Egor Lakomkin}, \bibinfo{person}{Ehab AlBadawy}, \bibinfo{person}{Elina Lobanova}, \bibinfo{person}{Emily Dinan}, \bibinfo{person}{Eric~Michael Smith}, \bibinfo{person}{Filip Radenovic}, \bibinfo{person}{Francisco Guzmán}, \bibinfo{person}{Frank Zhang}, \bibinfo{person}{Gabriel Synnaeve}, \bibinfo{person}{Gabrielle Lee},
  \bibinfo{person}{Georgia~Lewis Anderson}, \bibinfo{person}{Govind Thattai}, \bibinfo{person}{Graeme Nail}, \bibinfo{person}{Gregoire Mialon}, \bibinfo{person}{Guan Pang}, \bibinfo{person}{Guillem Cucurell}, \bibinfo{person}{Hailey Nguyen}, \bibinfo{person}{Hannah Korevaar}, \bibinfo{person}{Hu Xu}, \bibinfo{person}{Hugo Touvron}, \bibinfo{person}{Iliyan Zarov}, \bibinfo{person}{Imanol~Arrieta Ibarra}, \bibinfo{person}{Isabel Kloumann}, \bibinfo{person}{Ishan Misra}, \bibinfo{person}{Ivan Evtimov}, \bibinfo{person}{Jack Zhang}, \bibinfo{person}{Jade Copet}, \bibinfo{person}{Jaewon Lee}, \bibinfo{person}{Jan Geffert}, \bibinfo{person}{Jana Vranes}, \bibinfo{person}{Jason Park}, \bibinfo{person}{Jay Mahadeokar}, \bibinfo{person}{Jeet Shah}, \bibinfo{person}{Jelmer van~der Linde}, \bibinfo{person}{Jennifer Billock}, \bibinfo{person}{Jenny Hong}, \bibinfo{person}{Jenya Lee}, \bibinfo{person}{Jeremy Fu}, \bibinfo{person}{Jianfeng Chi}, \bibinfo{person}{Jianyu Huang}, \bibinfo{person}{Jiawen Liu},
  \bibinfo{person}{Jie Wang}, \bibinfo{person}{Jiecao Yu}, \bibinfo{person}{Joanna Bitton}, \bibinfo{person}{Joe Spisak}, \bibinfo{person}{Jongsoo Park}, \bibinfo{person}{Joseph Rocca}, \bibinfo{person}{Joshua Johnstun}, \bibinfo{person}{Joshua Saxe}, \bibinfo{person}{Junteng Jia}, \bibinfo{person}{Kalyan~Vasuden Alwala}, \bibinfo{person}{Karthik Prasad}, \bibinfo{person}{Kartikeya Upasani}, \bibinfo{person}{Kate Plawiak}, \bibinfo{person}{Ke Li}, \bibinfo{person}{Kenneth Heafield}, \bibinfo{person}{Kevin Stone}, \bibinfo{person}{Khalid El-Arini}, \bibinfo{person}{Krithika Iyer}, \bibinfo{person}{Kshitiz Malik}, \bibinfo{person}{Kuenley Chiu}, \bibinfo{person}{Kunal Bhalla}, \bibinfo{person}{Kushal Lakhotia}, \bibinfo{person}{Lauren Rantala-Yeary}, \bibinfo{person}{Laurens van~der Maaten}, \bibinfo{person}{Lawrence Chen}, \bibinfo{person}{Liang Tan}, \bibinfo{person}{Liz Jenkins}, \bibinfo{person}{Louis Martin}, \bibinfo{person}{Lovish Madaan}, \bibinfo{person}{Lubo Malo}, \bibinfo{person}{Lukas Blecher},
  \bibinfo{person}{Lukas Landzaat}, \bibinfo{person}{Luke de Oliveira}, \bibinfo{person}{Madeline Muzzi}, \bibinfo{person}{Mahesh Pasupuleti}, \bibinfo{person}{Mannat Singh}, \bibinfo{person}{Manohar Paluri}, \bibinfo{person}{Marcin Kardas}, \bibinfo{person}{Maria Tsimpoukelli}, \bibinfo{person}{Mathew Oldham}, \bibinfo{person}{Mathieu Rita}, \bibinfo{person}{Maya Pavlova}, \bibinfo{person}{Melanie Kambadur}, \bibinfo{person}{Mike Lewis}, \bibinfo{person}{Min Si}, \bibinfo{person}{Mitesh~Kumar Singh}, \bibinfo{person}{Mona Hassan}, \bibinfo{person}{Naman Goyal}, \bibinfo{person}{Narjes Torabi}, \bibinfo{person}{Nikolay Bashlykov}, \bibinfo{person}{Nikolay Bogoychev}, \bibinfo{person}{Niladri Chatterji}, \bibinfo{person}{Ning Zhang}, \bibinfo{person}{Olivier Duchenne}, \bibinfo{person}{Onur Çelebi}, \bibinfo{person}{Patrick Alrassy}, \bibinfo{person}{Pengchuan Zhang}, \bibinfo{person}{Pengwei Li}, \bibinfo{person}{Petar Vasic}, \bibinfo{person}{Peter Weng}, \bibinfo{person}{Prajjwal Bhargava},
  \bibinfo{person}{Pratik Dubal}, \bibinfo{person}{Praveen Krishnan}, \bibinfo{person}{Punit~Singh Koura}, \bibinfo{person}{Puxin Xu}, \bibinfo{person}{Qing He}, \bibinfo{person}{Qingxiao Dong}, \bibinfo{person}{Ragavan Srinivasan}, \bibinfo{person}{Raj Ganapathy}, \bibinfo{person}{Ramon Calderer}, \bibinfo{person}{Ricardo~Silveira Cabral}, \bibinfo{person}{Robert Stojnic}, \bibinfo{person}{Roberta Raileanu}, \bibinfo{person}{Rohan Maheswari}, \bibinfo{person}{Rohit Girdhar}, \bibinfo{person}{Rohit Patel}, \bibinfo{person}{Romain Sauvestre}, \bibinfo{person}{Ronnie Polidoro}, \bibinfo{person}{Roshan Sumbaly}, \bibinfo{person}{Ross Taylor}, \bibinfo{person}{Ruan Silva}, \bibinfo{person}{Rui Hou}, \bibinfo{person}{Rui Wang}, \bibinfo{person}{Saghar Hosseini}, \bibinfo{person}{Sahana Chennabasappa}, \bibinfo{person}{Sanjay Singh}, \bibinfo{person}{Sean Bell}, \bibinfo{person}{Seohyun~Sonia Kim}, \bibinfo{person}{Sergey Edunov}, \bibinfo{person}{Shaoliang Nie}, \bibinfo{person}{Sharan Narang},
  \bibinfo{person}{Sharath Raparthy}, \bibinfo{person}{Sheng Shen}, \bibinfo{person}{Shengye Wan}, \bibinfo{person}{Shruti Bhosale}, \bibinfo{person}{Shun Zhang}, \bibinfo{person}{Simon Vandenhende}, \bibinfo{person}{Soumya Batra}, \bibinfo{person}{Spencer Whitman}, \bibinfo{person}{Sten Sootla}, \bibinfo{person}{Stephane Collot}, \bibinfo{person}{Suchin Gururangan}, \bibinfo{person}{Sydney Borodinsky}, \bibinfo{person}{Tamar Herman}, \bibinfo{person}{Tara Fowler}, \bibinfo{person}{Tarek Sheasha}, \bibinfo{person}{Thomas Georgiou}, \bibinfo{person}{Thomas Scialom}, \bibinfo{person}{Tobias Speckbacher}, \bibinfo{person}{Todor Mihaylov}, \bibinfo{person}{Tong Xiao}, \bibinfo{person}{Ujjwal Karn}, \bibinfo{person}{Vedanuj Goswami}, \bibinfo{person}{Vibhor Gupta}, \bibinfo{person}{Vignesh Ramanathan}, \bibinfo{person}{Viktor Kerkez}, \bibinfo{person}{Vincent Gonguet}, \bibinfo{person}{Virginie Do}, \bibinfo{person}{Vish Vogeti}, \bibinfo{person}{Vítor Albiero}, \bibinfo{person}{Vladan Petrovic},
  \bibinfo{person}{Weiwei Chu}, \bibinfo{person}{Wenhan Xiong}, \bibinfo{person}{Wenyin Fu}, \bibinfo{person}{Whitney Meers}, \bibinfo{person}{Xavier Martinet}, \bibinfo{person}{Xiaodong Wang}, \bibinfo{person}{Xiaofang Wang}, \bibinfo{person}{Xiaoqing~Ellen Tan}, \bibinfo{person}{Xide Xia}, \bibinfo{person}{Xinfeng Xie}, \bibinfo{person}{Xuchao Jia}, \bibinfo{person}{Xuewei Wang}, \bibinfo{person}{Yaelle Goldschlag}, \bibinfo{person}{Yashesh Gaur}, \bibinfo{person}{Yasmine Babaei}, \bibinfo{person}{Yi Wen}, \bibinfo{person}{Yiwen Song}, \bibinfo{person}{Yuchen Zhang}, \bibinfo{person}{Yue Li}, \bibinfo{person}{Yuning Mao}, \bibinfo{person}{Zacharie~Delpierre Coudert}, \bibinfo{person}{Zheng Yan}, \bibinfo{person}{Zhengxing Chen}, \bibinfo{person}{Zoe Papakipos}, \bibinfo{person}{Aaditya Singh}, \bibinfo{person}{Aayushi Srivastava}, \bibinfo{person}{Abha Jain}, \bibinfo{person}{Adam Kelsey}, \bibinfo{person}{Adam Shajnfeld}, \bibinfo{person}{Adithya Gangidi}, \bibinfo{person}{Adolfo Victoria},
  \bibinfo{person}{Ahuva Goldstand}, \bibinfo{person}{Ajay Menon}, \bibinfo{person}{Ajay Sharma}, \bibinfo{person}{Alex Boesenberg}, \bibinfo{person}{Alexei Baevski}, \bibinfo{person}{Allie Feinstein}, \bibinfo{person}{Amanda Kallet}, \bibinfo{person}{Amit Sangani}, \bibinfo{person}{Amos Teo}, \bibinfo{person}{Anam Yunus}, \bibinfo{person}{Andrei Lupu}, \bibinfo{person}{Andres Alvarado}, \bibinfo{person}{Andrew Caples}, \bibinfo{person}{Andrew Gu}, \bibinfo{person}{Andrew Ho}, \bibinfo{person}{Andrew Poulton}, \bibinfo{person}{Andrew Ryan}, \bibinfo{person}{Ankit Ramchandani}, \bibinfo{person}{Annie Dong}, \bibinfo{person}{Annie Franco}, \bibinfo{person}{Anuj Goyal}, \bibinfo{person}{Aparajita Saraf}, \bibinfo{person}{Arkabandhu Chowdhury}, \bibinfo{person}{Ashley Gabriel}, \bibinfo{person}{Ashwin Bharambe}, \bibinfo{person}{Assaf Eisenman}, \bibinfo{person}{Azadeh Yazdan}, \bibinfo{person}{Beau James}, \bibinfo{person}{Ben Maurer}, \bibinfo{person}{Benjamin Leonhardi}, \bibinfo{person}{Bernie Huang},
  \bibinfo{person}{Beth Loyd}, \bibinfo{person}{Beto~De Paola}, \bibinfo{person}{Bhargavi Paranjape}, \bibinfo{person}{Bing Liu}, \bibinfo{person}{Bo Wu}, \bibinfo{person}{Boyu Ni}, \bibinfo{person}{Braden Hancock}, \bibinfo{person}{Bram Wasti}, \bibinfo{person}{Brandon Spence}, \bibinfo{person}{Brani Stojkovic}, \bibinfo{person}{Brian Gamido}, \bibinfo{person}{Britt Montalvo}, \bibinfo{person}{Carl Parker}, \bibinfo{person}{Carly Burton}, \bibinfo{person}{Catalina Mejia}, \bibinfo{person}{Ce Liu}, \bibinfo{person}{Changhan Wang}, \bibinfo{person}{Changkyu Kim}, \bibinfo{person}{Chao Zhou}, \bibinfo{person}{Chester Hu}, \bibinfo{person}{Ching-Hsiang Chu}, \bibinfo{person}{Chris Cai}, \bibinfo{person}{Chris Tindal}, \bibinfo{person}{Christoph Feichtenhofer}, \bibinfo{person}{Cynthia Gao}, \bibinfo{person}{Damon Civin}, \bibinfo{person}{Dana Beaty}, \bibinfo{person}{Daniel Kreymer}, \bibinfo{person}{Daniel Li}, \bibinfo{person}{David Adkins}, \bibinfo{person}{David Xu}, \bibinfo{person}{Davide Testuggine},
  \bibinfo{person}{Delia David}, \bibinfo{person}{Devi Parikh}, \bibinfo{person}{Diana Liskovich}, \bibinfo{person}{Didem Foss}, \bibinfo{person}{Dingkang Wang}, \bibinfo{person}{Duc Le}, \bibinfo{person}{Dustin Holland}, \bibinfo{person}{Edward Dowling}, \bibinfo{person}{Eissa Jamil}, \bibinfo{person}{Elaine Montgomery}, \bibinfo{person}{Eleonora Presani}, \bibinfo{person}{Emily Hahn}, \bibinfo{person}{Emily Wood}, \bibinfo{person}{Eric-Tuan Le}, \bibinfo{person}{Erik Brinkman}, \bibinfo{person}{Esteban Arcaute}, \bibinfo{person}{Evan Dunbar}, \bibinfo{person}{Evan Smothers}, \bibinfo{person}{Fei Sun}, \bibinfo{person}{Felix Kreuk}, \bibinfo{person}{Feng Tian}, \bibinfo{person}{Filippos Kokkinos}, \bibinfo{person}{Firat Ozgenel}, \bibinfo{person}{Francesco Caggioni}, \bibinfo{person}{Frank Kanayet}, \bibinfo{person}{Frank Seide}, \bibinfo{person}{Gabriela~Medina Florez}, \bibinfo{person}{Gabriella Schwarz}, \bibinfo{person}{Gada Badeer}, \bibinfo{person}{Georgia Swee}, \bibinfo{person}{Gil Halpern},
  \bibinfo{person}{Grant Herman}, \bibinfo{person}{Grigory Sizov}, \bibinfo{person}{Guangyi}, \bibinfo{person}{Zhang}, \bibinfo{person}{Guna Lakshminarayanan}, \bibinfo{person}{Hakan Inan}, \bibinfo{person}{Hamid Shojanazeri}, \bibinfo{person}{Han Zou}, \bibinfo{person}{Hannah Wang}, \bibinfo{person}{Hanwen Zha}, \bibinfo{person}{Haroun Habeeb}, \bibinfo{person}{Harrison Rudolph}, \bibinfo{person}{Helen Suk}, \bibinfo{person}{Henry Aspegren}, \bibinfo{person}{Hunter Goldman}, \bibinfo{person}{Hongyuan Zhan}, \bibinfo{person}{Ibrahim Damlaj}, \bibinfo{person}{Igor Molybog}, \bibinfo{person}{Igor Tufanov}, \bibinfo{person}{Ilias Leontiadis}, \bibinfo{person}{Irina-Elena Veliche}, \bibinfo{person}{Itai Gat}, \bibinfo{person}{Jake Weissman}, \bibinfo{person}{James Geboski}, \bibinfo{person}{James Kohli}, \bibinfo{person}{Janice Lam}, \bibinfo{person}{Japhet Asher}, \bibinfo{person}{Jean-Baptiste Gaya}, \bibinfo{person}{Jeff Marcus}, \bibinfo{person}{Jeff Tang}, \bibinfo{person}{Jennifer Chan},
  \bibinfo{person}{Jenny Zhen}, \bibinfo{person}{Jeremy Reizenstein}, \bibinfo{person}{Jeremy Teboul}, \bibinfo{person}{Jessica Zhong}, \bibinfo{person}{Jian Jin}, \bibinfo{person}{Jingyi Yang}, \bibinfo{person}{Joe Cummings}, \bibinfo{person}{Jon Carvill}, \bibinfo{person}{Jon Shepard}, \bibinfo{person}{Jonathan McPhie}, \bibinfo{person}{Jonathan Torres}, \bibinfo{person}{Josh Ginsburg}, \bibinfo{person}{Junjie Wang}, \bibinfo{person}{Kai Wu}, \bibinfo{person}{Kam~Hou U}, \bibinfo{person}{Karan Saxena}, \bibinfo{person}{Kartikay Khandelwal}, \bibinfo{person}{Katayoun Zand}, \bibinfo{person}{Kathy Matosich}, \bibinfo{person}{Kaushik Veeraraghavan}, \bibinfo{person}{Kelly Michelena}, \bibinfo{person}{Keqian Li}, \bibinfo{person}{Kiran Jagadeesh}, \bibinfo{person}{Kun Huang}, \bibinfo{person}{Kunal Chawla}, \bibinfo{person}{Kyle Huang}, \bibinfo{person}{Lailin Chen}, \bibinfo{person}{Lakshya Garg}, \bibinfo{person}{Lavender A}, \bibinfo{person}{Leandro Silva}, \bibinfo{person}{Lee Bell}, \bibinfo{person}{Lei
  Zhang}, \bibinfo{person}{Liangpeng Guo}, \bibinfo{person}{Licheng Yu}, \bibinfo{person}{Liron Moshkovich}, \bibinfo{person}{Luca Wehrstedt}, \bibinfo{person}{Madian Khabsa}, \bibinfo{person}{Manav Avalani}, \bibinfo{person}{Manish Bhatt}, \bibinfo{person}{Martynas Mankus}, \bibinfo{person}{Matan Hasson}, \bibinfo{person}{Matthew Lennie}, \bibinfo{person}{Matthias Reso}, \bibinfo{person}{Maxim Groshev}, \bibinfo{person}{Maxim Naumov}, \bibinfo{person}{Maya Lathi}, \bibinfo{person}{Meghan Keneally}, \bibinfo{person}{Miao Liu}, \bibinfo{person}{Michael~L. Seltzer}, \bibinfo{person}{Michal Valko}, \bibinfo{person}{Michelle Restrepo}, \bibinfo{person}{Mihir Patel}, \bibinfo{person}{Mik Vyatskov}, \bibinfo{person}{Mikayel Samvelyan}, \bibinfo{person}{Mike Clark}, \bibinfo{person}{Mike Macey}, \bibinfo{person}{Mike Wang}, \bibinfo{person}{Miquel~Jubert Hermoso}, \bibinfo{person}{Mo Metanat}, \bibinfo{person}{Mohammad Rastegari}, \bibinfo{person}{Munish Bansal}, \bibinfo{person}{Nandhini Santhanam},
  \bibinfo{person}{Natascha Parks}, \bibinfo{person}{Natasha White}, \bibinfo{person}{Navyata Bawa}, \bibinfo{person}{Nayan Singhal}, \bibinfo{person}{Nick Egebo}, \bibinfo{person}{Nicolas Usunier}, \bibinfo{person}{Nikhil Mehta}, \bibinfo{person}{Nikolay~Pavlovich Laptev}, \bibinfo{person}{Ning Dong}, \bibinfo{person}{Norman Cheng}, \bibinfo{person}{Oleg Chernoguz}, \bibinfo{person}{Olivia Hart}, \bibinfo{person}{Omkar Salpekar}, \bibinfo{person}{Ozlem Kalinli}, \bibinfo{person}{Parkin Kent}, \bibinfo{person}{Parth Parekh}, \bibinfo{person}{Paul Saab}, \bibinfo{person}{Pavan Balaji}, \bibinfo{person}{Pedro Rittner}, \bibinfo{person}{Philip Bontrager}, \bibinfo{person}{Pierre Roux}, \bibinfo{person}{Piotr Dollar}, \bibinfo{person}{Polina Zvyagina}, \bibinfo{person}{Prashant Ratanchandani}, \bibinfo{person}{Pritish Yuvraj}, \bibinfo{person}{Qian Liang}, \bibinfo{person}{Rachad Alao}, \bibinfo{person}{Rachel Rodriguez}, \bibinfo{person}{Rafi Ayub}, \bibinfo{person}{Raghotham Murthy}, \bibinfo{person}{Raghu
  Nayani}, \bibinfo{person}{Rahul Mitra}, \bibinfo{person}{Rangaprabhu Parthasarathy}, \bibinfo{person}{Raymond Li}, \bibinfo{person}{Rebekkah Hogan}, \bibinfo{person}{Robin Battey}, \bibinfo{person}{Rocky Wang}, \bibinfo{person}{Russ Howes}, \bibinfo{person}{Ruty Rinott}, \bibinfo{person}{Sachin Mehta}, \bibinfo{person}{Sachin Siby}, \bibinfo{person}{Sai~Jayesh Bondu}, \bibinfo{person}{Samyak Datta}, \bibinfo{person}{Sara Chugh}, \bibinfo{person}{Sara Hunt}, \bibinfo{person}{Sargun Dhillon}, \bibinfo{person}{Sasha Sidorov}, \bibinfo{person}{Satadru Pan}, \bibinfo{person}{Saurabh Mahajan}, \bibinfo{person}{Saurabh Verma}, \bibinfo{person}{Seiji Yamamoto}, \bibinfo{person}{Sharadh Ramaswamy}, \bibinfo{person}{Shaun Lindsay}, \bibinfo{person}{Shaun Lindsay}, \bibinfo{person}{Sheng Feng}, \bibinfo{person}{Shenghao Lin}, \bibinfo{person}{Shengxin~Cindy Zha}, \bibinfo{person}{Shishir Patil}, \bibinfo{person}{Shiva Shankar}, \bibinfo{person}{Shuqiang Zhang}, \bibinfo{person}{Shuqiang Zhang}, \bibinfo{person}{Sinong
  Wang}, \bibinfo{person}{Sneha Agarwal}, \bibinfo{person}{Soji Sajuyigbe}, \bibinfo{person}{Soumith Chintala}, \bibinfo{person}{Stephanie Max}, \bibinfo{person}{Stephen Chen}, \bibinfo{person}{Steve Kehoe}, \bibinfo{person}{Steve Satterfield}, \bibinfo{person}{Sudarshan Govindaprasad}, \bibinfo{person}{Sumit Gupta}, \bibinfo{person}{Summer Deng}, \bibinfo{person}{Sungmin Cho}, \bibinfo{person}{Sunny Virk}, \bibinfo{person}{Suraj Subramanian}, \bibinfo{person}{Sy Choudhury}, \bibinfo{person}{Sydney Goldman}, \bibinfo{person}{Tal Remez}, \bibinfo{person}{Tamar Glaser}, \bibinfo{person}{Tamara Best}, \bibinfo{person}{Thilo Koehler}, \bibinfo{person}{Thomas Robinson}, \bibinfo{person}{Tianhe Li}, \bibinfo{person}{Tianjun Zhang}, \bibinfo{person}{Tim Matthews}, \bibinfo{person}{Timothy Chou}, \bibinfo{person}{Tzook Shaked}, \bibinfo{person}{Varun Vontimitta}, \bibinfo{person}{Victoria Ajayi}, \bibinfo{person}{Victoria Montanez}, \bibinfo{person}{Vijai Mohan}, \bibinfo{person}{Vinay~Satish Kumar},
  \bibinfo{person}{Vishal Mangla}, \bibinfo{person}{Vlad Ionescu}, \bibinfo{person}{Vlad Poenaru}, \bibinfo{person}{Vlad~Tiberiu Mihailescu}, \bibinfo{person}{Vladimir Ivanov}, \bibinfo{person}{Wei Li}, \bibinfo{person}{Wenchen Wang}, \bibinfo{person}{Wenwen Jiang}, \bibinfo{person}{Wes Bouaziz}, \bibinfo{person}{Will Constable}, \bibinfo{person}{Xiaocheng Tang}, \bibinfo{person}{Xiaojian Wu}, \bibinfo{person}{Xiaolan Wang}, \bibinfo{person}{Xilun Wu}, \bibinfo{person}{Xinbo Gao}, \bibinfo{person}{Yaniv Kleinman}, \bibinfo{person}{Yanjun Chen}, \bibinfo{person}{Ye Hu}, \bibinfo{person}{Ye Jia}, \bibinfo{person}{Ye Qi}, \bibinfo{person}{Yenda Li}, \bibinfo{person}{Yilin Zhang}, \bibinfo{person}{Ying Zhang}, \bibinfo{person}{Yossi Adi}, \bibinfo{person}{Youngjin Nam}, \bibinfo{person}{Yu}, \bibinfo{person}{Wang}, \bibinfo{person}{Yu Zhao}, \bibinfo{person}{Yuchen Hao}, \bibinfo{person}{Yundi Qian}, \bibinfo{person}{Yunlu Li}, \bibinfo{person}{Yuzi He}, \bibinfo{person}{Zach Rait}, \bibinfo{person}{Zachary
  DeVito}, \bibinfo{person}{Zef Rosnbrick}, \bibinfo{person}{Zhaoduo Wen}, \bibinfo{person}{Zhenyu Yang}, \bibinfo{person}{Zhiwei Zhao}, {and} \bibinfo{person}{Zhiyu Ma}.} \bibinfo{year}{2024}\natexlab{}.
\newblock \bibinfo{title}{The Llama 3 Herd of Models}.
\newblock
\newblock
\showeprint[arxiv]{2407.21783}~[cs.AI]
\urldef\tempurl%
\url{https://arxiv.org/abs/2407.21783}
\showURL{%
\tempurl}


\bibitem[Huang et~al\mbox{.}(2024)]%
        {huang2024enovaautoscalingcosteffectivestable}
\bibfield{author}{\bibinfo{person}{Tao Huang}, \bibinfo{person}{Pengfei Chen}, \bibinfo{person}{Kyoka Gong}, \bibinfo{person}{Jocky Hawk}, \bibinfo{person}{Zachary Bright}, \bibinfo{person}{Wenxin Xie}, \bibinfo{person}{Kecheng Huang}, {and} \bibinfo{person}{Zhi Ji}.} \bibinfo{year}{2024}\natexlab{}.
\newblock \bibinfo{title}{ENOVA: Autoscaling towards Cost-effective and Stable Serverless LLM Serving}.
\newblock
\newblock
\showeprint[arxiv]{2407.09486}~[cs.DC]
\urldef\tempurl%
\url{https://arxiv.org/abs/2407.09486}
\showURL{%
\tempurl}


\bibitem[Jia(2024)]%
        {jia2024aime}
\bibfield{author}{\bibinfo{person}{Maxwell Jia}.} \bibinfo{year}{2024}\natexlab{}.
\newblock \bibinfo{title}{{AIME 2024 Dataset}}.
\newblock \bibinfo{howpublished}{\url{https://huggingface.co/datasets/Maxwell-Jia/AIME_2024}}.
\newblock
\newblock
\shownote{Accessed: 2025-04-10}.


\bibitem[Juravsky et~al\mbox{.}(2024)]%
        {juravsky2024hydragenhighthroughputllminference}
\bibfield{author}{\bibinfo{person}{Jordan Juravsky}, \bibinfo{person}{Bradley Brown}, \bibinfo{person}{Ryan Ehrlich}, \bibinfo{person}{Daniel~Y. Fu}, \bibinfo{person}{Christopher Ré}, {and} \bibinfo{person}{Azalia Mirhoseini}.} \bibinfo{year}{2024}\natexlab{}.
\newblock \bibinfo{title}{Hydragen: High-Throughput LLM Inference with Shared Prefixes}.
\newblock
\newblock
\showeprint[arxiv]{2402.05099}~[cs.LG]
\urldef\tempurl%
\url{https://arxiv.org/abs/2402.05099}
\showURL{%
\tempurl}


\bibitem[Kamahori et~al\mbox{.}(2025)]%
        {kamahori2025fiddlercpugpuorchestrationfast}
\bibfield{author}{\bibinfo{person}{Keisuke Kamahori}, \bibinfo{person}{Tian Tang}, \bibinfo{person}{Yile Gu}, \bibinfo{person}{Kan Zhu}, {and} \bibinfo{person}{Baris Kasikci}.} \bibinfo{year}{2025}\natexlab{}.
\newblock \bibinfo{title}{Fiddler: CPU-GPU Orchestration for Fast Inference of Mixture-of-Experts Models}.
\newblock
\newblock
\showeprint[arxiv]{2402.07033}~[cs.LG]
\urldef\tempurl%
\url{https://arxiv.org/abs/2402.07033}
\showURL{%
\tempurl}


\bibitem[Kim et~al\mbox{.}(2025)]%
        {kim2025adordesignexplorationframework}
\bibfield{author}{\bibinfo{person}{Junsoo Kim}, \bibinfo{person}{Hunjong Lee}, \bibinfo{person}{Geonwoo Ko}, \bibinfo{person}{Gyubin Choi}, \bibinfo{person}{Seri Ham}, \bibinfo{person}{Seongmin Hong}, {and} \bibinfo{person}{Joo-Young Kim}.} \bibinfo{year}{2025}\natexlab{}.
\newblock \bibinfo{title}{ADOR: A Design Exploration Framework for LLM Serving with Enhanced Latency and Throughput}.
\newblock
\newblock
\showeprint[arxiv]{2503.04253}~[cs.AR]
\urldef\tempurl%
\url{https://arxiv.org/abs/2503.04253}
\showURL{%
\tempurl}


\bibitem[Kwon et~al\mbox{.}(2023)]%
        {pagedattn}
\bibfield{author}{\bibinfo{person}{Woosuk Kwon}, \bibinfo{person}{Zhuohan Li}, \bibinfo{person}{Siyuan Zhuang}, \bibinfo{person}{Ying Sheng}, \bibinfo{person}{Lianmin Zheng}, \bibinfo{person}{Cody~Hao Yu}, \bibinfo{person}{Joseph~E. Gonzalez}, \bibinfo{person}{Hao Zhang}, {and} \bibinfo{person}{Ion Stoica}.} \bibinfo{year}{2023}\natexlab{}.
\newblock \bibinfo{title}{Efficient Memory Management for Large Language Model Serving with PagedAttention}.
\newblock
\newblock
\showeprint[arxiv]{2309.06180}~[cs.LG]
\urldef\tempurl%
\url{https://arxiv.org/abs/2309.06180}
\showURL{%
\tempurl}


\bibitem[Liang et~al\mbox{.}(2023)]%
        {liang2023holisticevaluationlanguagemodels}
\bibfield{author}{\bibinfo{person}{Percy Liang}, \bibinfo{person}{Rishi Bommasani}, \bibinfo{person}{Tony Lee}, \bibinfo{person}{Dimitris Tsipras}, \bibinfo{person}{Dilara Soylu}, \bibinfo{person}{Michihiro Yasunaga}, \bibinfo{person}{Yian Zhang}, \bibinfo{person}{Deepak Narayanan}, \bibinfo{person}{Yuhuai Wu}, \bibinfo{person}{Ananya Kumar}, \bibinfo{person}{Benjamin Newman}, \bibinfo{person}{Binhang Yuan}, \bibinfo{person}{Bobby Yan}, \bibinfo{person}{Ce Zhang}, \bibinfo{person}{Christian Cosgrove}, \bibinfo{person}{Christopher~D. Manning}, \bibinfo{person}{Christopher Ré}, \bibinfo{person}{Diana Acosta-Navas}, \bibinfo{person}{Drew~A. Hudson}, \bibinfo{person}{Eric Zelikman}, \bibinfo{person}{Esin Durmus}, \bibinfo{person}{Faisal Ladhak}, \bibinfo{person}{Frieda Rong}, \bibinfo{person}{Hongyu Ren}, \bibinfo{person}{Huaxiu Yao}, \bibinfo{person}{Jue Wang}, \bibinfo{person}{Keshav Santhanam}, \bibinfo{person}{Laurel Orr}, \bibinfo{person}{Lucia Zheng}, \bibinfo{person}{Mert Yuksekgonul},
  \bibinfo{person}{Mirac Suzgun}, \bibinfo{person}{Nathan Kim}, \bibinfo{person}{Neel Guha}, \bibinfo{person}{Niladri Chatterji}, \bibinfo{person}{Omar Khattab}, \bibinfo{person}{Peter Henderson}, \bibinfo{person}{Qian Huang}, \bibinfo{person}{Ryan Chi}, \bibinfo{person}{Sang~Michael Xie}, \bibinfo{person}{Shibani Santurkar}, \bibinfo{person}{Surya Ganguli}, \bibinfo{person}{Tatsunori Hashimoto}, \bibinfo{person}{Thomas Icard}, \bibinfo{person}{Tianyi Zhang}, \bibinfo{person}{Vishrav Chaudhary}, \bibinfo{person}{William Wang}, \bibinfo{person}{Xuechen Li}, \bibinfo{person}{Yifan Mai}, \bibinfo{person}{Yuhui Zhang}, {and} \bibinfo{person}{Yuta Koreeda}.} \bibinfo{year}{2023}\natexlab{}.
\newblock \bibinfo{title}{Holistic Evaluation of Language Models}.
\newblock
\newblock
\showeprint[arxiv]{2211.09110}~[cs.CL]
\urldef\tempurl%
\url{https://arxiv.org/abs/2211.09110}
\showURL{%
\tempurl}


\bibitem[Lin et~al\mbox{.}(2024)]%
        {MLSYS2024_42a452cb}
\bibfield{author}{\bibinfo{person}{Ji Lin}, \bibinfo{person}{Jiaming Tang}, \bibinfo{person}{Haotian Tang}, \bibinfo{person}{Shang Yang}, \bibinfo{person}{Wei-Ming Chen}, \bibinfo{person}{Wei-Chen Wang}, \bibinfo{person}{Guangxuan Xiao}, \bibinfo{person}{Xingyu Dang}, \bibinfo{person}{Chuang Gan}, {and} \bibinfo{person}{Song Han}.} \bibinfo{year}{2024}\natexlab{}.
\newblock \showarticletitle{AWQ: Activation-aware Weight Quantization for On-Device LLM Compression and Acceleration}. In \bibinfo{booktitle}{\emph{Proceedings of Machine Learning and Systems}}, \bibfield{editor}{\bibinfo{person}{P.~Gibbons}, \bibinfo{person}{G.~Pekhimenko}, {and} \bibinfo{person}{C.~De Sa}} (Eds.), Vol.~\bibinfo{volume}{6}. \bibinfo{pages}{87--100}.
\newblock
\urldef\tempurl%
\url{https://proceedings.mlsys.org/paper_files/paper/2024/file/42a452cbafa9dd64e9ba4aa95cc1ef21-Paper-Conference.pdf}
\showURL{%
\tempurl}


\bibitem[Liu et~al\mbox{.}(2024)]%
        {liu2024retrievalattentionacceleratinglongcontextllm}
\bibfield{author}{\bibinfo{person}{Di Liu}, \bibinfo{person}{Meng Chen}, \bibinfo{person}{Baotong Lu}, \bibinfo{person}{Huiqiang Jiang}, \bibinfo{person}{Zhenhua Han}, \bibinfo{person}{Qianxi Zhang}, \bibinfo{person}{Qi Chen}, \bibinfo{person}{Chengruidong Zhang}, \bibinfo{person}{Bailu Ding}, \bibinfo{person}{Kai Zhang}, \bibinfo{person}{Chen Chen}, \bibinfo{person}{Fan Yang}, \bibinfo{person}{Yuqing Yang}, {and} \bibinfo{person}{Lili Qiu}.} \bibinfo{year}{2024}\natexlab{}.
\newblock \bibinfo{title}{RetrievalAttention: Accelerating Long-Context LLM Inference via Vector Retrieval}.
\newblock
\newblock
\showeprint[arxiv]{2409.10516}~[cs.LG]
\urldef\tempurl%
\url{https://arxiv.org/abs/2409.10516}
\showURL{%
\tempurl}


\bibitem[Liu et~al\mbox{.}(2025)]%
        {liu2025optimizingllmqueriesrelational}
\bibfield{author}{\bibinfo{person}{Shu Liu}, \bibinfo{person}{Asim Biswal}, \bibinfo{person}{Amog Kamsetty}, \bibinfo{person}{Audrey Cheng}, \bibinfo{person}{Luis~Gaspar Schroeder}, \bibinfo{person}{Liana Patel}, \bibinfo{person}{Shiyi Cao}, \bibinfo{person}{Xiangxi Mo}, \bibinfo{person}{Ion Stoica}, \bibinfo{person}{Joseph~E. Gonzalez}, {and} \bibinfo{person}{Matei Zaharia}.} \bibinfo{year}{2025}\natexlab{}.
\newblock \bibinfo{title}{Optimizing LLM Queries in Relational Data Analytics Workloads}.
\newblock
\newblock
\showeprint[arxiv]{2403.05821}~[cs.LG]
\urldef\tempurl%
\url{https://arxiv.org/abs/2403.05821}
\showURL{%
\tempurl}


\bibitem[Narayan et~al\mbox{.}(2022)]%
        {narayan2022foundationmodelswrangledata}
\bibfield{author}{\bibinfo{person}{Avanika Narayan}, \bibinfo{person}{Ines Chami}, \bibinfo{person}{Laurel Orr}, \bibinfo{person}{Simran Arora}, {and} \bibinfo{person}{Christopher Ré}.} \bibinfo{year}{2022}\natexlab{}.
\newblock \bibinfo{title}{Can Foundation Models Wrangle Your Data?}
\newblock
\newblock
\showeprint[arxiv]{2205.09911}~[cs.LG]
\urldef\tempurl%
\url{https://arxiv.org/abs/2205.09911}
\showURL{%
\tempurl}


\bibitem[Patel et~al\mbox{.}(2024a)]%
        {patel2024asplos}
\bibfield{author}{\bibinfo{person}{Pratyush Patel}, \bibinfo{person}{Esha Choukse}, \bibinfo{person}{Chaojie Zhang}, \bibinfo{person}{\'{I}\~{n}igo Goiri}, \bibinfo{person}{Brijesh Warrier}, \bibinfo{person}{Nithish Mahalingam}, {and} \bibinfo{person}{Ricardo Bianchini}.} \bibinfo{year}{2024}\natexlab{a}.
\newblock \showarticletitle{Characterizing Power Management Opportunities for LLMs in the Cloud}. In \bibinfo{booktitle}{\emph{Proceedings of the 29th ACM International Conference on Architectural Support for Programming Languages and Operating Systems, Volume 3}} (La Jolla, CA, USA) \emph{(\bibinfo{series}{ASPLOS '24})}. \bibinfo{publisher}{Association for Computing Machinery}, \bibinfo{address}{New York, NY, USA}, \bibinfo{pages}{207–222}.
\newblock
\showISBNx{9798400703867}
\urldef\tempurl%
\url{https://doi.org/10.1145/3620666.3651329}
\showDOI{\tempurl}


\bibitem[Patel et~al\mbox{.}(2024b)]%
        {splitwise2024}
\bibfield{author}{\bibinfo{person}{Pratyush Patel}, \bibinfo{person}{Esha Choukse}, \bibinfo{person}{Chaojie Zhang}, \bibinfo{person}{Aashaka Shah}, \bibinfo{person}{Íñigo Goiri}, \bibinfo{person}{Saeed Maleki}, {and} \bibinfo{person}{Ricardo Bianchini}.} \bibinfo{year}{2024}\natexlab{b}.
\newblock \showarticletitle{Splitwise: Efficient Generative LLM Inference Using Phase Splitting}. In \bibinfo{booktitle}{\emph{2024 ACM/IEEE 51st Annual International Symposium on Computer Architecture (ISCA)}}. \bibinfo{pages}{118--132}.
\newblock
\urldef\tempurl%
\url{https://doi.org/10.1109/ISCA59077.2024.00019}
\showDOI{\tempurl}


\bibitem[Peng et~al\mbox{.}(2024)]%
        {peng2024harnessingdramssdsustainable}
\bibfield{author}{\bibinfo{person}{Jie Peng}, \bibinfo{person}{Zhang Cao}, \bibinfo{person}{Huaizhi Qu}, \bibinfo{person}{Zhengyu Zhang}, \bibinfo{person}{Chang Guo}, \bibinfo{person}{Yanyong Zhang}, \bibinfo{person}{Zhichao Cao}, {and} \bibinfo{person}{Tianlong Chen}.} \bibinfo{year}{2024}\natexlab{}.
\newblock \bibinfo{title}{Harnessing Your DRAM and SSD for Sustainable and Accessible LLM Inference with Mixed-Precision and Multi-level Caching}.
\newblock
\newblock
\showeprint[arxiv]{2410.14740}~[cs.LG]
\urldef\tempurl%
\url{https://arxiv.org/abs/2410.14740}
\showURL{%
\tempurl}


\bibitem[Qiao et~al\mbox{.}(2024)]%
        {qiao2024conserveharvestinggpuslowlatency}
\bibfield{author}{\bibinfo{person}{Yifan Qiao}, \bibinfo{person}{Shu Anzai}, \bibinfo{person}{Shan Yu}, \bibinfo{person}{Haoran Ma}, \bibinfo{person}{Yang Wang}, \bibinfo{person}{Miryung Kim}, {and} \bibinfo{person}{Harry Xu}.} \bibinfo{year}{2024}\natexlab{}.
\newblock \bibinfo{title}{ConServe: Harvesting GPUs for Low-Latency and High-Throughput Large Language Model Serving}.
\newblock
\newblock
\showeprint[arxiv]{2410.01228}~[cs.DC]
\urldef\tempurl%
\url{https://arxiv.org/abs/2410.01228}
\showURL{%
\tempurl}


\bibitem[Qin et~al\mbox{.}(2024)]%
        {qin2024mooncakekvcachecentricdisaggregatedarchitecture}
\bibfield{author}{\bibinfo{person}{Ruoyu Qin}, \bibinfo{person}{Zheming Li}, \bibinfo{person}{Weiran He}, \bibinfo{person}{Mingxing Zhang}, \bibinfo{person}{Yongwei Wu}, \bibinfo{person}{Weimin Zheng}, {and} \bibinfo{person}{Xinran Xu}.} \bibinfo{year}{2024}\natexlab{}.
\newblock \bibinfo{title}{Mooncake: A KVCache-centric Disaggregated Architecture for LLM Serving}.
\newblock
\newblock
\showeprint[arxiv]{2407.00079}~[cs.DC]
\urldef\tempurl%
\url{https://arxiv.org/abs/2407.00079}
\showURL{%
\tempurl}


\bibitem[Shazeer et~al\mbox{.}(2017)]%
        {shazeer2017outrageouslylargeneuralnetworks}
\bibfield{author}{\bibinfo{person}{Noam Shazeer}, \bibinfo{person}{Azalia Mirhoseini}, \bibinfo{person}{Krzysztof Maziarz}, \bibinfo{person}{Andy Davis}, \bibinfo{person}{Quoc Le}, \bibinfo{person}{Geoffrey Hinton}, {and} \bibinfo{person}{Jeff Dean}.} \bibinfo{year}{2017}\natexlab{}.
\newblock \bibinfo{title}{Outrageously Large Neural Networks: The Sparsely-Gated Mixture-of-Experts Layer}.
\newblock
\newblock
\showeprint[arxiv]{1701.06538}~[cs.LG]
\urldef\tempurl%
\url{https://arxiv.org/abs/1701.06538}
\showURL{%
\tempurl}


\bibitem[Sheng et~al\mbox{.}(2023)]%
        {sheng2023flexgenhighthroughputgenerativeinference}
\bibfield{author}{\bibinfo{person}{Ying Sheng}, \bibinfo{person}{Lianmin Zheng}, \bibinfo{person}{Binhang Yuan}, \bibinfo{person}{Zhuohan Li}, \bibinfo{person}{Max Ryabinin}, \bibinfo{person}{Daniel~Y. Fu}, \bibinfo{person}{Zhiqiang Xie}, \bibinfo{person}{Beidi Chen}, \bibinfo{person}{Clark Barrett}, \bibinfo{person}{Joseph~E. Gonzalez}, \bibinfo{person}{Percy Liang}, \bibinfo{person}{Christopher Ré}, \bibinfo{person}{Ion Stoica}, {and} \bibinfo{person}{Ce Zhang}.} \bibinfo{year}{2023}\natexlab{}.
\newblock \bibinfo{title}{FlexGen: High-Throughput Generative Inference of Large Language Models with a Single GPU}.
\newblock
\newblock
\showeprint[arxiv]{2303.06865}~[cs.LG]
\urldef\tempurl%
\url{https://arxiv.org/abs/2303.06865}
\showURL{%
\tempurl}


\bibitem[Song et~al\mbox{.}(2024)]%
        {song2024powerinferfastlargelanguage}
\bibfield{author}{\bibinfo{person}{Yixin Song}, \bibinfo{person}{Zeyu Mi}, \bibinfo{person}{Haotong Xie}, {and} \bibinfo{person}{Haibo Chen}.} \bibinfo{year}{2024}\natexlab{}.
\newblock \bibinfo{title}{PowerInfer: Fast Large Language Model Serving with a Consumer-grade GPU}.
\newblock
\newblock
\showeprint[arxiv]{2312.12456}~[cs.LG]
\urldef\tempurl%
\url{https://arxiv.org/abs/2312.12456}
\showURL{%
\tempurl}


\bibitem[Stojkovic et~al\mbox{.}(2025)]%
        {stojkovic2025tapasthermalpowerawarescheduling}
\bibfield{author}{\bibinfo{person}{Jovan Stojkovic}, \bibinfo{person}{Chaojie Zhang}, \bibinfo{person}{Íñigo Goiri}, \bibinfo{person}{Esha Choukse}, \bibinfo{person}{Haoran Qiu}, \bibinfo{person}{Rodrigo Fonseca}, \bibinfo{person}{Josep Torrellas}, {and} \bibinfo{person}{Ricardo Bianchini}.} \bibinfo{year}{2025}\natexlab{}.
\newblock \bibinfo{title}{TAPAS: Thermal- and Power-Aware Scheduling for LLM Inference in Cloud Platforms}.
\newblock
\newblock
\showeprint[arxiv]{2501.02600}~[cs.DC]
\urldef\tempurl%
\url{https://arxiv.org/abs/2501.02600}
\showURL{%
\tempurl}


\bibitem[Stojkovic et~al\mbox{.}(2024)]%
        {stojkovic2024dynamollmdesigningllminference}
\bibfield{author}{\bibinfo{person}{Jovan Stojkovic}, \bibinfo{person}{Chaojie Zhang}, \bibinfo{person}{Íñigo Goiri}, \bibinfo{person}{Josep Torrellas}, {and} \bibinfo{person}{Esha Choukse}.} \bibinfo{year}{2024}\natexlab{}.
\newblock \bibinfo{title}{DynamoLLM: Designing LLM Inference Clusters for Performance and Energy Efficiency}.
\newblock
\newblock
\showeprint[arxiv]{2408.00741}~[cs.AI]
\urldef\tempurl%
\url{https://arxiv.org/abs/2408.00741}
\showURL{%
\tempurl}


\bibitem[Sun et~al\mbox{.}(2024)]%
        {llumnix2024}
\bibfield{author}{\bibinfo{person}{Biao Sun}, \bibinfo{person}{Ziming Huang}, \bibinfo{person}{Hanyu Zhao}, \bibinfo{person}{Wencong Xiao}, \bibinfo{person}{Xinyi Zhang}, \bibinfo{person}{Yong Li}, {and} \bibinfo{person}{Wei Lin}.} \bibinfo{year}{2024}\natexlab{}.
\newblock \showarticletitle{Llumnix: Dynamic Scheduling for Large Language Model Serving}. In \bibinfo{booktitle}{\emph{18th USENIX Symposium on Operating Systems Design and Implementation (OSDI 24)}}. \bibinfo{publisher}{USENIX Association}, \bibinfo{address}{Santa Clara, CA}, \bibinfo{pages}{173--191}.
\newblock
\showISBNx{978-1-939133-40-3}
\urldef\tempurl%
\url{https://www.usenix.org/conference/osdi24/presentation/sun-biao}
\showURL{%
\tempurl}


\bibitem[Team et~al\mbox{.}(2025)]%
        {theaibrixteam2025aibrixscalablecosteffectivelarge}
\bibfield{author}{\bibinfo{person}{The~AIBrix Team}, \bibinfo{person}{Jiaxin Shan}, \bibinfo{person}{Varun Gupta}, \bibinfo{person}{Le Xu}, \bibinfo{person}{Haiyang Shi}, \bibinfo{person}{Jingyuan Zhang}, \bibinfo{person}{Ning Wang}, \bibinfo{person}{Linhui Xu}, \bibinfo{person}{Rong Kang}, \bibinfo{person}{Tongping Liu}, \bibinfo{person}{Yifei Zhang}, \bibinfo{person}{Yiqing Zhu}, \bibinfo{person}{Shuowei Jin}, \bibinfo{person}{Gangmuk Lim}, \bibinfo{person}{Binbin Chen}, \bibinfo{person}{Zuzhi Chen}, \bibinfo{person}{Xiao Liu}, \bibinfo{person}{Xin Chen}, \bibinfo{person}{Kante Yin}, \bibinfo{person}{Chak-Pong Chung}, \bibinfo{person}{Chenyu Jiang}, \bibinfo{person}{Yicheng Lu}, \bibinfo{person}{Jianjun Chen}, \bibinfo{person}{Caixue Lin}, \bibinfo{person}{Wu Xiang}, \bibinfo{person}{Rui Shi}, {and} \bibinfo{person}{Liguang Xie}.} \bibinfo{year}{2025}\natexlab{}.
\newblock \bibinfo{title}{AIBrix: Towards Scalable, Cost-Effective Large Language Model Inference Infrastructure}.
\newblock
\newblock
\showeprint[arxiv]{2504.03648}~[cs.DC]
\urldef\tempurl%
\url{https://arxiv.org/abs/2504.03648}
\showURL{%
\tempurl}


\bibitem[{vLLM Team}(2023)]%
        {vllm_github}
\bibfield{author}{\bibinfo{person}{{vLLM Team}}.} \bibinfo{year}{2023}\natexlab{}.
\newblock \bibinfo{title}{{vLLM: A High-Throughput and Memory-Efficient Inference Engine for LLMs}}.
\newblock \bibinfo{howpublished}{\url{https://github.com/vllm-project/vllm}}.
\newblock
\newblock
\shownote{Accessed: 2025-04-11}.


\bibitem[Xu et~al\mbox{.}(2025)]%
        {xu2025moegenhighthroughputmoeinference}
\bibfield{author}{\bibinfo{person}{Tairan Xu}, \bibinfo{person}{Leyang Xue}, \bibinfo{person}{Zhan Lu}, \bibinfo{person}{Adrian Jackson}, {and} \bibinfo{person}{Luo Mai}.} \bibinfo{year}{2025}\natexlab{}.
\newblock \bibinfo{title}{MoE-Gen: High-Throughput MoE Inference on a Single GPU with Module-Based Batching}.
\newblock
\newblock
\showeprint[arxiv]{2503.09716}~[cs.DC]
\urldef\tempurl%
\url{https://arxiv.org/abs/2503.09716}
\showURL{%
\tempurl}


\bibitem[Xue et~al\mbox{.}(2025)]%
        {xue2025moeinfinityefficientmoeinference}
\bibfield{author}{\bibinfo{person}{Leyang Xue}, \bibinfo{person}{Yao Fu}, \bibinfo{person}{Zhan Lu}, \bibinfo{person}{Luo Mai}, {and} \bibinfo{person}{Mahesh Marina}.} \bibinfo{year}{2025}\natexlab{}.
\newblock \bibinfo{title}{MoE-Infinity: Efficient MoE Inference on Personal Machines with Sparsity-Aware Expert Cache}.
\newblock
\newblock
\showeprint[arxiv]{2401.14361}~[cs.LG]
\urldef\tempurl%
\url{https://arxiv.org/abs/2401.14361}
\showURL{%
\tempurl}


\bibitem[Yu et~al\mbox{.}(2022)]%
        {orca2022}
\bibfield{author}{\bibinfo{person}{Gyeong-In Yu}, \bibinfo{person}{Joo~Seong Jeong}, \bibinfo{person}{Geon-Woo Kim}, \bibinfo{person}{Soojeong Kim}, {and} \bibinfo{person}{Byung-Gon Chun}.} \bibinfo{year}{2022}\natexlab{}.
\newblock \showarticletitle{Orca: A Distributed Serving System for {Transformer-Based} Generative Models}. In \bibinfo{booktitle}{\emph{16th USENIX Symposium on Operating Systems Design and Implementation (OSDI 22)}}. \bibinfo{publisher}{USENIX Association}, \bibinfo{address}{Carlsbad, CA}, \bibinfo{pages}{521--538}.
\newblock
\showISBNx{978-1-939133-28-1}
\urldef\tempurl%
\url{https://www.usenix.org/conference/osdi22/presentation/yu}
\showURL{%
\tempurl}


\bibitem[Yuan et~al\mbox{.}(2025)]%
        {yuan2025nativesparseattentionhardwarealigned}
\bibfield{author}{\bibinfo{person}{Jingyang Yuan}, \bibinfo{person}{Huazuo Gao}, \bibinfo{person}{Damai Dai}, \bibinfo{person}{Junyu Luo}, \bibinfo{person}{Liang Zhao}, \bibinfo{person}{Zhengyan Zhang}, \bibinfo{person}{Zhenda Xie}, \bibinfo{person}{Y.~X. Wei}, \bibinfo{person}{Lean Wang}, \bibinfo{person}{Zhiping Xiao}, \bibinfo{person}{Yuqing Wang}, \bibinfo{person}{Chong Ruan}, \bibinfo{person}{Ming Zhang}, \bibinfo{person}{Wenfeng Liang}, {and} \bibinfo{person}{Wangding Zeng}.} \bibinfo{year}{2025}\natexlab{}.
\newblock \bibinfo{title}{Native Sparse Attention: Hardware-Aligned and Natively Trainable Sparse Attention}.
\newblock
\newblock
\showeprint[arxiv]{2502.11089}~[cs.CL]
\urldef\tempurl%
\url{https://arxiv.org/abs/2502.11089}
\showURL{%
\tempurl}


\bibitem[Zhang et~al\mbox{.}(2023)]%
        {zhang2023h2oheavyhitteroracleefficient}
\bibfield{author}{\bibinfo{person}{Zhenyu Zhang}, \bibinfo{person}{Ying Sheng}, \bibinfo{person}{Tianyi Zhou}, \bibinfo{person}{Tianlong Chen}, \bibinfo{person}{Lianmin Zheng}, \bibinfo{person}{Ruisi Cai}, \bibinfo{person}{Zhao Song}, \bibinfo{person}{Yuandong Tian}, \bibinfo{person}{Christopher Ré}, \bibinfo{person}{Clark Barrett}, \bibinfo{person}{Zhangyang Wang}, {and} \bibinfo{person}{Beidi Chen}.} \bibinfo{year}{2023}\natexlab{}.
\newblock \bibinfo{title}{H$_2$O: Heavy-Hitter Oracle for Efficient Generative Inference of Large Language Models}.
\newblock
\newblock
\showeprint[arxiv]{2306.14048}~[cs.LG]
\urldef\tempurl%
\url{https://arxiv.org/abs/2306.14048}
\showURL{%
\tempurl}


\bibitem[Zhao et~al\mbox{.}(2024)]%
        {MLSYS2024_5edb57c0}
\bibfield{author}{\bibinfo{person}{Yilong Zhao}, \bibinfo{person}{Chien-Yu Lin}, \bibinfo{person}{Kan Zhu}, \bibinfo{person}{Zihao Ye}, \bibinfo{person}{Lequn Chen}, \bibinfo{person}{Size Zheng}, \bibinfo{person}{Luis Ceze}, \bibinfo{person}{Arvind Krishnamurthy}, \bibinfo{person}{Tianqi Chen}, {and} \bibinfo{person}{Baris Kasikci}.} \bibinfo{year}{2024}\natexlab{}.
\newblock \showarticletitle{Atom: Low-Bit Quantization for Efficient and Accurate LLM Serving}. In \bibinfo{booktitle}{\emph{Proceedings of Machine Learning and Systems}}, \bibfield{editor}{\bibinfo{person}{P.~Gibbons}, \bibinfo{person}{G.~Pekhimenko}, {and} \bibinfo{person}{C.~De Sa}} (Eds.), Vol.~\bibinfo{volume}{6}. \bibinfo{pages}{196--209}.
\newblock
\urldef\tempurl%
\url{https://proceedings.mlsys.org/paper_files/paper/2024/file/5edb57c05c81d04beb716ef1d542fe9e-Paper-Conference.pdf}
\showURL{%
\tempurl}


\bibitem[Zheng et~al\mbox{.}(2024)]%
        {zheng2024sglangefficientexecutionstructured}
\bibfield{author}{\bibinfo{person}{Lianmin Zheng}, \bibinfo{person}{Liangsheng Yin}, \bibinfo{person}{Zhiqiang Xie}, \bibinfo{person}{Chuyue Sun}, \bibinfo{person}{Jeff Huang}, \bibinfo{person}{Cody~Hao Yu}, \bibinfo{person}{Shiyi Cao}, \bibinfo{person}{Christos Kozyrakis}, \bibinfo{person}{Ion Stoica}, \bibinfo{person}{Joseph~E. Gonzalez}, \bibinfo{person}{Clark Barrett}, {and} \bibinfo{person}{Ying Sheng}.} \bibinfo{year}{2024}\natexlab{}.
\newblock \bibinfo{title}{SGLang: Efficient Execution of Structured Language Model Programs}.
\newblock
\newblock
\showeprint[arxiv]{2312.07104}~[cs.AI]
\urldef\tempurl%
\url{https://arxiv.org/abs/2312.07104}
\showURL{%
\tempurl}


\bibitem[Zhong et~al\mbox{.}(2024)]%
        {distserve}
\bibfield{author}{\bibinfo{person}{Yinmin Zhong}, \bibinfo{person}{Shengyu Liu}, \bibinfo{person}{Junda Chen}, \bibinfo{person}{Jianbo Hu}, \bibinfo{person}{Yibo Zhu}, \bibinfo{person}{Xuanzhe Liu}, \bibinfo{person}{Xin Jin}, {and} \bibinfo{person}{Hao Zhang}.} \bibinfo{year}{2024}\natexlab{}.
\newblock \showarticletitle{{DistServe}: Disaggregating Prefill and Decoding for Goodput-optimized Large Language Model Serving}. In \bibinfo{booktitle}{\emph{18th USENIX Symposium on Operating Systems Design and Implementation (OSDI 24)}}. \bibinfo{publisher}{USENIX Association}, \bibinfo{address}{Santa Clara, CA}, \bibinfo{pages}{193--210}.
\newblock
\showISBNx{978-1-939133-40-3}
\urldef\tempurl%
\url{https://www.usenix.org/conference/osdi24/presentation/zhong-yinmin}
\showURL{%
\tempurl}


\end{thebibliography}

\end{document}